\documentclass{aa}  
\usepackage{natbib,twoopt}
\usepackage[breaklinks=true]{hyperref} 
\bibpunct{(}{)}{;}{a}{}{,}             
\makeatletter
  \newcommandtwoopt{\citeads}[3][][]{\href{http://adsabs.harvard.edu/abs/#3}%
    {\def\hyper@linkstart##1##2{}%
     \let\hyper@linkend\@empty\citealp[#1][#2]{#3}}}
  \newcommandtwoopt{\citepads}[3][][]{\href{http://adsabs.harvard.edu/abs/#3}%
    {\def\hyper@linkstart##1##2{}%
     \let\hyper@linkend\@empty\citep[#1][#2]{#3}}}
  \newcommandtwoopt{\citetads}[3][][]{\href{http://adsabs.harvard.edu/abs/#3}%
    {\def\hyper@linkstart##1##2{}%
     \let\hyper@linkend\@empty\citet[#1][#2]{#3}}}
  \newcommandtwoopt{\citeyearads}[3][][]%
    {\href{http://adsabs.harvard.edu/abs/#3}
    {\def\hyper@linkstart##1##2{}%
     \let\hyper@linkend\@empty\citeyear[#1][#2]{#3}}}
\makeatother

\usepackage{adjustbox}
\usepackage{graphicx}
\usepackage{txfonts}
\usepackage{tabularx, blindtext}
\usepackage{rotating}
\usepackage{pdflscape}
\usepackage{inputenc}
\usepackage{caption}
\usepackage{longtable}
\usepackage{lscape}
\usepackage{subcaption}
\DeclareCaptionFormat{continued}{#1 (continued)#2#3\par}

\begin{document} 
\title{Inspection of 19 globular cluster candidates in the Galactic bulge with the VVV survey}
   \author{E. R. Garro \inst{1}
          \and
          D.  Minniti\inst{1,2}
          \and
          M.  Gómez\inst{1}
          \and
           J.  Alonso-García\inst{3,4}    
         \and
         V. Ripepi\inst{5}
         \and
         J.~G.~Fernández-Trincado\inst{6}
         \and
         F.~Vivanco~Cádiz\inst{1}
          }
   \institute{Departamento de Ciencias Físicas, Facultad de Ciencias Exactas, Universidad Andres Bello, Fernández Concha 700, Las Condes, Santiago, Chile
   \and
 Vatican Observatory, Vatican City State, V-00120, Italy
 \and
 Centro de Astronomía (CITEVA), Universidad de Antofagasta, Av. Angamos 601, Antofagasta, Chile
 \and
 Millennium Institute of Astrophysics , Nuncio Monse\~nor Sotero Sanz 100, Of. 104, Providencia, Santiago, Chile
 \and
 INAF-Osservatorio Astronomico di Capodimonte, Salita Moiariello 16, 80131 Naples, Italy
 \and
Instituto de Astronomía, Universidad Católica del Norte, Av. Angamos 0610, Antofagasta, Chile
}
  \date{Received July 17,  2021 ; Accepted November 11, 2021}

 
  \abstract
   {
The census of the globular clusters (GCs) in the Milky Way (MW) is still a work in progress.  It has been possible to discover many new star clusters both in the Galactic disk and bulge thanks to the advent of new deep surveys. Unfortunately, many of these new candidates are not yet studied in detail,  leaving a veil on their real physical nature.}
   {
   We explore the nature of 19 new GC candidates in the Galactic bulge,  based on the analysis of their colour-magnitude diagrams (CMDs) in the near-IR,  using the VISTA Variables in the Via Láctea Survey (VVV) database. We estimate their main astrophysical parameters: reddening and extinction, distance, total luminosity, mean
cluster proper motions (PMs),  metallicity and age.
   }
   {
We obtain the cluster catalogues including the likely cluster members by applying a decontamination procedure on the observed CMDs, based upon the vector PM diagrams from VIRAC2.  We adopt near-IR reddening maps in order to calculate the reddening and extinction for each cluster, and then estimate the distance moduli and heliocentric distances. Metallicities and ages are evaluated by fitting theoretical stellar isochrones. We calculate also their luminosities in comparison with known Galactic GCs. 
   }
   {
We estimate a wide reddening range of the $0.25 \leqslant  E(J-K_s) \leqslant  2.0$ mag and extinction $0.11 \leqslant A_{Ks} \leqslant  0.86$ mag for the sample clusters as expected in the bulge regions.  The range of heliocentric distances is $6.8\leqslant D\leqslant  11.4$ kpc.  This allows us to place these clusters between 0.56 and 3.25 kpc from the Galactic centre, assuming $R_{\odot}=8.2$ kpc.  Also, their PMs are kinematically similar to the typical motion of the Galactic bulge,  apart from VVV-CL160, which shows different PMs.  We also derive their metallicities and ages, finding $-1.40 \leqslant$ [Fe/H] $\leqslant 0.0$ dex and $t\approx 8-13$ Gyr respectively.  The luminosities are calculated both in $K_s-$ and $V-$bands,  recovering $-3.4 \leqslant M_V \leqslant -7.5$.  We also examine the possible RR Lyrae members found in the cluster fields.
}
{
Based on their positions,  kinematics, metallicities and ages and comparing our results with  the literature,  we conclude that 9 candidates are real GCs,  7 need more observations to be fully confirmed as GCs, whereas 3 candidates are discarded for being younger open clusters.}

\titlerunning {Inspection of 19 globular cluster candidates in the Galactic bulge with the VVV Survey}
\authorrunning {E.R. Garro et al.}
   \maketitle

\section{Introduction}
\label{introduction}
As suggested by the $\Lambda$CDM model,  globular clusters (GCs) are the first stellar associations formed in the early Universe \citep{phipps_khochfar_lisa_varri_2019},  thus their physical properties are strongly linked with those of their host galaxies \citep{Brodie2006}.  These ancient star clusters have survived destruction over long galactic histories to become the fossil relics of the earliest epoch of galaxy formation \citep{Kerber2019, Ferraro2021}. Consequently, they represent powerful tools to investigate on the formation and evolution of galaxies.  However,  the information coming from the analysis of handful of Milky Way (MW) GCs may still be incomplete, either because some of them are very faint, too small or widespread, or embedded by dust and thus difficult to find.  Indeed,  most of the ``missing'' GCs are
likely located in obscured regions,  close to the Galactic centre or near the Galactic plane on the far side of the Galaxy \citep{Minniti_2017a,Garro_2020,Garro2021}.\\
 
Our long term main goal is to search for the hidden of these GCs in order to reconstruct the formation history of our Galaxy, focussing on the Galactic bulge.  For this purpose, in this paper we deal with the characterization of 19 GC candidates located towards the MW bulge.  This represents a challenge since these regions are affected by differential reddening and high stellar density.  
The best way to overcome these problems is to use near-infrared (IR) observations, given that red giant stars have their emission peak in the near-IR and the interstellar dust becomes nearly transparent at these wavelengths.  The identification of new candidates and their subsequent confirmation as real GCs can be facilitated at these spectral regions (for instance UKS 1 and VVV-CL001; \citealt{Fernandez_Trincado_2020} and \citealt{Fernandez_Trincado_2021}, just to mention a few).  A step forwards has been possible thanks to the VISTA Variables in the Via Láctea (VVV) survey and its eXtension (VVVX).  More than 300-star cluster candidates have been discovered in the Galactic bulge and disk and only a fraction of them have been analysed (\citealt{Froebrich2007,Borissova2014,Minniti_2017,Camargo_2018,Palma2019,CamargoMinniti_2019, Garro_2020,Garro2021, Minniti2021_cl160,Obasi2021},  just to mention a few).\\

A variety of techniques have been used in the literature to identify potential star clusters against the background fields in order to unveil their real nature.  For example, \cite{Minniti_2017a} built density maps using only the red giants, and identified the apparent over-densities to mark the
location of GC candidates. The identification of the over-densities was done by visual inspection, based on the size of each over-density and considering the typical size of known Galactic GCs ($\sim 2-5'$).  Furthermore, they compared the CMD of potential candidates with those of well-characterised Galactic GCs as well as their respective background fields, in order to verify that the cluster RGB appear tighter than that observed in the background.  
Another method used to discriminate field stars from the cluster candidate is the statistical decontamination, as done by \cite{PALMA201650},  \cite{Minniti2011} and \cite{Garro2021}, which subtracts from the cluster CMD the stars that fall in the same intervals of
colour and magnitude of the background CMDs.  \\
However, not all over-densities are true clusters. They may be simply grouping of stars or statistical fluctuations of the projected stellar density in the plane of the sky \citep{Gran2019,Palma2019,Minniti2021c}. \\
Hence,  one of the most reliable methods capable of confirming or discarding the cluster nature is the kinematical analysis. The coherent stellar motions ensure the cluster membership, and allows to separate the stars belonging to an association from those randomly distributed in the field.  Several works (\citealt{Zoccali2002,Sariya2015,Baumgardt2018, Garro_2020,Garro2021,Obasi2021,Minniti2021_cl160}, just to mention a few) have approached this technique, especially exploiting the high-precision of Gaia proper motions (PMs), confirming, in many cases, the GC nature of some objects.  A complementary proof used to identify the cluster nature is the presence of RR Lyrae stars, which are good tracers of old stellar populations (e.g., \citealt{Minniti_2017}).\\

In Section \ref{observationaldataset}, we briefly describe the datasets used in this work. In Section \ref{pmdecontaminationprocedure}, we delineate the performed decontamination procedure, and two test done in order to confirm the existence of the cluster.  The methods to estimate the main physical parameters are explained in Section \ref{estimationparameters}. In Section \ref{searchingRRL}, we searched for RR Lyrae stars belonging to the clusters in order to confirm the GC nature.  Comparison with the literature and final notes are reported in Section \ref{individualnotes}.  Summary and conclusions are presented in Section \ref{summaryconclusion}.

\section{Observational dataset}
\label{observationaldataset}
\begin{figure*}
\centering
\includegraphics[width=6cm, height=6cm]{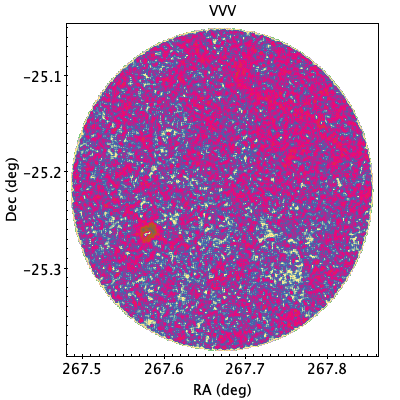} 
\includegraphics[width=6cm, height=6cm]{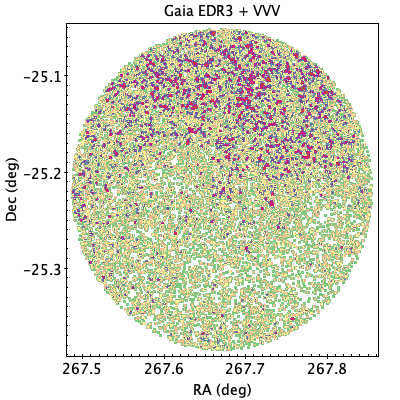} 
\includegraphics[width=6cm, height=6cm]{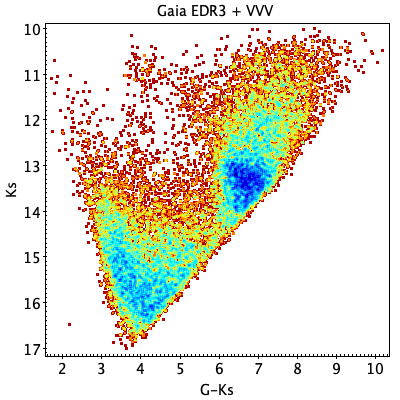} 
\caption{VVV (on the left) and Gaia+VVV (in the middle) density maps for a $r=10'$ field centred on VVV-CL150, used as representative cluster. The redder areas are representative of over-densities, while the yellower areas are lower densities.  The  Gaia EDR3 + VVV CMD (on the right) is displayed for the same field. }
\label{example}
\end{figure*}

We use the near-IR dataset from the VVV survey \citep{Minniti2010,Saito2012}, acquired with the VISTA InfraRed CAMera (VIRCAM) at the 4.1m wide-field Visible and Infrared Survey Telescope for Astronomy (VISTA; \citealt{Emerson2010}) at ESO Paranal Observatory.  The VVV data are reduced at the Cambridge Astronomical Survey Unit \citep{Irwin2004} and further processing and archiving is performed with the VISTA Data Flow System \citep{Cross2012} by the Wide-Field Astronomy Unit and made available at the VISTA Science Archive.  We use preliminary data from VIRAC version 2 (VIRAC-2; \citealt{Smith2018}), described in detail in Smith et al. (2021, in preparation). In summary, VIRAC-2 identifies the sources in the VVV images and extracts their photometry through point spread function (PSF)-fitting techniques using DoPhot \citep{Schechter1993, 2012AJ....143...70A,2018A&A...619A...4A}. Their astrometry is calibrated to the Gaia DR 2 \citep{2018A&A...616A...1G} astrometric reference system, and their photometry are calibrated into the VISTA magnitude system \citep{Gonzalez_Fernandez2018} against the Two Micron All Sky Survey (2MASS; \citealt{2006AJ....131.1163S}) using a globally optimised model of frame-by-frame zero points plus an illumination correction.  \\
We use also the 2MASS catalogue in order to include brighter stars with $K_s<11$ mag, since these stars are saturated in the VVV images.  Additionally,  we transformed the 2MASS photometry into the VISTA magnitude scale \citep{Gonzalez_Fernandez2018}, 
since the magnitude scale is different in the two photometric systems. \\
Moreover,  we also explore the optical Gaia Early Data Release 3 (EDR3; \citealt{Gaia2020}) for eight GC candidates: FSR~0009,  FSR~1775, VVV-CL131, VVV-CL143, ESO 393-12,  ESO 456-09, DB044 and Kronberger 49, for which we obtain suitable CMDs and reliable physical parameters.  Nevertheless, we cannot use the Gaia EDR3 data for the other targets because the CMDs are largely affected by the differential reddening, which distorts all optical evolutionary sequences.  For this purpose we use the contaminated VVV-CL150 field as representative, making a comparison between the near-IR and optical wavelengths, in order to show how the dust is opaque in the optical image and does not allow seeing beyond, also it distorts the VVV-CL150 CMD, as demonstrated by Fig. \ref{example}.  For that reason, we cannot use the optical Gaia EDR3 dataset for the other star clusters.

\begin{figure*}
\centering
\includegraphics[width=14cm, height=10cm]{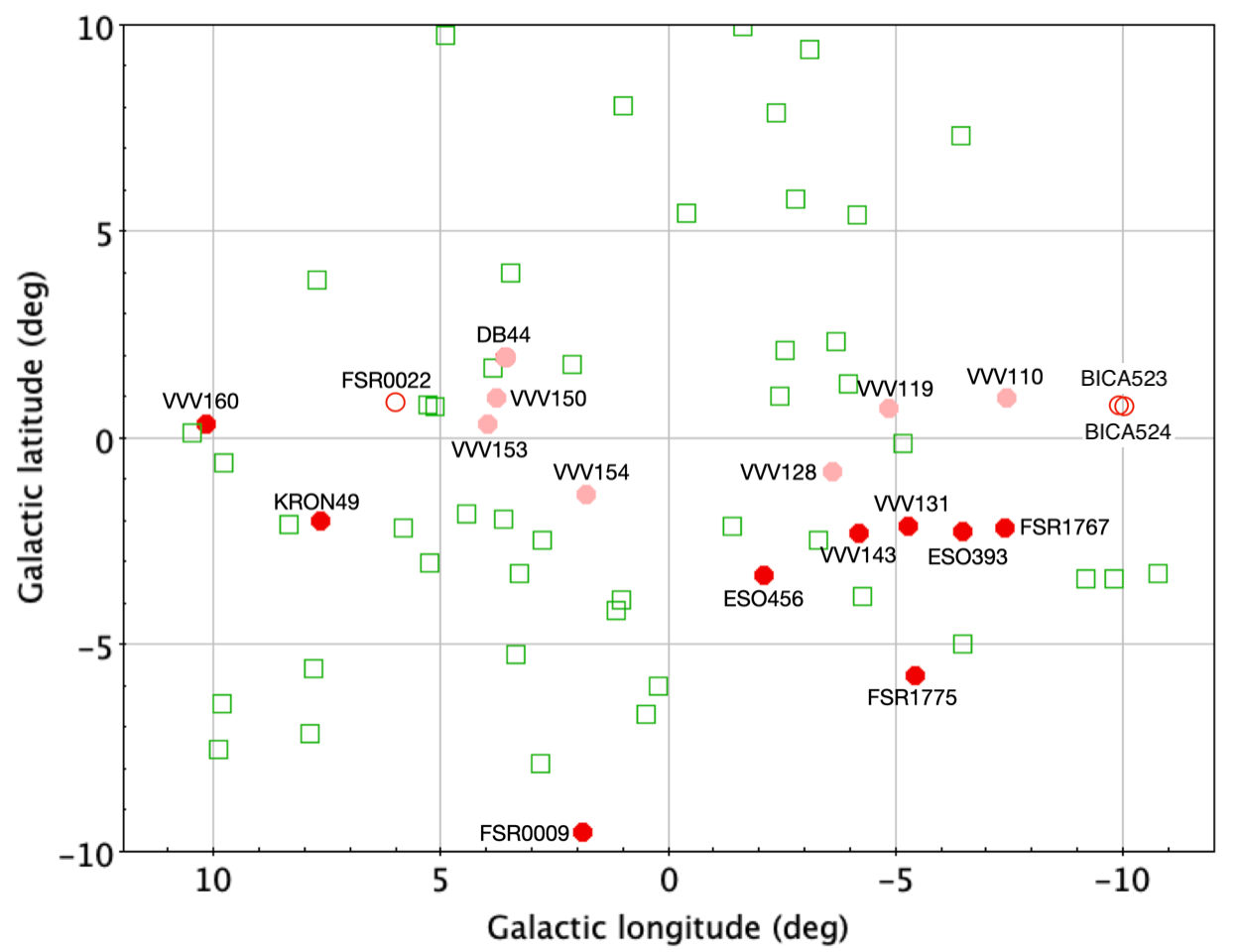} 
\caption{Map depicting Galactic latitude versus longitude for our sample. The confirmed GCs are highlighted as red points, whereas transparent-red points are used to point out the \textit{unconfirmed} GCs. We also mark with open red circles the open clusters excluded from our analysis.  We show known GGs from \cite{VasilievBaumgardt2021} with open green squares, for comparison. We use shorter cluster identifications only for our convenience.}
\label{position}
\end{figure*}

\section{PM-decontamination procedure}
\label{pmdecontaminationprocedure}
We investigate 19 GC candidates located towards the Galactic bulge, which are situated within a projected angular distance of $10^{\circ}$ from the Galactic centre,  as shown in Fig. \ref{position}. \\
For the VVV, 2MASS and Gaia EDR3 data, we perform a decontamination procedure in order to build a clean, decontaminated catalogue of probable cluster members following the same procedure applied by \cite{Garro_2020,Garro2021}.  

\begin{figure*}
\centering
\includegraphics[width=18cm, height=5cm]{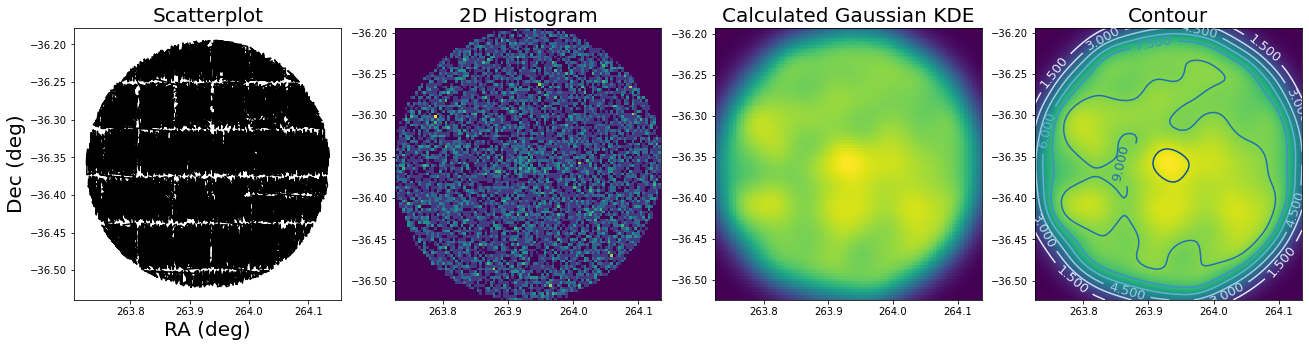} 
\includegraphics[width=18cm, height=5cm]{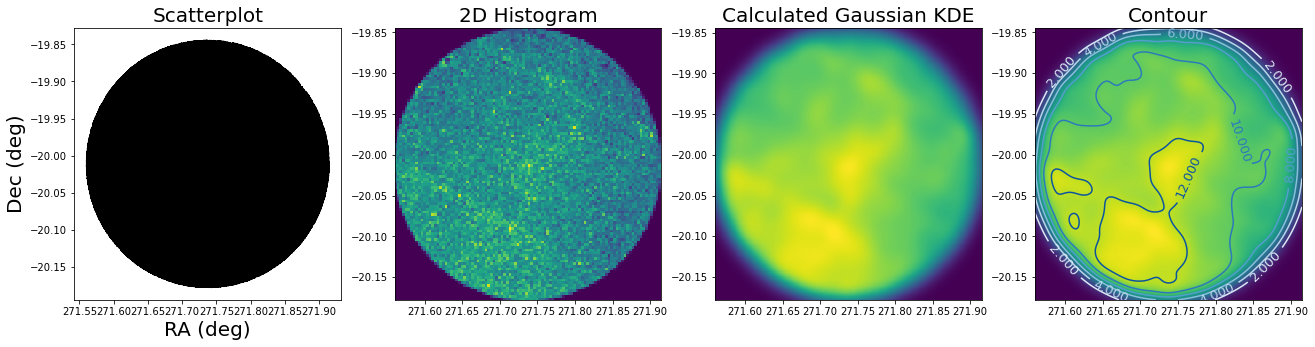} 
\caption{KDE technique applied to select the likely size of a cluster during the decontamination procedure. From left to right: a scatterplot, 2D Histogram, Gaussian KDE and 2D density with contours are shown for FSR 1767 (top panels) and VVV-CL160 (bottom panels) fields, used as representative. Green and yellow areas are illustrative of over-densities, while the blue areas show
lower densities.}
\label{kde}
\end{figure*}

\begin{figure}
\centering
\includegraphics[width=8cm, height=8cm]{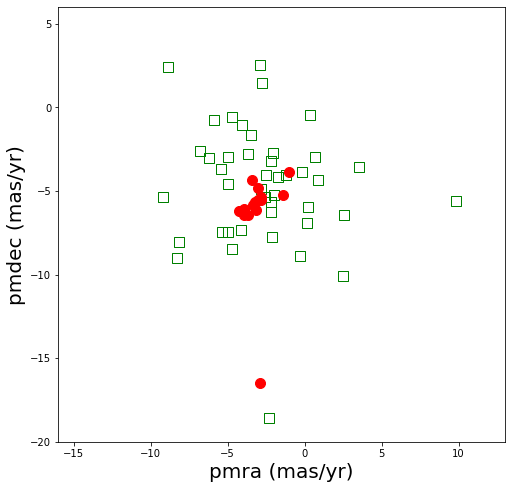} 
\caption{Vector PMs diagram is shown to demonstrate that the kinematics of the star clusters here analysed are consistent with the known GCs located in the direction of the MW bulge.  Our targets are highlight with red points,  whereas the known GCs, located in the bulge within $-10<l\ (deg) <10$ and $-10<b\ (deg) <10$, are represented by open green squares. }
\label{pm_comp}
\end{figure}

Briefly, we underline the main steps taken to obtain the final cluster catalogues.  First of all,  we include all stars with parallax values smaller than 0.5 mas, discarding all nearby stars.  It is very difficult to derive the size of the clusters, especially in such crowded regions and for very weak clusters.  For that reason,  we derive the cluster size adopting two methods: \textit{(i)} we inspect the density diagrams as a function of sky position, in order to visually identify the cluster dimensions,  \textit{(ii)} we apply the Gaussian Kernel Density Estimate (KDE; e.g.  \citealt{Rosenblatt1956,Parzen1962}).  Summarizing, considering the scatterplot on the left hand side of Fig. \ref{kde}, overlapping points make the figure hard to read.  Even worse, it is impossible to determine how many data points are in each position. In this case, a possible solution is to cut the plotting window into several bins (100-300 bins), and colour-code the number of data points in each bin. Following the shape of the bin, this makes 2D histogram. Then, it is possible to obtain a smoother result using the Gaussian KDE. Its representation is called a 2D density plot, and we add a contour to denote each density step. We use both density maps and KDE contours to identify the optimal cluster size, selecting the higher density area around the cluster centre.  Therefore, we select all stars within the typical GC radius $r<3'$ from the cluster centre,  however we adopt a smaller radius $r<1.8'$ for the cluster VVV-CL153, and a larger radius for the clusters FSR~0009 and VVV-CL154,  $r<3.6'$ and $r<4.2'$ respectively.  Then,  we construct and inspect the vector PM diagram (VPM),  as 2D histogram, of the relative cluster PMs, as shown in Fig. \ref{vpm}.  Given that our decontamination procedure is based on the PM-selection, we also construct histograms of PM in RA ($\mu_{\alpha_{\ast}}$,  blue histograms) and in Dec ($\mu_{\delta}$, red histograms) to clearly recognise the peak (against the noise) of the two distributions. Then, we visually identify the cluster peak in the VPM diagram, estimating the mean cluster PMs using the $\sigma-$clipping technique.  All mean cluster PMs are kinematically in agreement with that of the Galactic bulge, except VVV-CL160 which deviates from the typical bulge values \citep{Minniti2021_cl160}.  The mean PMs for our GC sample are shown in Figure \ref{pm_comp} (highlighted with red circles), along with the mean PMs of all other known MW bulge GCs measured by \cite{VasilievBaumgardt2021}.  Indeed, this figure illustrates that the new GCs show kinematics that are akin with the rest of the MW GCs.  Finally, we select all stars within 1.0, 1.5 or 2.0 mas yr$^{-1}$ from the mean cluster PMs. The choices to adopt different radii and PM selections ($\sigma_{PM}$) for all star clusters arise from holding only stars that are likely members of the clusters, minimizing the likelihood of including outer stars as well as of excluding member stars.  A preliminary step, made exclusively for clusters analysed in the optical passband, is to join the Gaia EDR3 and the VVV datasets (using a $0.5''$ matching radius) before the quoted above decontamination procedure.\\
The positions, radii and PMs selections are summarised in Table \ref{DP} for every candidate.\\

\begin{figure*}
     \centering
     \begin{subfigure}[b]{0.4\textwidth}
         \centering
         \includegraphics[width=\textwidth, height=2.5cm]{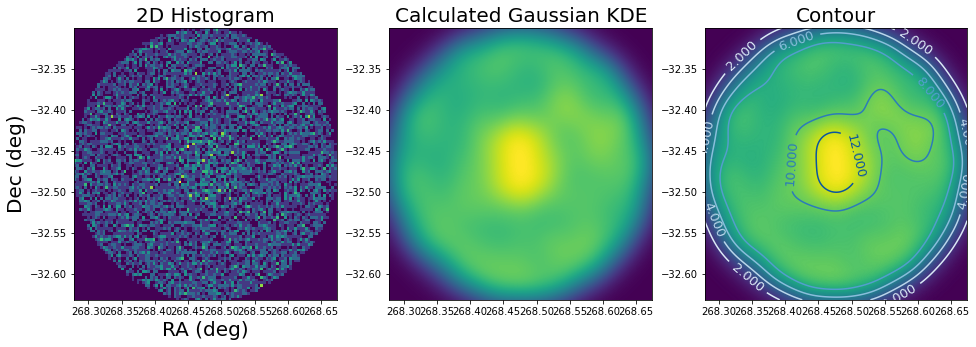} 
		\includegraphics[width=\textwidth, height=2.5cm]{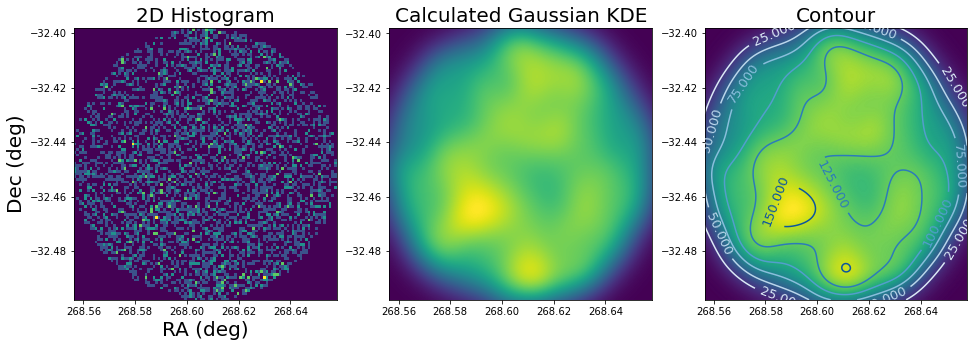} 
         \caption{ESO 456-09}
         \label{fig:ESO45609}
     \end{subfigure}
     \hfill
     \begin{subfigure}[b]{0.4\textwidth}
         \centering
         \includegraphics[width=\textwidth, height=2.5cm]{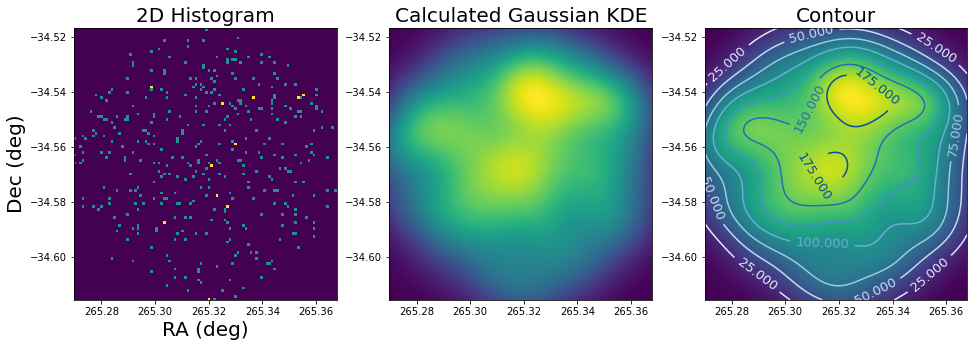}
          \includegraphics[width=\textwidth, height=2.5cm]{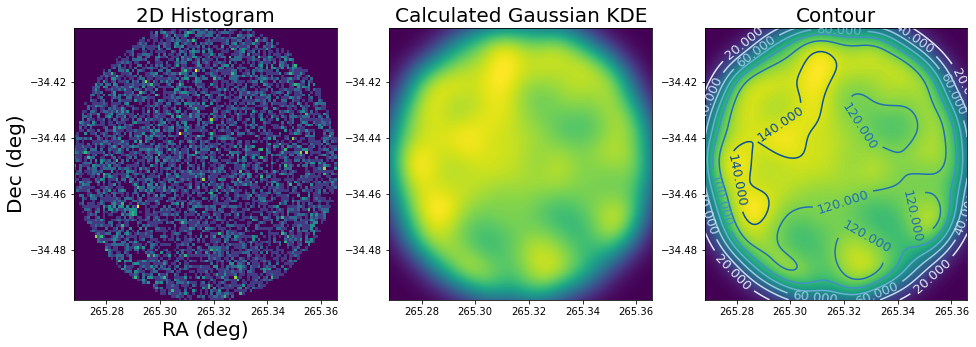}
         \caption{VVV-CL131}
         \label{fig:VVVCL131}
     \end{subfigure}
     \hfill
     \begin{subfigure}[b]{0.4\textwidth}
         \centering
         \includegraphics[width=\textwidth, height=2.5cm]{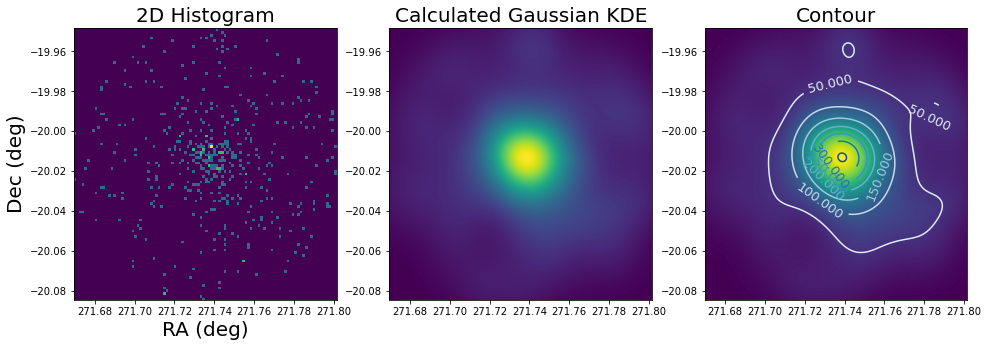}
          \includegraphics[width=\textwidth, height=2.5cm]{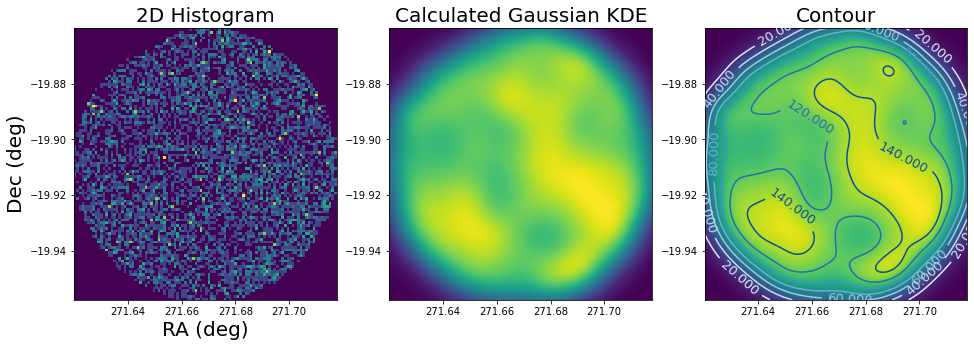}
         \caption{VVV-CL160}
         \label{fig:VVV-CL160}
     \end{subfigure}
     \hfill
     \begin{subfigure}[b]{0.4\textwidth}
         \centering
         \includegraphics[width=\textwidth, height=2.5cm]{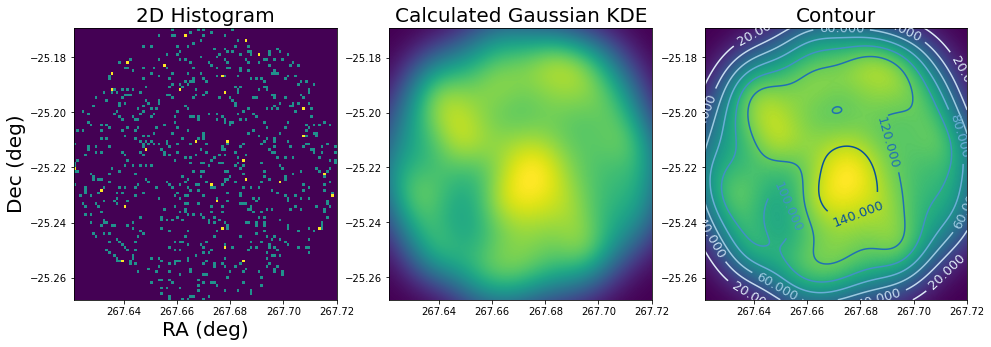}
          \includegraphics[width=\textwidth, height=2.5cm]{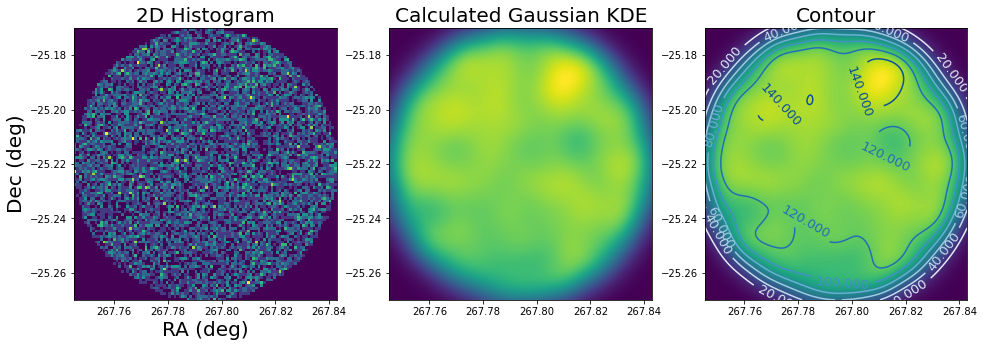}
         \caption{VVV-CL150}
         \label{fig:VVVCL150}
     \end{subfigure}
        \caption{Spatial distribution for four clusters (on the top) and their relative fields (on the 					bottom) selected at $ 5' \lesssim r\lesssim 8'$ from the cluster centre, taken as representative. We use the KDE technique, as in Fig. \ref{kde}, in order to better distinguish the over-densities (yellow/green colours) from lower-densities (blue colours). }
        \label{spatial}
\end{figure*}

Finally, we perform two tests in order to confirm that our targets are real clusters, even if they have previously been catalogued as such in the literature (see Sect. \ref{individualnotes}). \\
First, we compare the spatial distribution (Fig. \ref{spatial}) of our clusters (using the decontaminated catalogues) with those of a sample of field stars (cleaned by nearby stars) selected at $ 5' \lesssim r \lesssim 8' $ from the relative centre of the cluster. We apply again the KDE technique in order to highlight over-densities.  Clearly, the spatial distributions between the clusters and the fields are very different, since a clear central over-density is visible for each cluster, which decrease towards the cluster outskirts; on the other hand, a lumpy and non-homogeneous distribution is seen for the fields, as expected.  However, as we can appreciate from Fig. \ref{spatial}, the central over-density is unequivocal for ESO 456-09 (Fig. \ref{fig:ESO45609}), VVV-CL160 (Fig. \ref{fig:VVV-CL160}) and VVV-CL150 (Fig . \ref{fig:VVVCL150}), but it is less clear for VVVCL-131 (Fig. \ref{fig:VVVCL131}). This is related to the fact that some of these clusters have a sparse distribution, as they probably survived dynamical processes, typical of regions of high density such as the Galactic bulge. \\
As a second test, we compare our sample with field stars constructing both PMs histograms and CMDs (Fig. \ref{cl+field}).  We expect that field samples include bulge+disk stars, but it is very difficult to split them and to treat them separately, since bulge and disk stars may have similar motions.  As done for the clusters, we perform the $\sigma$-clipping technique in order to derive the mean field PMs, listed in Table  \ref{comparison}. We find that the means of the PMs do not differ by much with respect to clusters, but they show much larger dispersions than the cluster sample, as expected. In this case, the CMDs represent powerful tools to indicate if a star cluster is present or not. Indeed, comparing the clusters and field CMDs, many features, such as RC, narrow RGB, RGB-bump (RGBB), are clear in many cluster CMDs, while they are not visible in any field CMD.  Thus, all these features strongly support the real cluster nature of our objects.

\begin{figure*}
     \centering
     \begin{subfigure}[b]{0.3\textwidth}
         \centering
         \includegraphics[width=\textwidth]{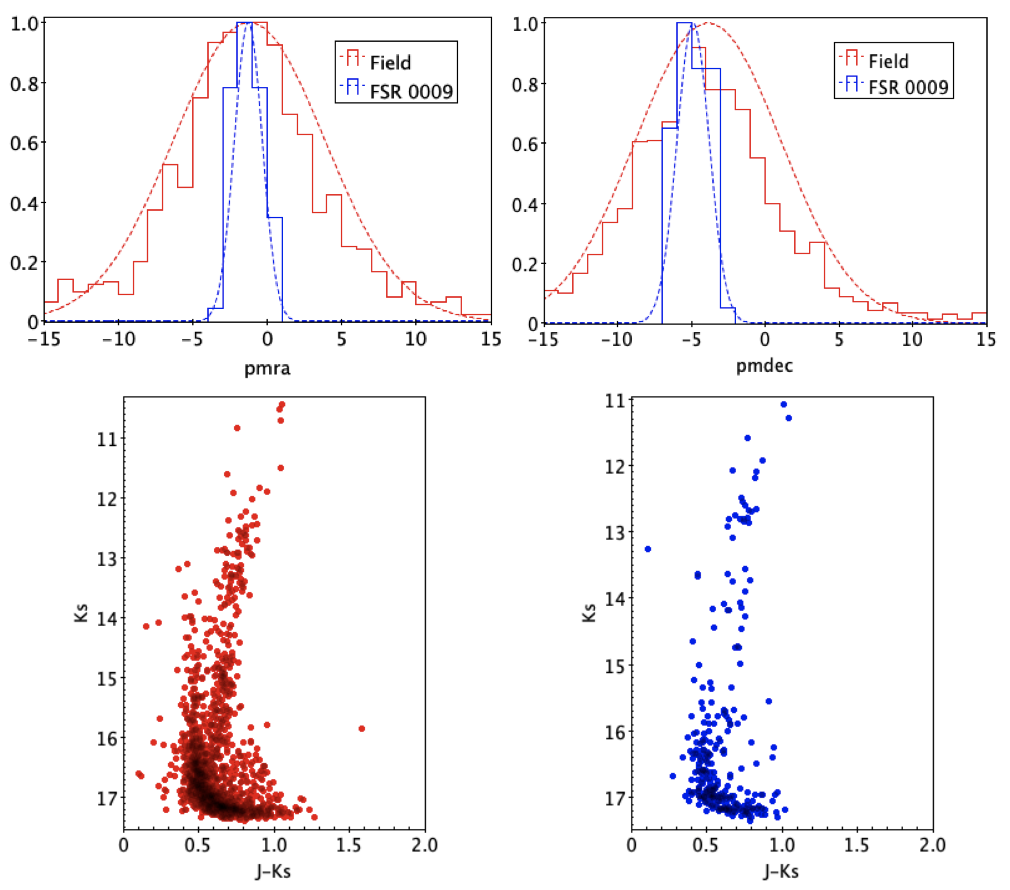} 
         \caption{FSR 0009}
         \label{fig:FSR0009+f}
     \end{subfigure}
     \hfill
     \begin{subfigure}[b]{0.3\textwidth}
         \centering
         \includegraphics[width=\textwidth]{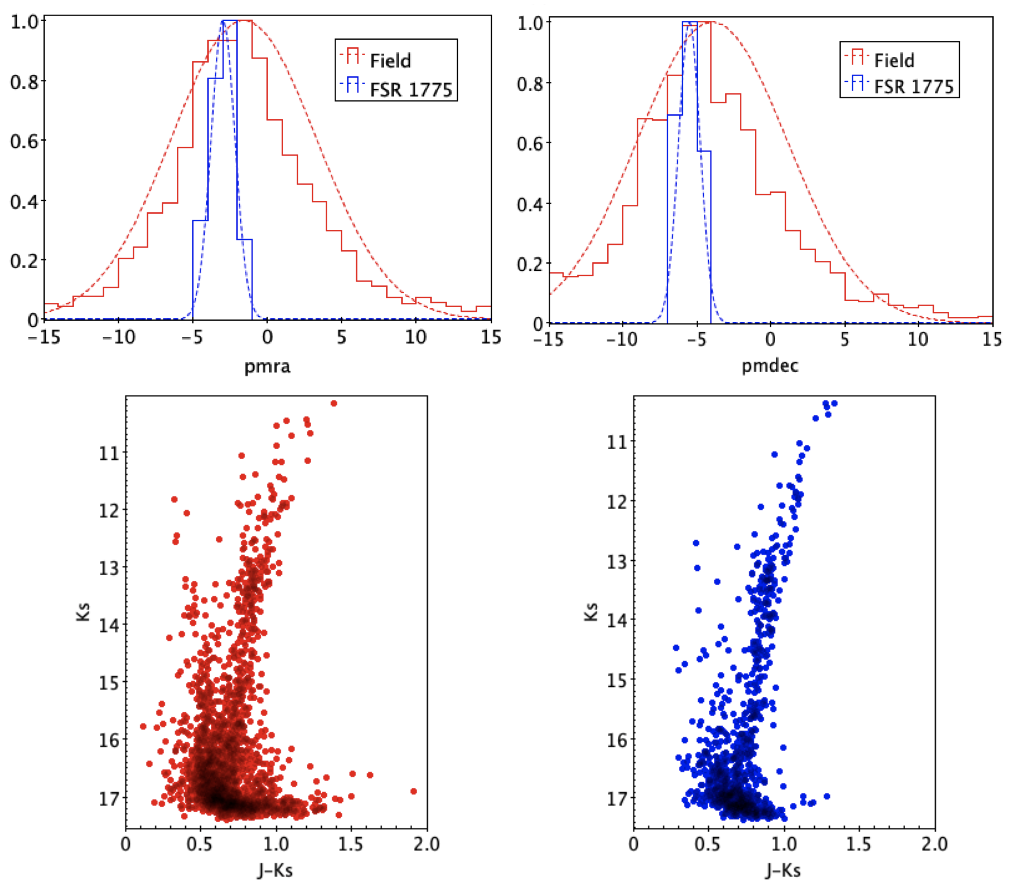} 
         \caption{FSR 1775}
         \label{fig:FSR1775+f}
     \end{subfigure}
      \hfill
     \begin{subfigure}[b]{0.3\textwidth}
         \centering
         \includegraphics[width=\textwidth]{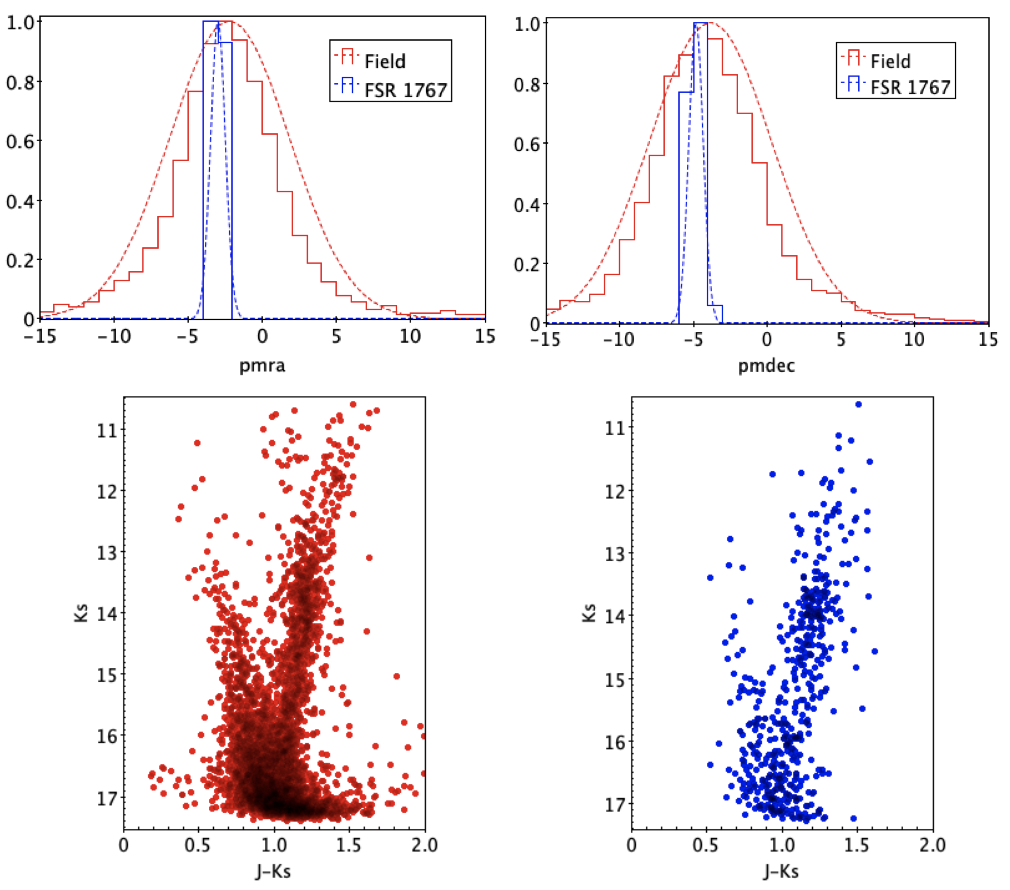} 
         \caption{FSR 1767}
         \label{fig:FSR1767+f}
     \end{subfigure}
      \hfill
     \begin{subfigure}[b]{0.3\textwidth}
         \centering
         \includegraphics[width=\textwidth]{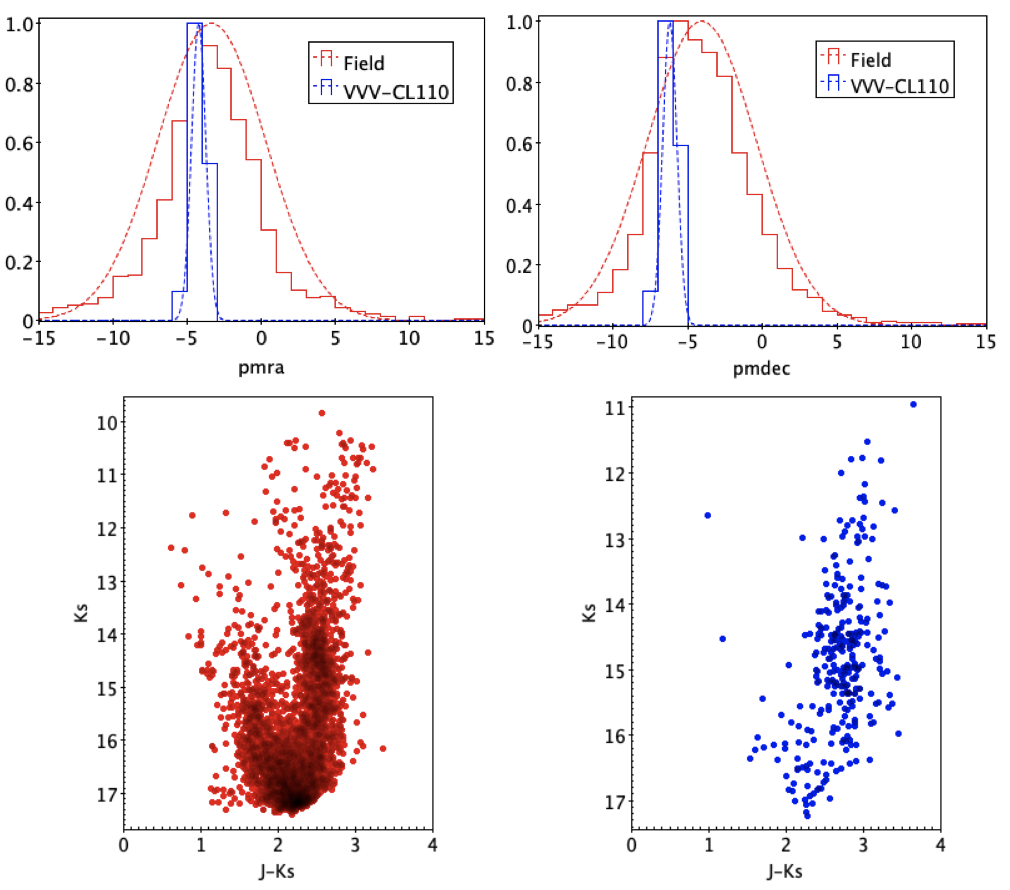} 
         \caption{VVV-CL110}
         \label{fig:VVVCL110+f}
     \end{subfigure}
     \hfill
     \begin{subfigure}[b]{0.3\textwidth}
         \centering
         \includegraphics[width=\textwidth]{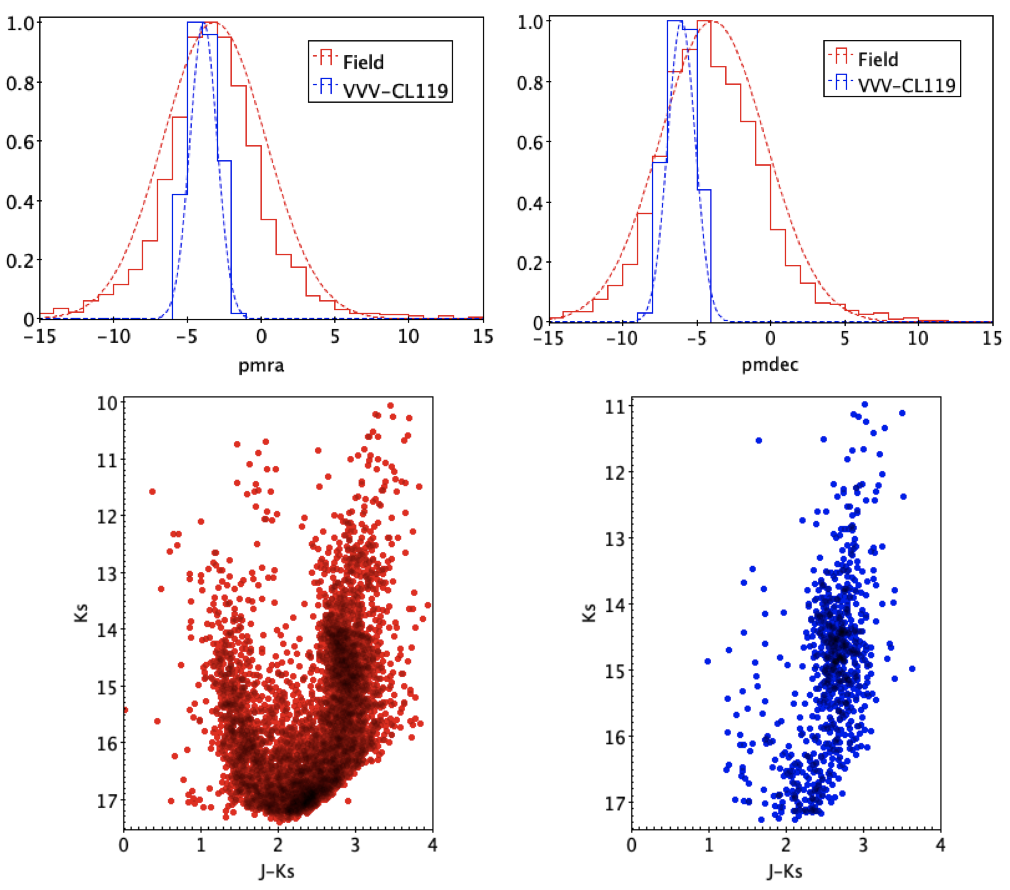} 
         \caption{VVV-CL119}
         \label{fig:VVVCL119+f}
     \end{subfigure}
               \hfill
     \begin{subfigure}[b]{0.3\textwidth}
         \centering
         \includegraphics[width=\textwidth]{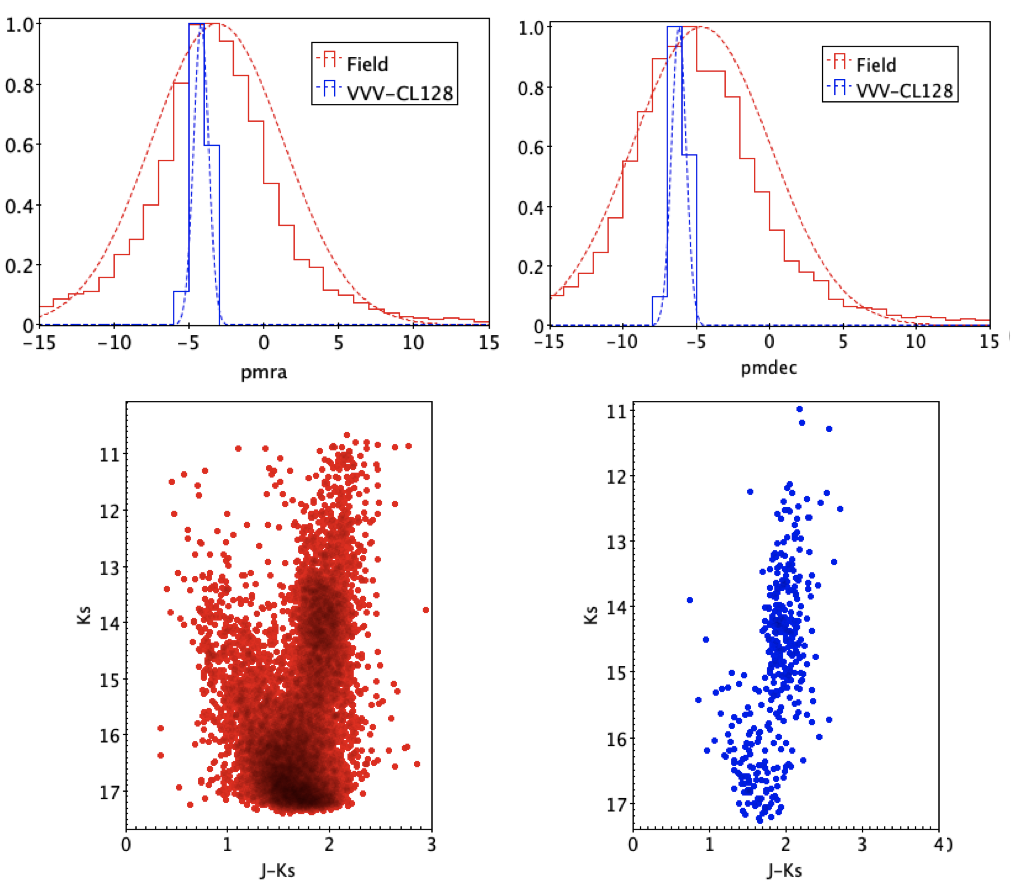} 
         \caption{VVV-CL128}
         \label{fig:VVVCL128+f}
     \end{subfigure}
          \hfill
     \begin{subfigure}[b]{0.3\textwidth}
         \centering
         \includegraphics[width=\textwidth]{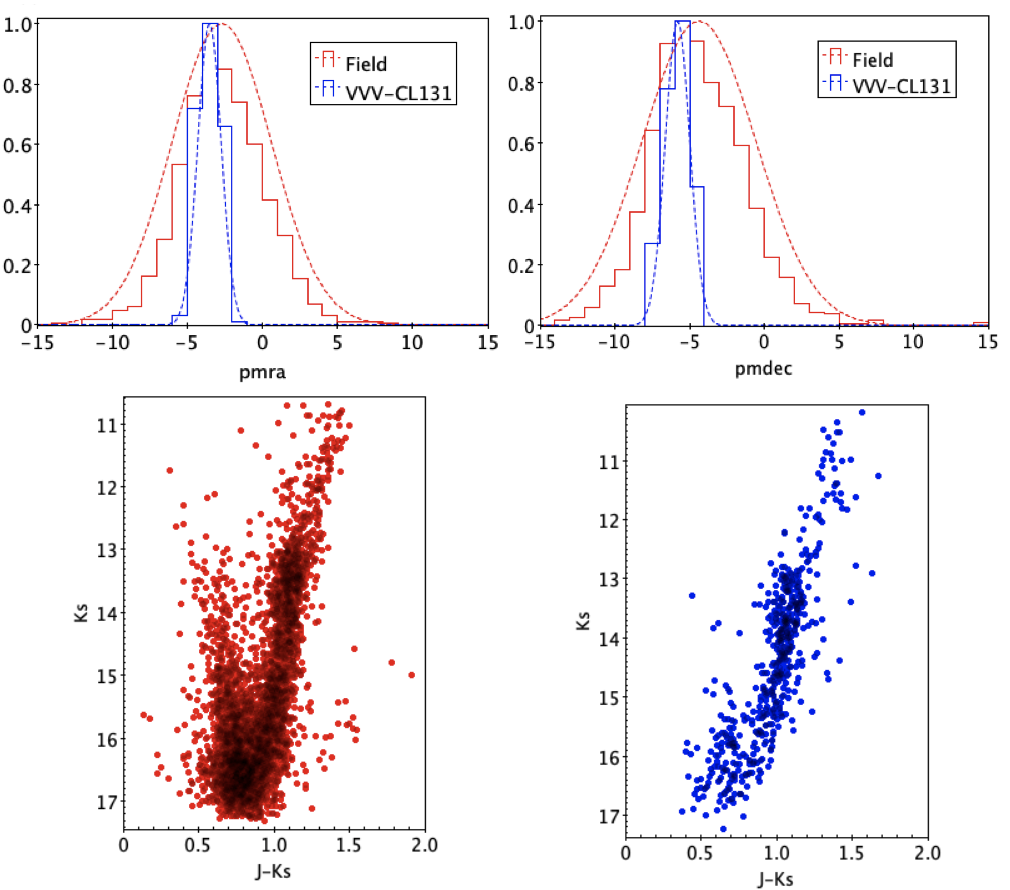} 
         \caption{VVV-CL131}
         \label{fig:VVVCL131+f}
     \end{subfigure}
               \hfill
     \begin{subfigure}[b]{0.3\textwidth}
         \centering
         \includegraphics[width=\textwidth]{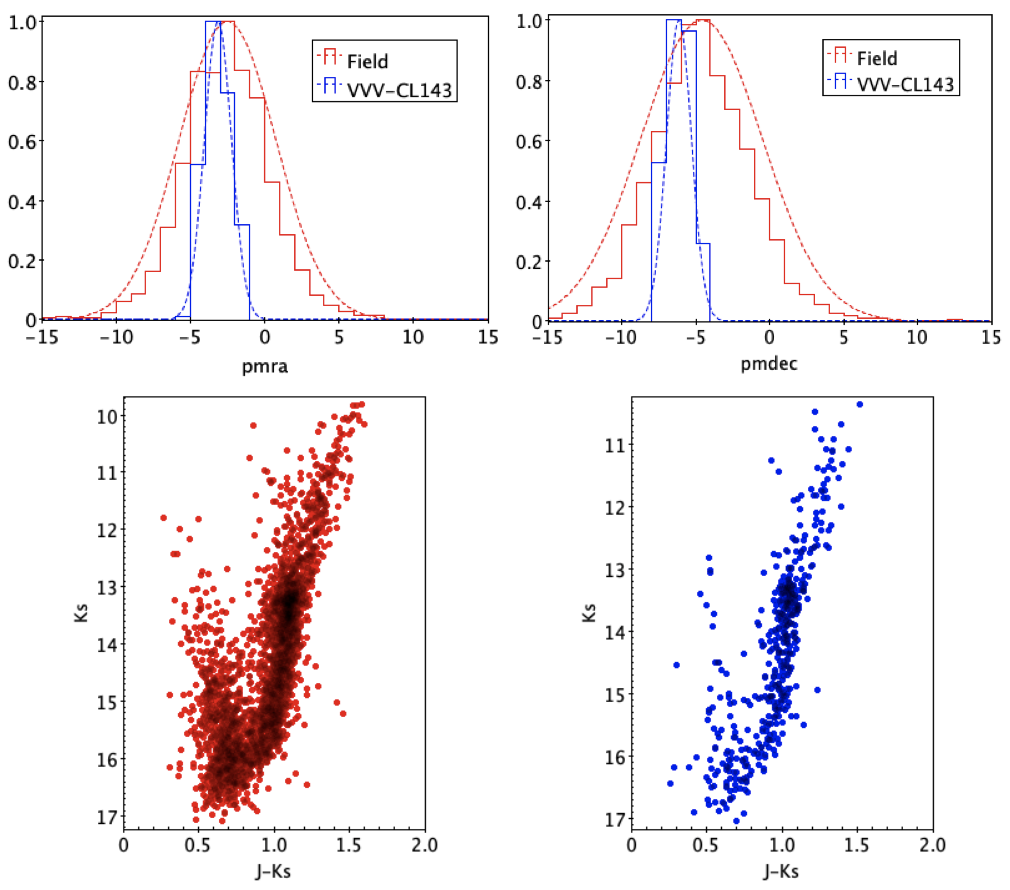} 
         \caption{VVV-CL143}
         \label{fig:VVVCL143+f}
     \end{subfigure}
         \hfill
      \begin{subfigure}[b]{0.3\textwidth}
         \centering
         \includegraphics[width=\textwidth]{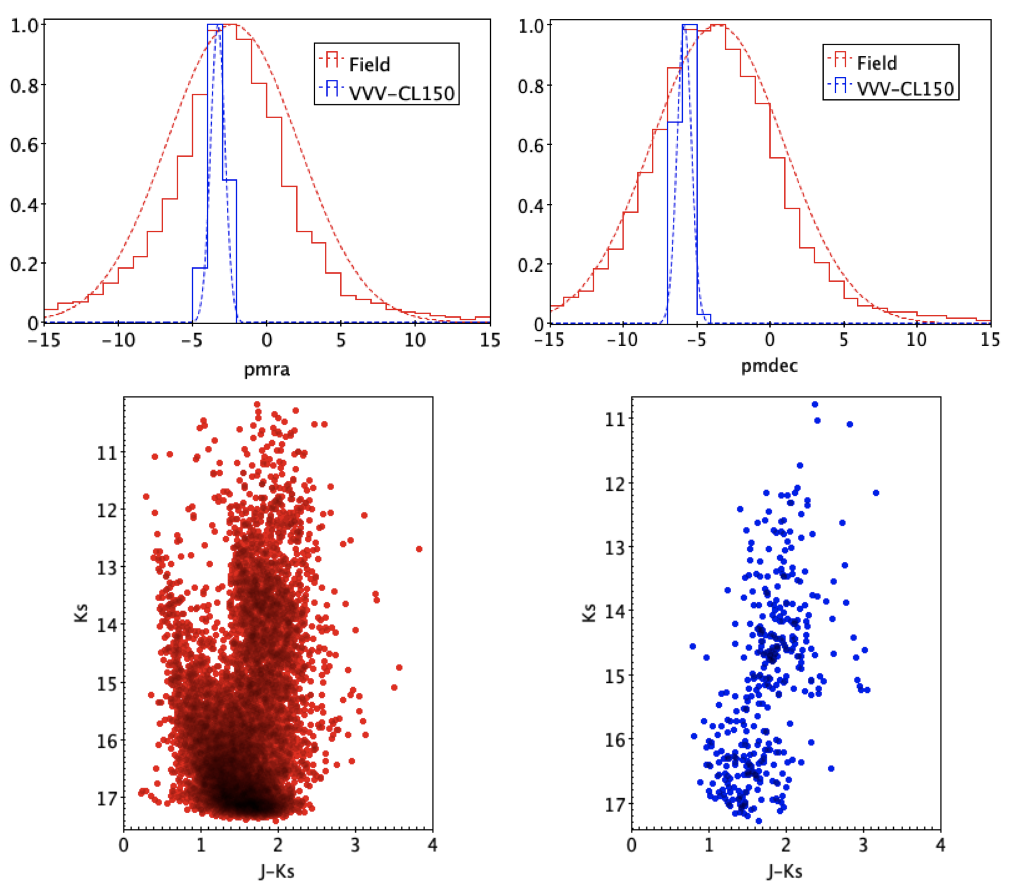} 
         \caption{VVV-CL150}
         \label{fig:VVVCL150+f}
     \end{subfigure}
        \caption{Normalised PMs histograms and CMDs for field (in red) and cluster (in blue) stars.  We fit the Gaussian distributions (dashed lines) using means and standard deviations listed in Table \ref{comparison}. }
        \label{cl+field}
\end{figure*}

\begin{figure*}
     \centering
	\ContinuedFloat
	      \begin{subfigure}[b]{0.3\textwidth}
         \centering
         \includegraphics[width=\textwidth]{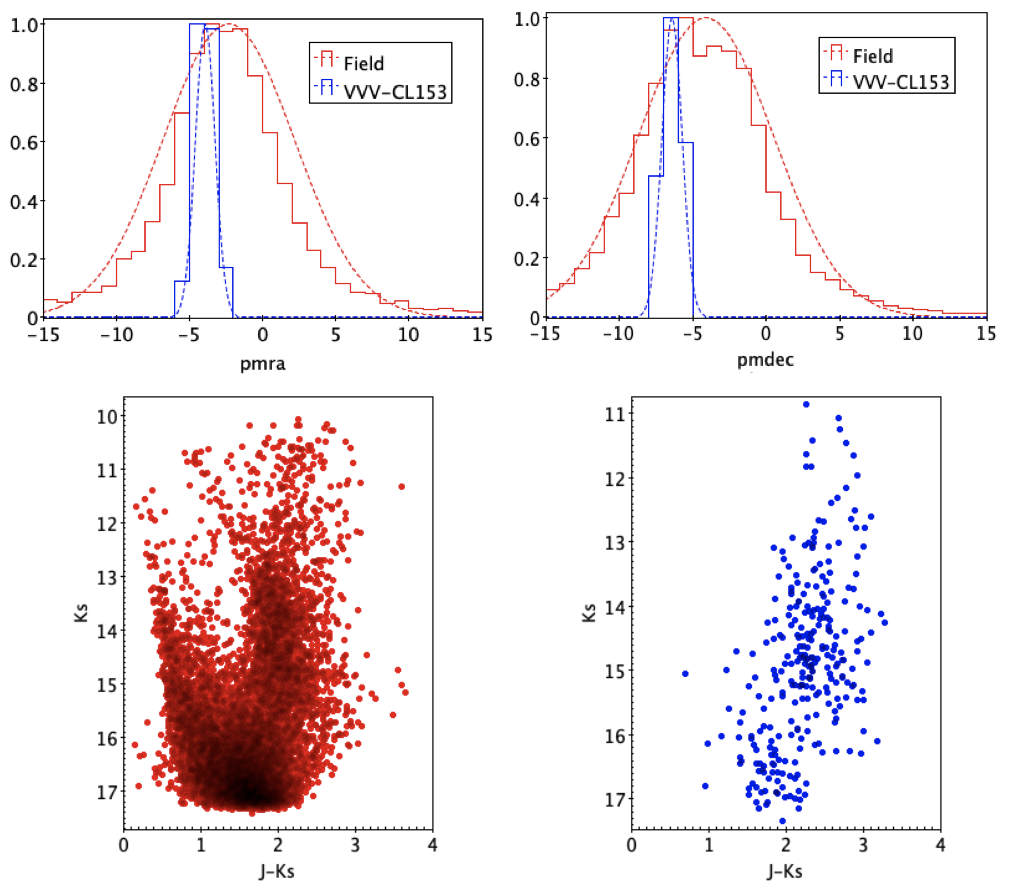} 
         \caption{VVV-CL153}
         \label{fig:VVVCL153+f}
     \end{subfigure}
     \hfill
      \begin{subfigure}[b]{0.3\textwidth}
         \centering
         \includegraphics[width=\textwidth]{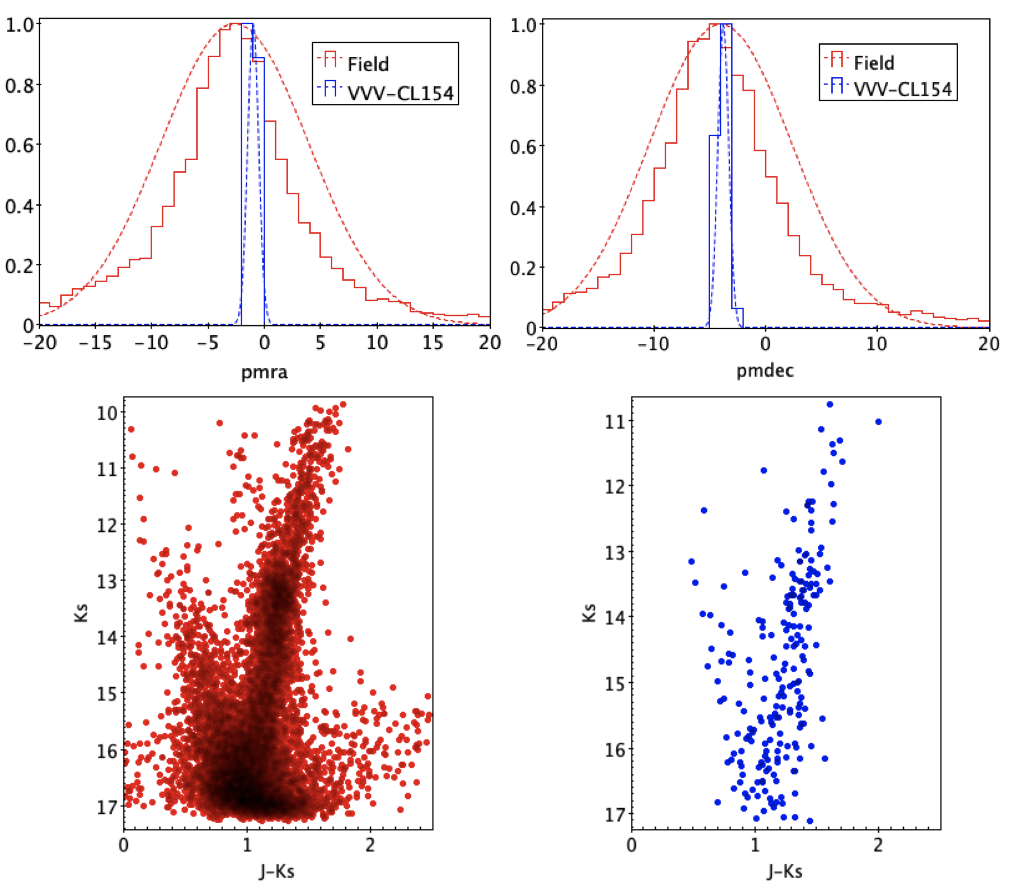} 
         \caption{VVV-CL154}
         \label{fig:VVVCL154+f}
     \end{subfigure}
     \hfill
           \begin{subfigure}[b]{0.3\textwidth}
         \centering
         \includegraphics[width=\textwidth]{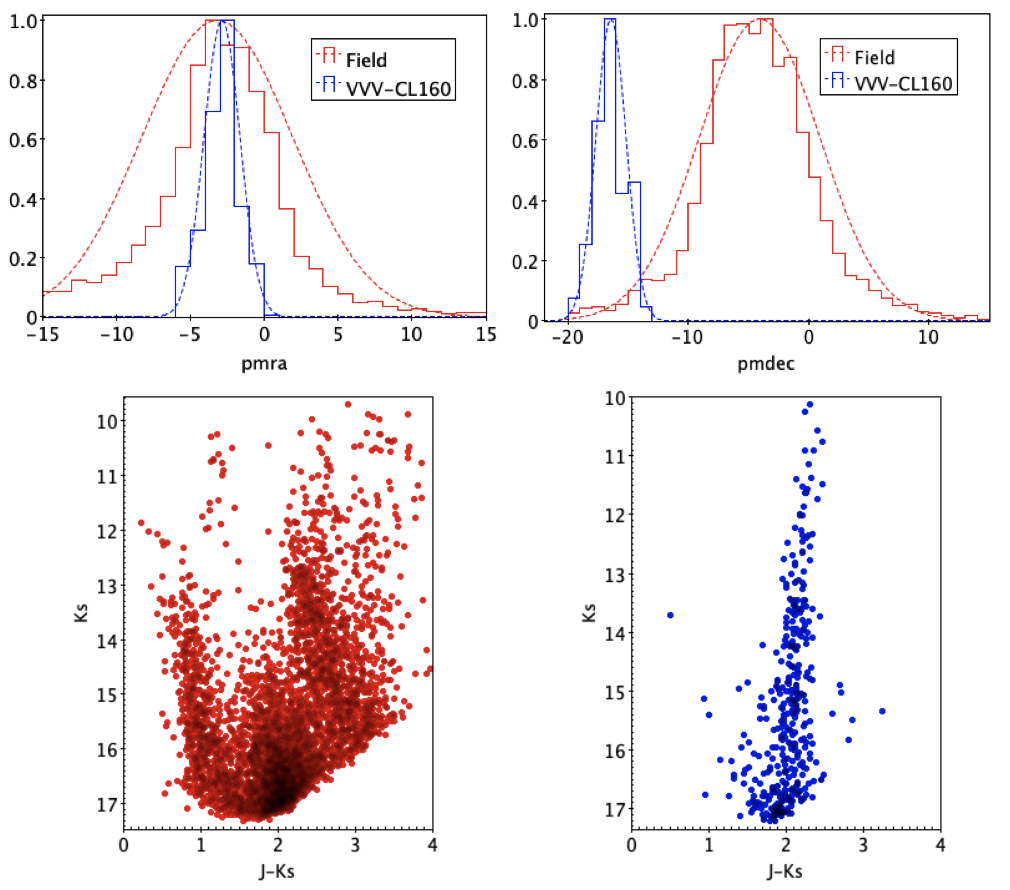} 
         \caption{VVV-CL160}
         \label{fig:VVVCL160+f}
     \end{subfigure}
      \hfill
           \begin{subfigure}[b]{0.3\textwidth}
         \centering
         \includegraphics[width=\textwidth]{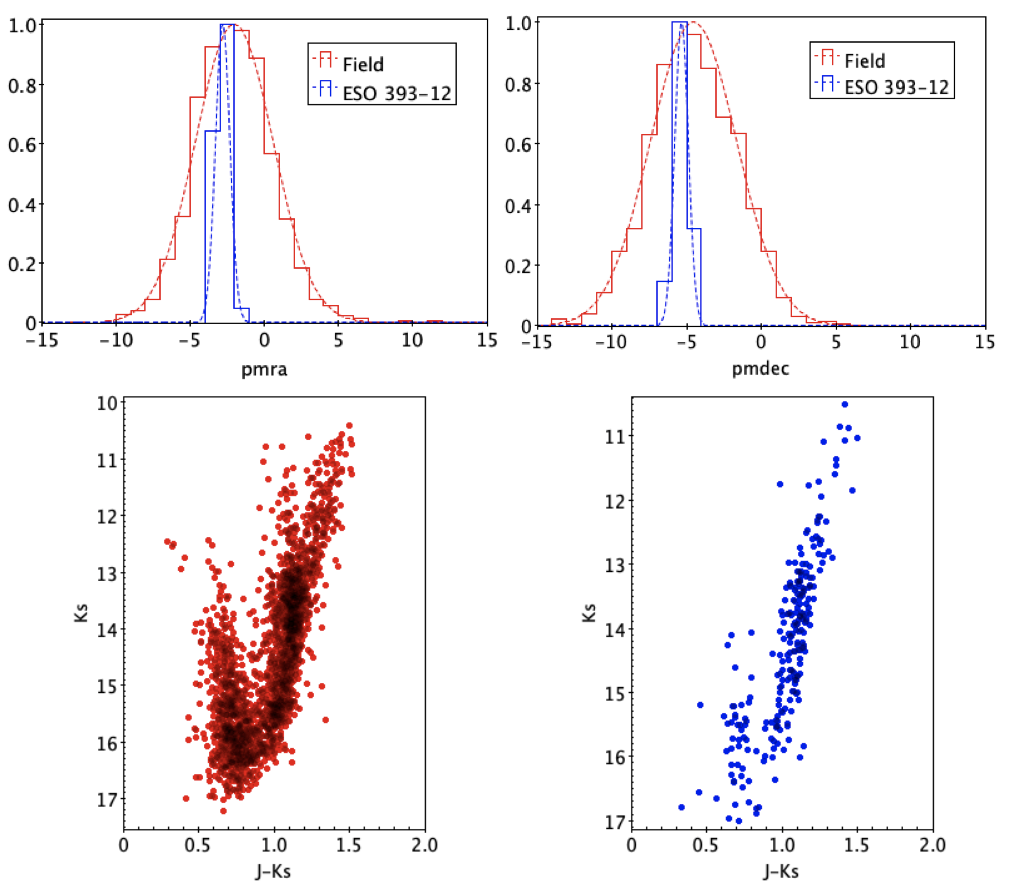} 
         \caption{ESO 393-12}
         \label{fig:ESO393+f}
     \end{subfigure}
           \hfill
           \begin{subfigure}[b]{0.3\textwidth}
         \centering
         \includegraphics[width=\textwidth]{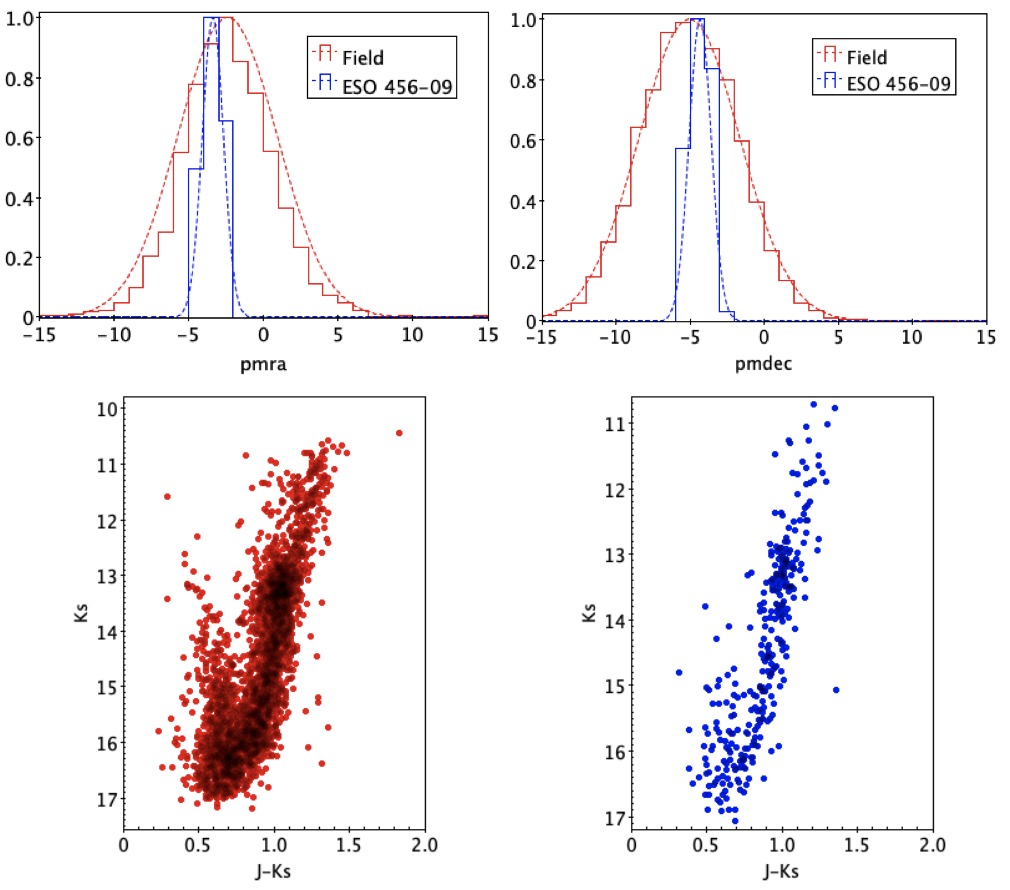} 
         \caption{ESO 456-09}
         \label{fig:ESO456+f}
     \end{subfigure}
                \hfill
           \begin{subfigure}[b]{0.3\textwidth}
         \centering
         \includegraphics[width=\textwidth]{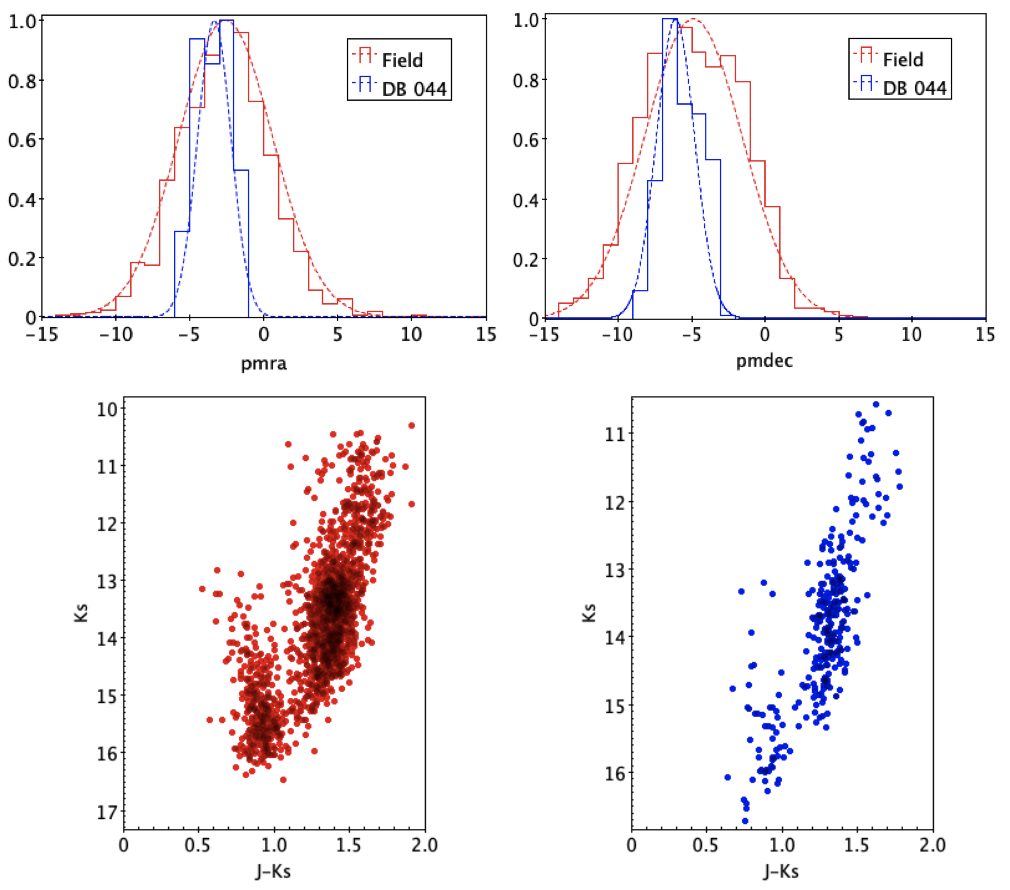} 
         \caption{DB 044}
         \label{fig:DB044+f}
     \end{subfigure}
                     \hfill
           \begin{subfigure}[b]{0.3\textwidth}
         \centering
         \includegraphics[width=\textwidth]{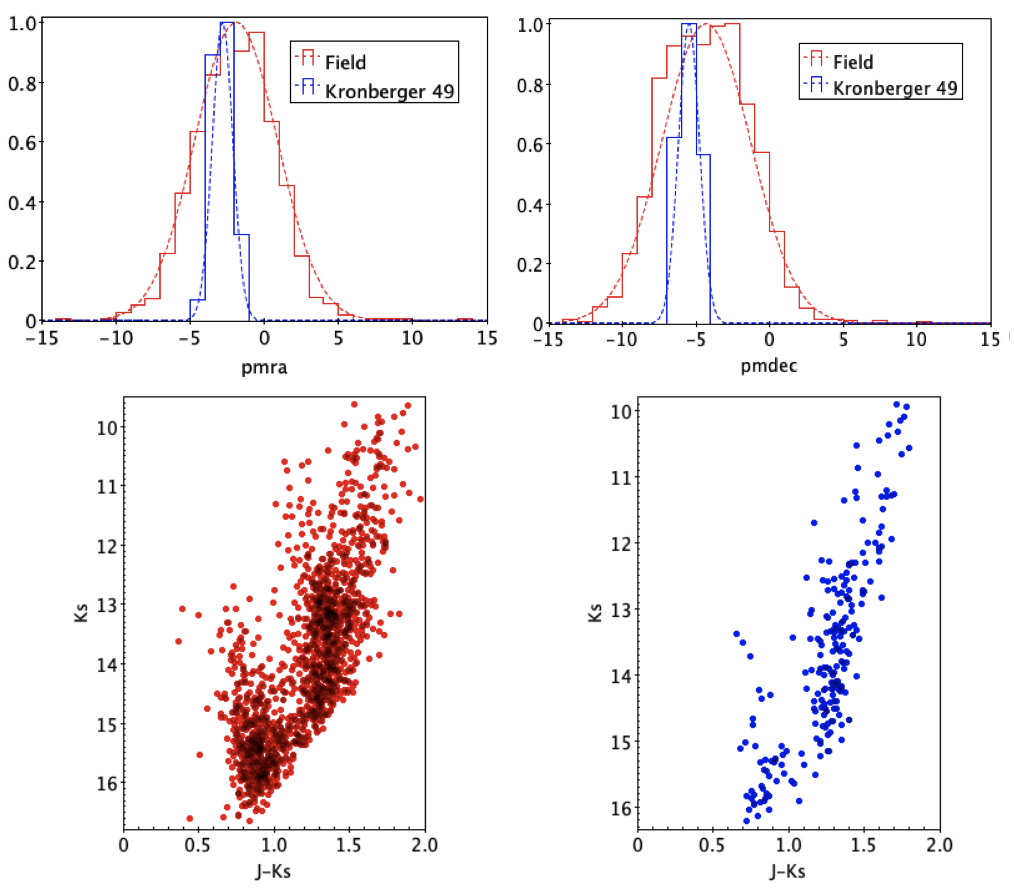} 
         \caption{Kronberger 49}
         \label{fig:Kron49+f}
     \end{subfigure}
     	      \caption{(continued)}
        \label{cl+field}
\end{figure*}

\begin{figure*}[!htb]
\centering
\includegraphics[width=4cm, height=3cm]{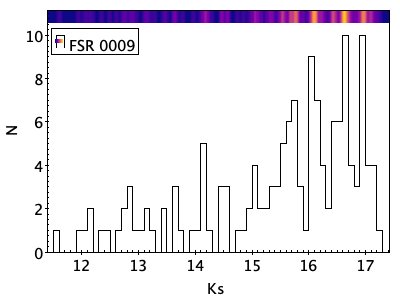} 
\includegraphics[width=4cm, height=3cm]{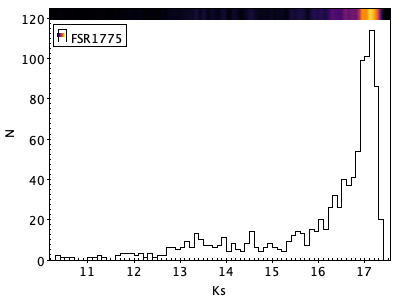} 
\includegraphics[width=4cm, height=3cm]{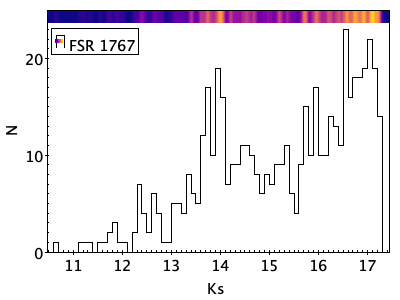} 
\includegraphics[width=4cm, height=3cm]{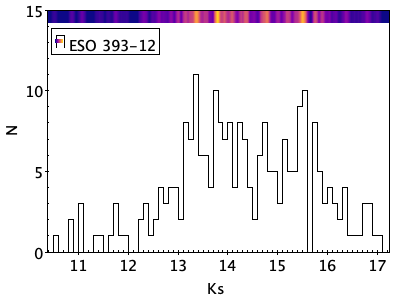} 
\includegraphics[width=4cm, height=3cm]{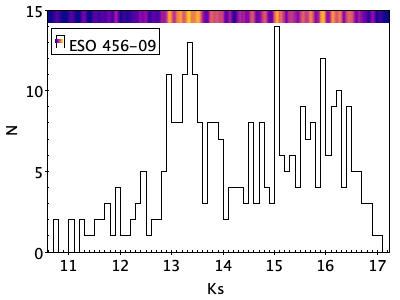} 
\includegraphics[width=4cm, height=3cm]{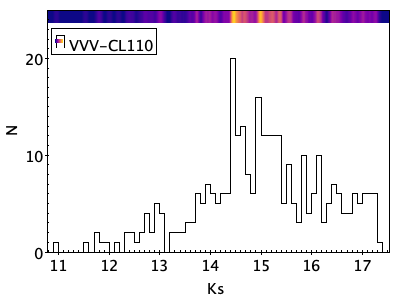} 
\includegraphics[width=4cm, height=3cm]{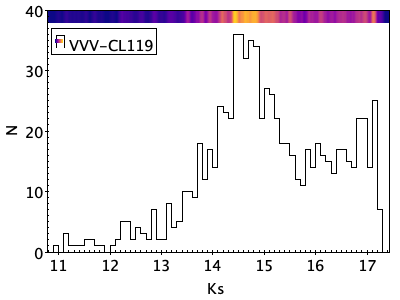}
\includegraphics[width=4cm, height=3cm]{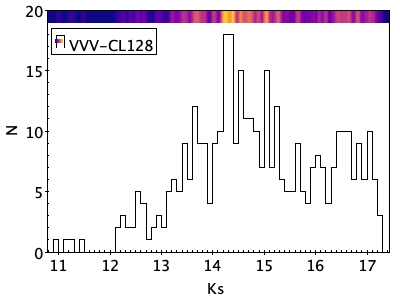} 
\includegraphics[width=4cm, height=3cm]{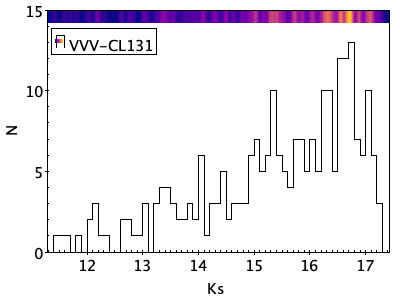} 
\includegraphics[width=4cm, height=3cm]{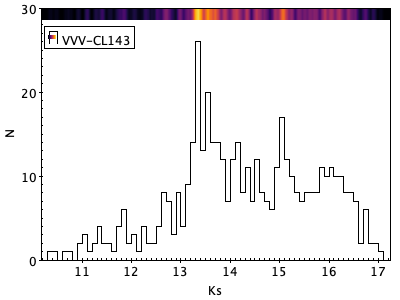} 
\includegraphics[width=4cm, height=3cm]{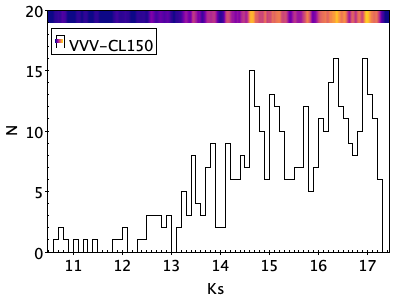} 
\includegraphics[width=4cm, height=3cm]{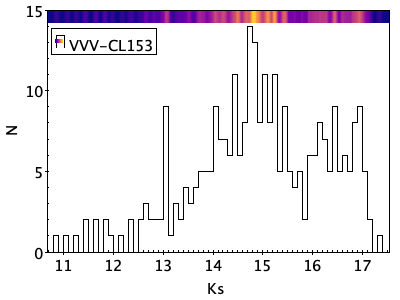} 
\includegraphics[width=4cm, height=3cm]{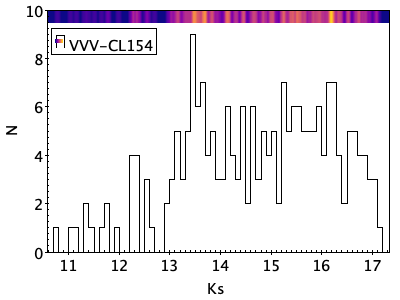} 
\includegraphics[width=4cm, height=3cm]{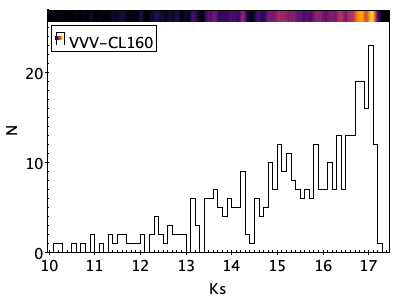} 
\includegraphics[width=4cm, height=3cm]{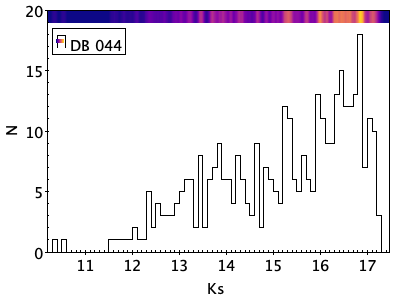} 
\includegraphics[width=4cm, height=3cm]{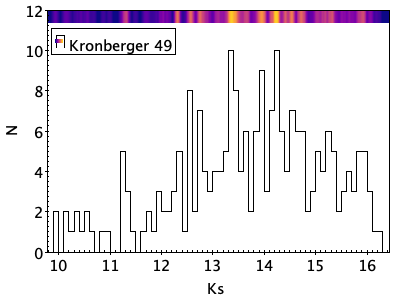} 
\caption{Luminosity functions in $K_s-$band for each star cluster. \textit{On the top} a densogram is used in order to recognise the RC easier and avoid the binning dependence. The yellow colour depicts the highest density that decreasing becomes magenta and dark blue. We use a bin size of 0.1.}
\label{LF}
\end{figure*}

\section{Estimation of physical parameters}
\label{estimationparameters}
\begin{figure*}
\centering
\includegraphics[width=7cm, height=6cm]{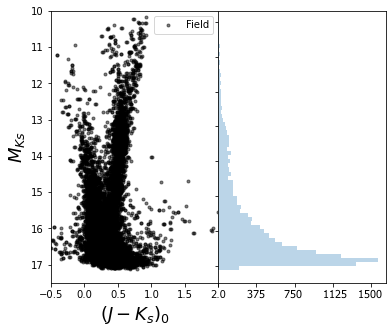} 
\includegraphics[width=7cm, height=6cm]{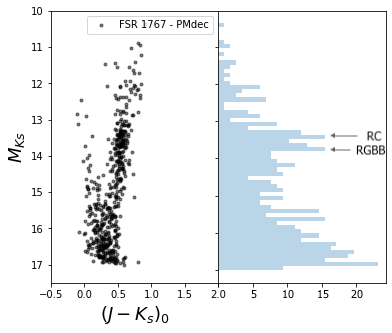}  
\caption{De-reddened CMDs for field stars (on the left) located at $5' \lesssim r  \lesssim8'$ from the FSR 1767 centre and for the FSR 1767 GC (on the right).  We also construct the $M_{Ks}$ luminosity function as blue histograms.  }
\label{deCMDs}
\end{figure*}

As shown in Fig. \ref{position}, the closest cluster is VVV-CL154 situated within 2.5$^{\circ}$ to the Galactic centre, while the most distant clusters are FSR~0009, BICA 523 and 524.  These regions are strongly affected by stellar crowding and interstellar extinction, which impact the CMDs.  Therefore, in order to overcome some of these difficulties, we estimate the reddening and the extinction towards the clusters following \cite{2018A&A...609A.116R}.  Indeed, we first construct the cluster luminosity functions (Fig. \ref{LF}) in order to distinguish clearly the RC (red clump) position,as the peak of the distribution, and then we assign the RC magnitude error as the average of the $K_s$ magnitude errors at the RC level.  After that,  we use the mean magnitude of RC stars in the near-IR from \cite{2018A&A...609A.116R}.  Their absolute magnitude in the $K_s-$band is $M_{Ks}=-1.605\pm 0.009$ mag and the intrinsic colour is $(J-K_s)_{0}= 0.66 \pm 0.02$ mag,  whereas for the optical wavebands their intrinsic RC magnitude is $M_G=0.459\pm 0.009$ mag.  Additionally,  we adopt  the following relations for the extinctions and reddenings: $A_{Ks}=0.428\times E(J-K_s)$ (since our targets are located in the inner galactic regions,  \citealt{Alonso_Garc_a_2017_AKs}), $A_{Ks}= 0.11 \times A_{V}$, $A_{G}=0.86\times A_{V}$ and $A_{G}=2.0\times E(BP-RP)$ \citep{Wang_2019}.  Also, we derive the colour excess for the optical passband comparing the cluster colours with the absolute colour by \cite{Babusiaux2018}.  As expected, the cluster fields exhibit a large range of extinction $0.11\leq A_{Ks} \leq 0.86$ mag as well as a large range of reddening $0.25 \leq E(J-K_s) \leq 2$ mag in the near-IR, demonstrating again that the reddening is not uniform,  which translates into uncertainties in the calculation of the main physical parameters.  For example,  VVV-CL128, CL150 and CL154 have wide red giant branches (RGBs) due to differential reddening and/or residual contamination, as we can appreciate in our CMDs (Fig. \ref{plots}).  The final reddenings obtained from the CMDs are listed in Table \ref{parameter}.  We also compare our reddenings with the high-resolution reddening map by \cite{Surot2021}, finding an excellent agreement ($0.2 \lesssim E(J-K_s)\lesssim 2.0$ mag). Fig.~\ref{deCMDs} shows the de-reddened CMDs for the field stars and FSR 1767, which we take as a representative sample cluster.  We selected: \textit{(i)} field stars in an adjacent area ($5' \lesssim r  \lesssim8'$ from the cluster centre),  cleaned by nearby stars; and \textit{(ii)} PM-members of FSR 1767. Comparing these two samples,  we do not appreciate a very large spread in colour. However, we can see a well-defined RGB sequence, and also the RC and the RGB-bump are visible in the FSR 1767 CMD, which disappear in the field CMD,  a feature that is commonly observed among GCs. Clearly, the position of the RC and RGBB could depend on the bin size in the $M_{Ks}$ magnitude, however in the FSR 1767 CMD (see Fig. \ref{plots} a double over-density is clear at the same magnitudes.\\

\begin{figure}
\centering
\includegraphics[width=9cm, height=7.7cm]{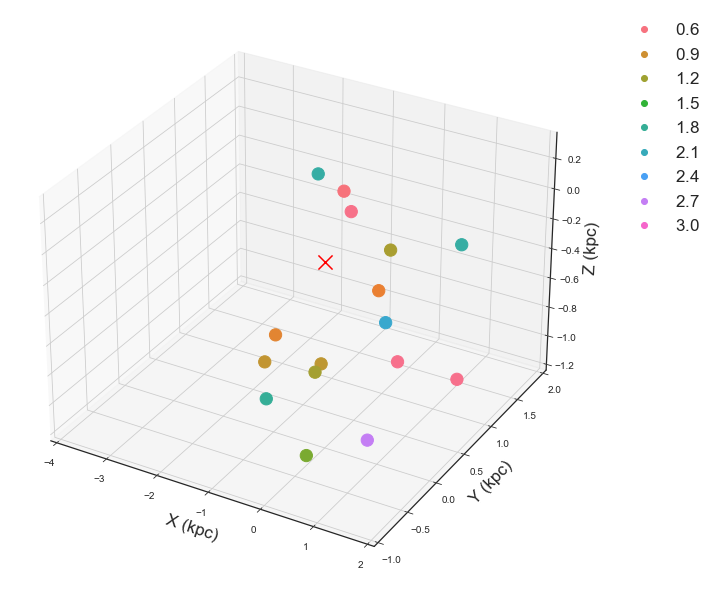} 
\caption{3D distribution of the star clusters  analysed in this work towards the Galactic bulge, using the Galactocentric coordinates (X,Y,Z). The red cross depicts the Galactic centre position at (0,0,0).  The colour of points indicates the Galactocentric distance ($R_{G}$ in kpc) as specified by the legend. }
\label{posgalcent}
\end{figure}

\begin{figure}
\centering
\includegraphics[width=8cm, height=8cm]{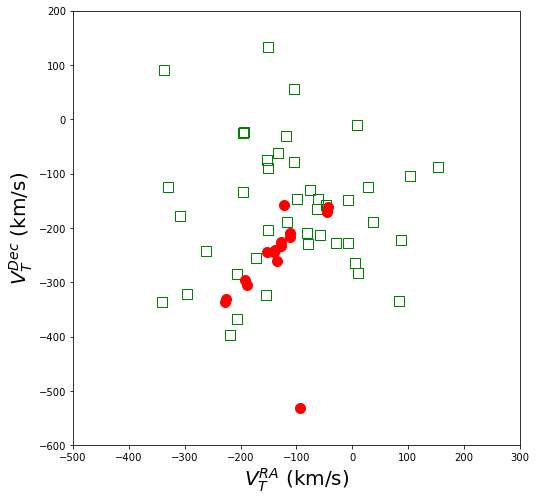} 
\caption{As Fig. \ref{pm_comp}, we show the star clusters tangential velocities to demonstrate that the kinematics of the star clusters here analysed are consistent with the known GCs from \cite{VasilievBaumgardt2021} located in the direction of the MW bulge. }
\label{tang_vel}
\end{figure}

Once reddening and extinction are obtained and taking into account of the RC absolute magnitude by \cite{2018A&A...609A.116R}, we are able to derive the GCs distance moduli and thus their heliocentric distances.  We find that the distance varies from $6.8$ to $11.4$ kpc from the Sun.  In order to confirm our VVV results, we also report in Table \ref{optparam} the optical results using the Gaia EDR3 photometry for those clusters for which we obtain suitable Gaia EDR3 data.  Furthermore,  we use the VVV distances to place our candidates at their respective distances $R_G$ from the Galactic centre, assuming that the Sun is located at $R_{\odot}=8.2$ kpc \citep{Gravity2019}.  We find that they are located at Galactocentric distances ranging between $0.56$ kpc (for the VVV-CL150) and $3.26$ kpc (for the VVV-CL28), as listed in Table \ref{parameter}.  Their 3D distribution adopting the Galactocentric coordinates (X,Y,Z) in the Galactic bulge is shown in Fig. \ref{posgalcent}.  \\
Once distances and PMs are known,  we can derive their tangential velocities ($V_T$), as shown in Fig. \ref{tang_vel}. It is clear that their tangential velocities are consistent with those measured by \cite{VasilievBaumgardt2021} for the known GCs located towards the MW bulge. This is a further confirmation that they are part of the bulge component. 

Using the reddening and distance estimates, we fit the alpha-enriched PARSEC isochrones \citep{2012MNRAS.427..127B,2017ApJ...835...77M}, which reproduce the evolutionary sequences in the observed CMDs slightly better than solar-scaled model,  in order to derive both metallicity and age for our targets.  We perform  isochrone-fitting, reproducing in particular the position of the RC, blue horizontal branch stars (BHB) and the RGB (Fig. \ref{plots}).  We compare our evolutionary sequences in the CMDs with the isochrones generated for different ages and metallicities and visually selecting the best fit. The derivation of the metallicity is simpler than the age, since the metallicity affects especially the evolved sequences (i.e., RC, HB, RGB).  Therefore, first of all we fix the metallicity for each candidate, and then we derive the ages.  We find 11 metal-rich GCs with [Fe/H] $=0.0$ to $-0.6$,  while 3 are metal-poor GCs with [Fe/H] $=-1.2$ and $-1.4$ and 2 show intermediate values of [Fe/H]$=-0.7$ and $-1.1$, with a mean errors of $\pm0.2$ dex.  On the other hand, estimating ages is a challenging task,  as one usually needs to reach $\sim 2$ mag below the main-sequence turn-off (MSTO) to determine their absolute values. Yet, we are unable to go down $K_s>17.5$ mag (and $G>17.0$ mag), meaning that the MSTO lies below the detection limit.  Despite this, we can obtain a robust lower limit for ages, considering three observables from the CMDs:
\begin{itemize}
\item firstly, we use the vertical method, estimating the difference in magnitude between the HB and the MSTO. In our cases, the $\Delta K_s$(HB-MSTO) $\gtrsim 2$ mag, thus ensuring that our candidates are older than $\sim 8$ Gyr;
\item secondly,  a large extension of the giant branch suggests an intermediate age system, indicating that bright and red stars are present especially along the asymptotic giant branch \citep{Freedman2020};
\item also, the presence of RR Lyrae variable stars  guarantee that the ages are older than 10 Gyr (Section 5).
\end{itemize} 
The most of the analysed clusters are consistent with the old ages expected for a GC, except for  BICA~523 located at RA=17:14:58.0,  Dec=-36:48:28,  BICA~524 situated at RA=17:15:08.0, Dec=-36:47:12, and FSR~0022 placed at RA=17:56:23.7, Dec=-23:11:35,  which are likely open clusters with $Age<6$ Gyr (i.e. \citealt{Salaris2004}),  therefore we exclude them from this work since our main long term goal is to complete the census of the GCs in the MW. \\
Furthermore, we derive the global metal content for the sample from the slope of the RGB in order to minimise the effect of age-metallicity degeneracy (e.g., \citealt{Ferraro2000,Valenti2004}).  We use [Fe/H]-slope$_{JK}^{RGB}$ relation by \cite{Cohen_2015}, finding an excellent agreement with our original estimates, within the errors (see Table \ref{parameter}).\\
After finding the best-fitting age and metallicity simultaneously,  we also give an estimation of the age and metallicity errors by varying the parameters, until the isochrones were not adequate to reproduce the evolutionary sequences as shown by Fig. \ref{plots}.  However,  it is clear that the correct determination of the age is still one of the most arduous tasks to perform, especially in environments with high stellar density and with differential reddening, and our ages are only rough estimates.  \\

\begin{figure}[!htb]
\centering
\includegraphics[width=5cm, height=5cm]{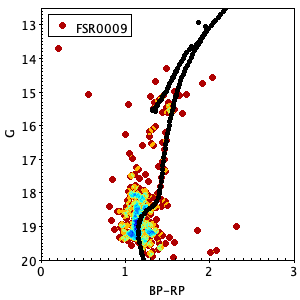} 
\includegraphics[width=5cm, height=5cm]{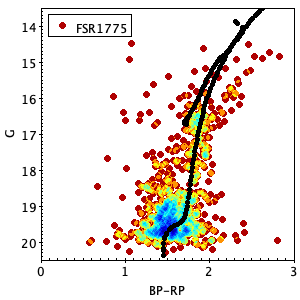} 
\includegraphics[width=5cm, height=5cm]{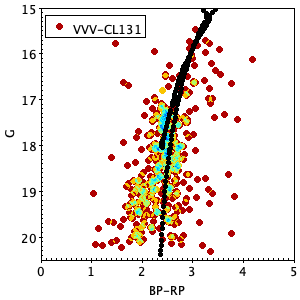} 
\includegraphics[width=5cm, height=5cm]{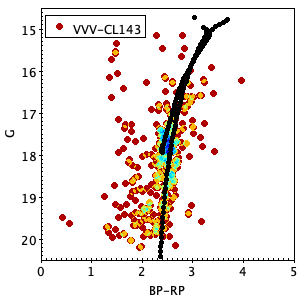} 
\includegraphics[width=5cm, height=5cm]{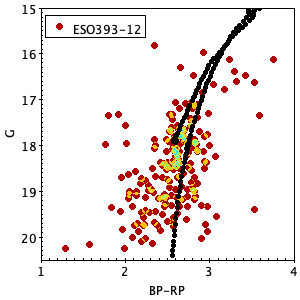} 
\includegraphics[width=5cm, height=5cm]{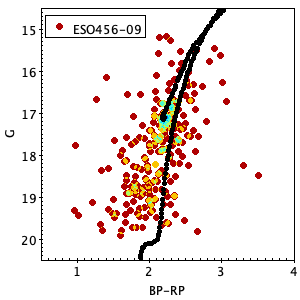} 
\includegraphics[width=5cm, height=5cm]{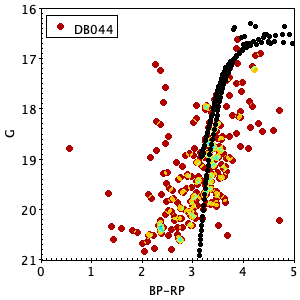} 
\includegraphics[width=5cm, height=5cm]{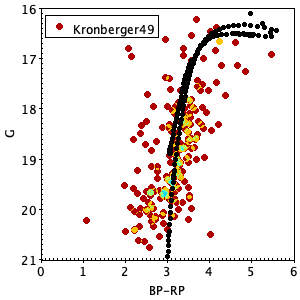} 
\caption{Gaia CMDs for the eight clusters for which we obtain reliable results in the optical passband. The black dotted line depicts the isochrone with the same age and metallicity as the VVV photometry and listed in Table \ref{parameter}.}
\label{gaiaCMD }
\end{figure}

Additionally, we show the Gaia optical CMDs in Fig. \ref{gaiaCMD } and list the physical parameters in Table \ref{optparam} for the eight GCs for which we have carried out the photometric analysis also in the optical passbands. Although it is difficult to reproduce simultaneously all the evolutionary sequences, we notice that both optical and near-IR analysis are in good agreement within the errors, confirming the VVV results. \\

Finally, the total luminosities in the $K_s-$band ($M_{Ks}$) for all the GCs in our sample are derived by measuring first the flux for each star and then the total flux for the cluster. Then, we converted the latter into the $M_{Ks}$, listed in Table \ref{parameter}.  As mentioned before, faint stars below the MSTO are missing in our catalogues, so the integrated luminosities of our clusters should be considered as lower limits.  For this purpose, we compared our GC $M_{Ks}$ with other known and well-characterised Galactic GCs, listed in Table \ref{tablelum}, where we recovered their reddening and distance modulus.  We follow the same approach as \cite{Garro2021b:submitted-a}. The main goal is to quantify the fraction of luminosity that comes from the faintest stars, in order to better estimate the total luminosity of each cluster.  
We obtain an empirical correction to our cluster's luminosities under the assumption of similarity among Galactic GCs.  We find that most of them are low-luminosity GCs, with $M_V$ fainter than the MW globular cluster luminosity function (GCLF) peak ($M_V=-7.4\pm0.2$ mag; \citealt{Harris_1991,ashman_1998}).  On the other hand, there are two GCs with total luminosities comparable with the MW GCLF peak:  VVV-CL119 ($M_V=-7.5$) and VVV-CL154 ($M_V=-7.2$).  \\

All their main physical parameters are listed in Table \ref{parameter}, where we summarised: the cluster identification (ID),  the RC position,  the reddening and extinction,  the distance,  the luminosity,  the mean cluster PMs,  the metallicity,  the age and the classification type for each candidate.   \\


\section{Searching for RR Lyrae stars}
\label{searchingRRL}
We also searched for RR Lyrae stars within a radius of $12'$ centred on the clusters (Table \ref{RRL_table}),  as these variables are good tracers of old populations and are also excellent distance indicators.   Our selection includes samples from the Optical Gravitational Lensing Experiment (OGLE; \citealt{2014AcA....64..177S}),  the VVV \citep{cite-key} and the Gaia \citep{2019A&A...622A..60C} catalogues. We have detected RR Lyrae stars in the fields of
 FSR 1775 ($N_{RRLS}^{12'}=8$),  FSR 1767 ($N_{RRLS}^{12'}=7$), Kronberger 49 ($N_{RRLS}^{12'}=9$),  ESO 393-12 ($N_{RRLS}^{12'}=15$),  ESO 456-09 ($N_{RRLS}^{12'}=30$), DB044 ($N_{RRLS}^{12'}=32$),  VVV-CL131 ($N_{RRLS}^{12'}=16$),  VVV-CL143 ($N_{RRLS}^{12'}=22$), and VVV-CL154 ($N_{RRLS}^{12'}=58$) fields. The number density of field bulge RRab stars has been estimated by \cite{Navarro2021}.\\

\begin{figure}[!htb]
\centering
\includegraphics[width=5cm, height=5cm]{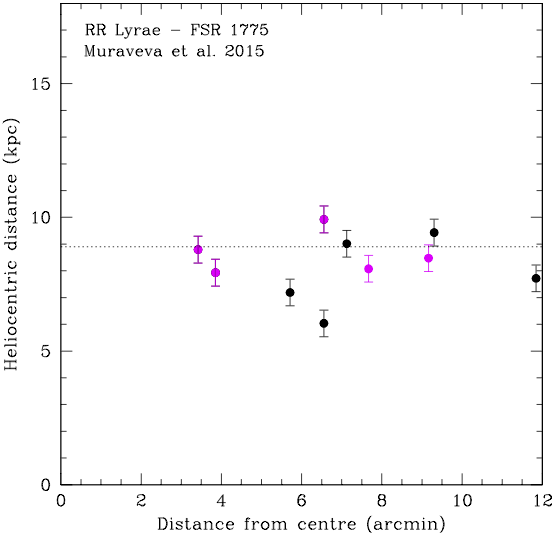} 
\includegraphics[width=5cm, height=5cm]{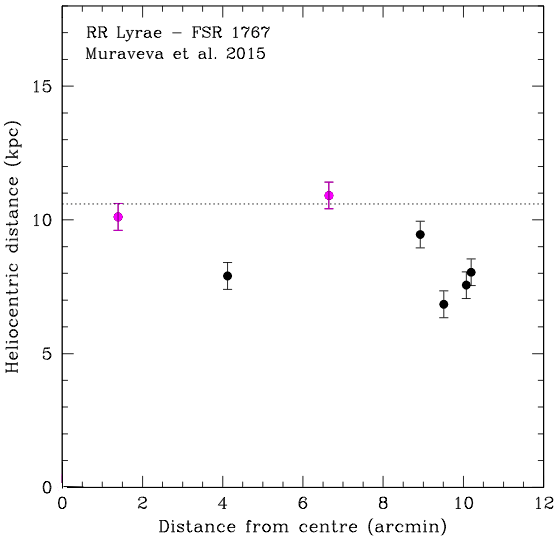} 
\includegraphics[width=5cm, height=5cm]{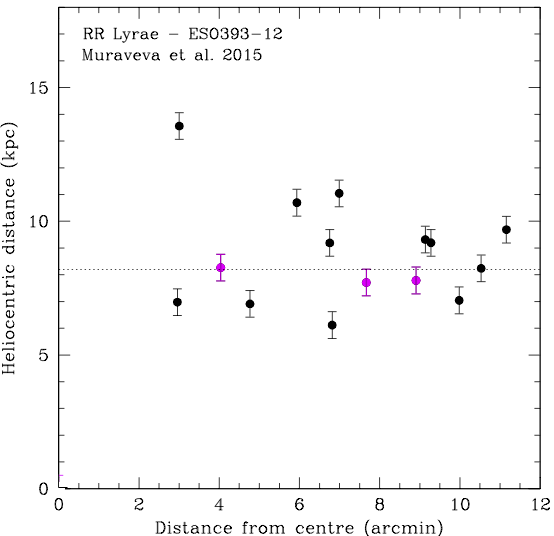} 
\includegraphics[width=5cm, height=5cm]{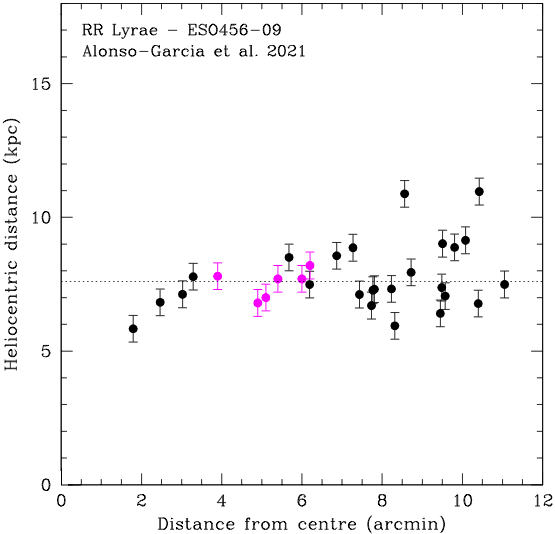} 
\includegraphics[width=5cm, height=5cm]{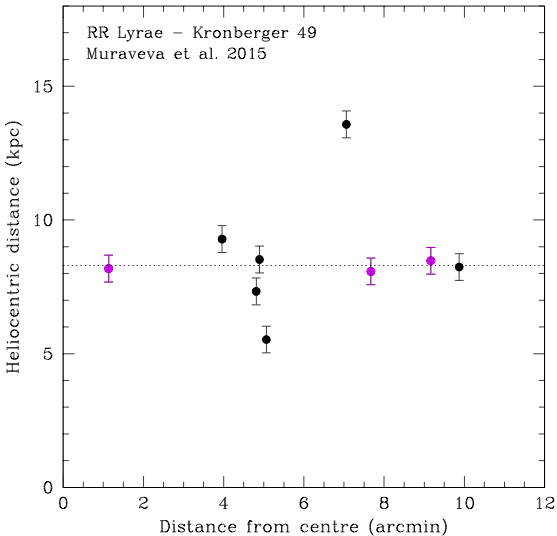} 
\includegraphics[width=5cm, height=5cm]{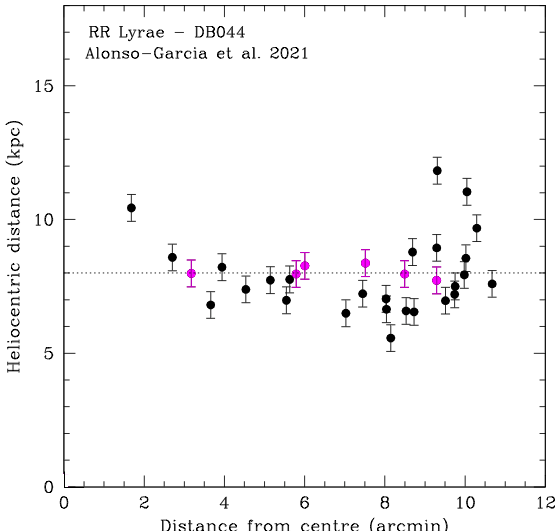} 
\includegraphics[width=5cm, height=5cm]{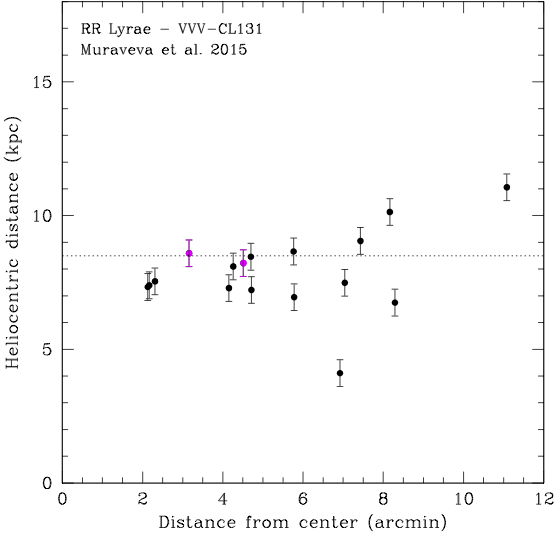} 
\includegraphics[width=5cm, height=5cm]{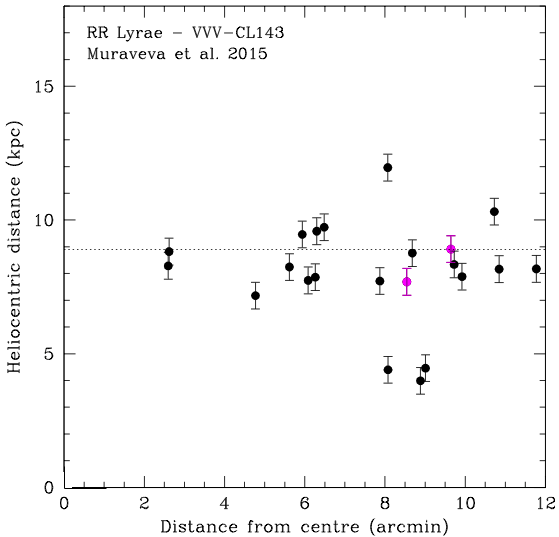} 
\includegraphics[width=5cm, height=5cm]{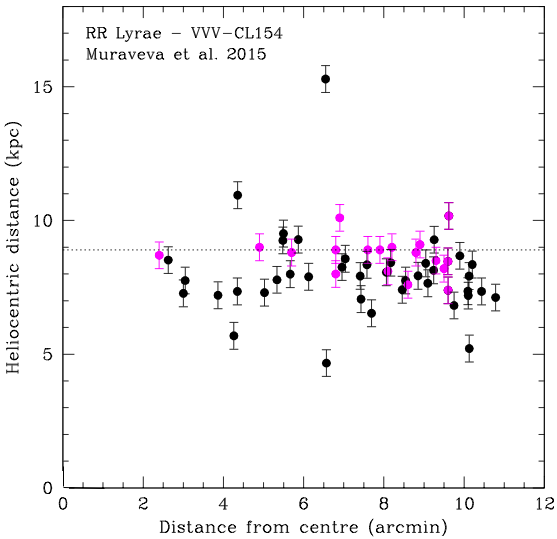} 
\caption{Spatial distribution of RR Lyrae variable stars detected within $12'$ from GC centres. The magenta points represent the RR Lyrae considered cluster members. The distance errorbars are of 0.5 kpc for each point. The dotted line depicts the heliocentric distance derived from the VVV photometry.}
\label{RRLplot }
\end{figure}

In order to distinguish RR Lyrae stars belonging to the GCs we mainly rely on two parameters: heliocentric distances and PMs.
We use the period-luminosity-metallicity (PLZ) relation in the near-IR following \cite{Muraveva_2015} (with RRc stars ``fundamentalized'' by adding 0.127 to the logarithm of the period) to derive their heliocentric distances ($D_{RRLS}$).  After that, we measure their projected distances from the GC centres ($D_c$),  and we construct a $D_{RRLS}-D_c$ diagram as shown in Fig. \ref{RRLplot }.  Then, we compare the RR Lyrae PMs with the cluster PMs as displayed in Fig. \ref{plots} (left panels), keeping only those stars that have PMs matching the mean cluster PMs within the errors. According to PMs and heliocentric distances, we strictly identify 
 $N_{RRLS}$(FSR1775) = 2,  $N_{RRLS}$(FSR1767) = 2,  $N_{RRLS}$(Kron49) = 3,  $N_{RRLS}$(ESO393-12) = 3,  $N_{RRLS}$(ESO456-09) = 6, $N_{RRLS}$(DB044) = 6, $N_{RRLS}$(VVV-CL131) = 2, $N_{RRLS}$(VVV-CL143) = 4,  and $N_{RRLS}$(VVV-CL154) = 15, as GC members.
We verified our distances using the PL relations from \cite{2017A&A...604A.120N} and \cite{2017A&A...605A..79G}, yielding slightly larger distance values, and \cite{AlonsoGarcia2021}, yielding slightly shorter distance values in the mean.  Specifically for the latter,  we transformed the metallicities, listed in Table \ref{parameter}, into log Z using the relation shown in \cite{2017A&A...604A.120N}, assuming $f=10^{[\alpha /FeH]} =3$ and $Z_{\odot}=0.017$.  However, we find that \cite{AlonsoGarcia2021} follows the DB044 and ESO 456-09 RR Lyrae trends better then \cite{Muraveva_2015}, so we prefer using it for these clusters.  However,  these relations are still within the mean errors of $\sim 0.5$ kpc obtained by the comparison between these four PLZ relations. \\ 

Although these variables are usually found in metal-poor populations, naming a few examples Terzan 10 ([Fe/H]$=-1.59$) and NGC 6656 ([Fe/H]$=-1.70$), there is evidence (e.g., \citealt{AlonsoGarcia2021}) of RR Lyrae stars located in metal-rich GCs, such as NGC 6441 ([Fe/H]$=-0.46$) and Patchick 99 ([Fe/H]$=-0.20$ -- \citealt{Garro2021}), as well as in intermediate metallicity GCs, like 2MASS-GC 02 ([Fe/H]$=-1.08$) and NGC 6569 ([Fe/H]$=-0.76$). In our case, we note that the GCs containing RR Lyrae show a metallicity range between $-0.6$ and $-1.2$.  However,  Kronberger 49 with [Fe/H] $=-0.2$ and DB044 with [Fe/H]$=-0.1$ could include 3 and 6 RR Lyrae members, respectively.  Therefore, we caution that Kronberger 49 membership because no RR Lyrae PMs match with the cluster PMs within $\sigma<1$ mas yr$^{-1}$.  Also, another caveat is that some variables are situated beyond $5'$ from the centre. This means that their position could be beyond the cluster core/tidal radius,  suggesting that their membership is less likely.  It is the case for example of VVV-CL143, since the RR Lyrae, which could be considered as members,  are located so far away from the cluster centre that their cluster membership is really unlikely.  Regardless, it is very unlikely that cluster with high metallicity host RR Lyrae stars.  Hence, we need spectroscopic observations in order to confirm their nature. \\

Finally,  we use these variables to determine the cluster distance independently.  We find a good agreement between the distances found by the VVV photometry (see Table \ref{parameter}) and the RR Lyrae estimates by \cite{Muraveva_2015} and \cite{AlonsoGarcia2021} relations: $D_{RRLS}$(VVV-CL154) = 8.1 kpc,  $D_{RRLS}$(VVV-CL131) = 9.2 kpc,  $D_{RRLS}$(VVV-CL143) = 8.8 kpc,  
$D_{RRLS}$(FSR1775) = 9.4 kpc, $D_{RRLS}$(FSR1767) = 9.4 kpc, $D_{RRLS}$(ESO393-12) = 7.9 kpc and  $D_{RRLS}$(ESO456-09) = 7.5 kpc. Anyway, if we include the Kronberger 49's variables we find for it a $D_{RRLS}$(Kron49) = 8.3 kpc,  and also DB044's RR Lyrae we obtain a $D_{RRLS}$(DB044) = 8.0 kpc, both in agreement with the VVV distances.\\
The magenta squares/points in the CMDs, VPM diagrams and distance diagrams (Figs. \ref{plots} and \ref{RRLplot }) depict the RR Lyrae members of each GC.

\section{Notes on Individual Clusters}
\label{individualnotes}
Below we report on the individual clusters. 

\subsection{FSR catalogue}
FSR 0009,  called also MWSC 2921, was identified as an open galactic cluster in the list of \cite{Buckner2013}. They placed it at equatorial coordinates (J2000) RA=18:28:32.9, Dec=-31:54:18  and galactic coordinates $l=1.86$,  $b=-9.52$,  $\sim 30''$ away from our centre. They concluded that it is a poorly-populated cluster with 73 members, its PMs are $\mu_{\alpha_{\ast}}=1.25$ mas yr$^{-1}$ and $\mu_{\delta}=-6.83$ mas yr$^{-1}$, and its distance is about $3.4$ kpc.  These results differ from ours,  because we find a GC with $\sim 150$ stars, with different PMs (see Table \ref{DP}) \\


The nature of FSR 1767 was a topic of discussion in the past. \cite{Froebrich2007} concluded that it was not a star cluster.  Whereas, with 2MASS photometry and PMs,  \cite{Bonatto2009} proved that FSR 1767 was a globular cluster located at equatorial coordinates (J2000) RA=17:35:44.8, Dec=-36:21:42, and galactic coordinates $l=352.60$,  $b=-02.17$,  $\sim 26''$ away from our centre.  They suggested that it appears to be a detached post-collapse core, similar to those of other nearby low-luminosity GCs projected towards the bulge.   On the other hand,  \cite{Buckner2013} classified FSR 1767 as an open cluster with $\mu_{\alpha_{\ast}}=0.97$ mas yr$^{-1}$ and $\mu_{\delta}=-3.57$ mas yr$^{-1}$, at a distance about $3.6$ kpc.  They counted 121 members.  Our results rather agree with \cite{Bonatto2009} since we classify it as GC, containing $\sim 530$ stars,  and we place this cluster at a  distance of $\sim 10.6\pm 0.2$ kpc. \\

FSR 1775, named also MWSC 2750,  was catalogued as a possible open cluster poorly-populated (52 members),  by  \cite{Buckner2013} at equatorial coordinates (J2000) RA=17:56:07.9, Dec=-36:33:54 and galactic coordinates $l=354.55$,  $b=-5.79$,  $\sim 30''$ away from our position space.They derived PMs of $\mu_{\alpha_{\ast}}=-9.53$ mas yr$^{-1}$ and $\mu_{\delta}=-1.00$ mas yr$^{-1}$ and a distance of $4.6$ kpc.  Again we obtained different PMs and distances based on $\sim 260$ members (see Table \ref{DP}).\\

In addition, we exclude the open cluster nature because both FSR 1767 and 1775 contain RR Lyrae stars, suggesting rather old ages.

\subsection{VVV catalogue}
All clusters named VVV-CL in this work were discovered in the VVV survey by \cite{Borissova2014}.  We report below our comments on these individual candidates.\\

VVV-CL110,  CL128, CL131, CL153 are located in VVV tiles b342, b317, b302,  b336, respectively.  \cite{Borissova2014} commented that VVV-CL110,  CL128 and CL131 could be GCs or old OCs. Indeed,  we confirm their nature as real GCs.  They suggested that VVV-CL153 could be a prominent bright cluster.  Selecting a region of $r\leq 1.8'$ from the GC centre and including all stars within 1.5 mas yr$^{-1}$ from the cluster PMs,  we found a not so bright cluster ($M_V\approx -6.8$ mag) as indicated by \cite{Borissova2014}.\\

VVV-CL119.  \cite{Borissova2014} found it in VVV tile b344 and classified it as a possible old open cluster.  Constructing the CMDs, they located the RC and the TO point at $K_s=13.8\pm 0.2$ mag and $K_s=17.3\pm 0.3$ mag, respectively.  Their main physical parameters are: a mean reddening of $E(J-K_s)=2.03\pm 0.4$, a distance modulus of $(M-m)_0=14.17\pm 0.3$ ($6.8$ kpc),  an age of $5\pm 1.2$ Gyr,  and a mean metallicity, calculated from five spectroscopically observed stars of [Fe/H] $=-0.3\pm 0.18$.  The main differences that we notice with respect to this work are related to the position of the RC and TO.  We identify the RC at $\sim 0.7$ mag fainter than theirs and we do not reach to the MSTO. Although the reddenings are in agreement, we place VVV-CL119 at 11.3 kpc from the Sun,  so at $\sim 4.5$ kpc further out than their distance. Finally, the metallicities obtained are in agreement within the errors. \\

VVV-CL143.  This star cluster lies in VVV tile b302. \cite{Borissova2014}  identified a defined RGB,  several RC stars at $K_s=13.3\pm 0.2$ mag,  and the TO point at $K_s=16.62\pm 0.3$ mag. They estimated a reddening of $E(J-K_s)=0.58\pm 0.2$,  a distance modulus of $(M-m)_0=14.45 \pm 0.6$ ($7.8$ kpc),  an age of $4 \pm 0.7$ Gyr, and a mean metallicity of [Fe/H] $=-0.62\pm 0.52$, which was spectroscopically evaluated from 10-12 observed stars.  Despite their analysis, they were unable to identify the real nature of VVV-CL143, since it could be either a populated old open cluster or a young globular cluster, especially when relatively low metallicity and CMD morphology are taken into account.  We confirm their results since this is a well-populated GC, including $470$ stars, with a metallicity of [Fe/H] $=-0.6\pm 0.2$ at $D=8.9 \pm 0.5$ kpc and an age of $\sim 10$ Gyr. \\


VVV-CL150.  \cite{Borissova2014} recognized this cluster as a possible new bulge globular cluster, lying in VVV tile b350. Their CMD is well populated and they place the RC at $K_s=13.10\pm 0.2$, the TO at $K_s=17.2\pm 0.4$ mag. They calculated reddening of $E(J-K_s)=1.2 \pm 0.1$, distance modulus of $(M-m)_0=14.22\pm 0.7$ (6.98 kpc),  age of $10 \pm 0.8$ Gyr and metallicity of $[Fe/H]=-0.75\pm0.11$. Also in this case we obtain with similar results. We count $383$ stars belonging to VVV-CL150. We do not identify the RC, not even from the inspection of the cluster LF, concluding that it is a metal-poor GC with [Fe/H] $=-1.30$. Also, we find a similar age (12 Gyr) within the errors. \\

VVV-CL154.  Located in VVV tile b321,  it was classified as an old open cluster in close proximity to the Sun, since \cite{Borissova2014} derived from its CMD analysis a reddening of $E(J-K_s)=1.2\pm 0.1$,  a distance modulus of $(M-m)_0=10.1\pm0.3$ ($1.05$ kpc), an age of $8\pm 0.2$ Gyr and metallicity of $[Fe/H] =-0.69\pm 0.4$.  Aside from the metallicity our estimates are different for this cluster. Our findings indicate an old and distant ($ \sim 8.9 $ kpc) GC, located near the Galactic centre ($R_G=0.78$ kpc). \\

VVV-CL160.  This cluster lies in VVV tile b340. The RGB and MS are well defined in the Borissova's CMD, where the RC and TO are identified at $K_s=13.1 \pm 0.3$ mag and $K_s=14.8 \pm 0.4$ mag, respectively.  Their calculated physical parameters are: reddening $E(J-K_s)=1.72\pm 0.1$, distance modulus $(M-m)_0=13.60\pm0.3$ ($5.25$ kpc), age $1.6\pm 0.5$ Gyr and metallicity $[Fe/H]=-0.72\pm 0.21$.  They concluded that VVV-CL160 may be an old metal-poor open cluster.  Recently,  \cite{Minniti2021_cl160} found a reddening of $E(J-K_s)= 1.95$ mag,  a distance modulus of $(m-M)_0 =13.01$ mag ($4.0$ kpc), metallicity of [Fe/H] $=-1.40$, age of 12 Gyr and luminosity of $M_V =-5.1$.   Interestingly,  \cite{Minniti2021_cl160} concluded that VVV-CL160 may belong to a disrupted dwarf galaxy,  since the kinematics are similar to the known GC  NGC 6544 and the Hrid halo stream.  Our parameters are in reasonable agreement with both works, since we find reddening value similar to that of \cite{Borissova2014}, while we agree with \cite{Minniti2021_cl160} about the metallicity,  luminosity and age, confirming the GC nature.  However,  we place this cluster farther than in previous works, at $6.8$ kpc from the Sun.\\

Finally, VVV-CL119, CL143, CL150 were also catalogued as star cluster candidates in the list of \cite{Minniti_2017a}.

\subsection{ESO star cluster candidates}
Listed in the \cite{Kharchenko2013} catalogue,  ESO 393-12 and ESO 456-9 were recognized as possible globular clusters, located at RA=17:38:40, Dec=-35:39.0:00 and RA=17:53:56.4, Dec= -32:27:54,   $\sim27.40''$ and $26.90''$ from our positions, respectively.  In this work,  the PM values are $\mu_{\alpha_{\ast}}=4.16$ mas yr$^{-1}$ and $\mu_{\delta}=-4.85$ mas yr$^{-1}$ for ESO 393-12,  while $\mu_{\alpha_{\ast}}=3.83$ mas yr$^{-1}$ and $\mu_{\delta}=-10.60$ mas yr$^{-1}$ for ESO 456-09.  On the other hand, \cite{Dias2014} catalogued them as open galactic clusters. Our results are on the same line as \cite{Kharchenko2013},  since we find two GCs,  including RR Lyrae variable stars, which suggests the old age ($t>10$ Gyr) for these clusters. 

\subsection{DB044}
DB044 was discovered by \cite{Dutra2000} classified this as a possible star cluster. It was further studied by \cite{Kharchenko2013},  as a possible open cluster, with PMs of $\mu_{\alpha_{\ast}}=-6.38$ mas yr$^{-1}$ and $\mu_{\delta}=-11.70$ mas yr$^{-1}$. We obtain very different cluster PMs, and we exclude the open cluster nature for two reasons: firstly, because it may include 6 RR Lyrae stars in its field,  and secondly because we derive an age of $\sim$11 Gyr. Admittedly, this is not a clear case.

\subsection{Kronberger 49}
Kronberger 49, designed as DSH J1810.3-2320,  was discovered by \cite{Kronberger2006}.  Inspections of DSS/XDSS and 2MASS images exhibited a compact object, suggesting that it could be a GC core.  Further, \cite{Ortolani_2006} observed Kronberger 49 at the Telescopio Nazionale Galileo (TNG) using $V$, $I$ and Gunn $z$ photometry,  combining them with additional photometry from 2MASS,  DENIS, WISE and VVV survey. They analysed the optical decontaminated CMD, where prominent features are a well-defined TO and an RGB with a turnover, suggesting high metallicity. Indeed,  the main parameters are a distance modulus of $(m-M)_V=18.8$,  $(m_M)_I = 16.9$, and a reddening of $E(V-I)=1.8$ (equivalent to $E(B-V)=1.35$), thus a heliocentric distance of $d=8\pm 1$ kpc.  The absolute magnitude was estimated following two methods based on 2MASS photometry,  they calculated $M_V=-5.4\pm 0.5$ and $-4.8\pm 0.5$.
Moreover, they placed Kronberger 49 at $1.2$ kpc from the Galactic centre (assuming $R_{\odot}=8.28$ kpc), indicating that it would be an inner bulge cluster. In conclusion, constructing a radial density profile they proposed that this cluster might be a core collapse GC, similar to many others in the central part of the MW.  They concluded that this is a metal-rich object with $[Fe/H]\approx -0.1$, with an age $t\gtrsim 10$ Gyr.  We find similarities between the present and previous works, since we place Kronberger 49 at $R_G=1.14$ from the Galactic centre and  at $D=8.3$ kpc from the Sun, in agreement with \cite{Ortolani_2006}.  Also, we confirm the metal-rich content and the old age, but we derive a slightly brighter luminosity.

\section{Summary and conclusions}
\label{summaryconclusion}
In recent decades many star clusters have been discovered towards the Galactic bulge. However, most of them have not yet been well studied.  In this work we study 19 star clusters, recovering their main physical parameters.  Reddenings  and extinctions along each cluster field are calculated adopting reddening maps in the near-IR and measuring the RC position.  We also measure their distances using the VVV photometry,  including also the Gaia photometry in some cases. We also search for potential RR Lyrae stars members in order to derive distances as independent method.  Using the isochrone-fitting method with the PARSEC isochrone models we derive their metallicities and ages. We find 11 metal-rich and 3 metal-poor GCs, whereas two GCs show intermediate metallicity.  We find that the majority of our targets are low-luminosity GCs with $M_V$ fainter than the MW GCLF peak.\\

Dedicated photometric studies spread across the bulge concluded that its mean age is of $\sim 10$ Gyr, particularly setting a lower limit on ages higher than 5 Gyr \citep{Clarkson2011}.  In this context,  we confirm that the clusters studied in this work are members of the Galactic bulge,  according to their kinematics, positions, ages ($t>8$ Gyr) and metallicity range ($0.0>$ [Fe/H] $>-1.40$). We also identify three open clusters: FSR 0022,  BICA 523, BICA 524.\\

In conclusion,  we confirm the nature of 9 candidates as bona fide GCs: VVV-CL131, VVV-CL143, VVV-CL160, FSR 0009,  FSR 1767, FSR 1775, ESO 393-12, ESO 456-09, Kronberger 49  On the other hand, the remaining 7 candidates deserve further study to be fully confirmed as GCs, since the effect of the dust strongly reddens their CMDs.  This can be appreciated, especially,  in the CMDs of VVV-CL110, CL119, CL128, CL150,CL153,  which show a wider RGB, which is not typical for bulge GCs, and a not well-defined RC.  Therefore,  follow-up observations are needed, including spectroscopy in order to measure their kinematics/dynamics and chemical composition, and deeper imaging to derive their absolute age using the MSTO and calculate their structural parameters.

\begin{acknowledgements}
We gratefully acknowledge the use of data from the ESO Public Survey program IDs 179.B-2002 and 198.B-2004 taken with the VISTA telescope and data products from the Cambridge Astronomical Survey Unit. ERG acknowledges support from an UNAB PhD scholarship and ANID PhD scholarship No. 21210330. D.M. acknowledges support by the BASAL Center for Astrophysics and Associated Technologies (CATA) through grant FB 210003.  J.A.-G. acknowledges support from Fondecyt Regular 1201490 and from ANID – Millennium Science Initiative Program – ICN12\_009 awarded to the Millennium Institute of Astrophysics MAS.
\end{acknowledgements}

\bibliographystyle{aa.bst}
\bibliography{bibliopaper}
\newpage
\begin{table}[th]
\onecolumn
\centering 
\caption{Position, selection cuts for the PM-decontamination procedure and mean cluster PMs for the GC candidates.}
\begin{adjustbox}{max width=\textwidth}
\begin{tabular}{lcccccc}
\hline\hline
Cluster ID & RA & Dec & r & $\mu_{\alpha_{\ast}}$ & $\mu_{\delta}$ & $\sigma_{PM}$ \\
			    & [hh:mm:ss] & [dd:mm:ss] & [arcmin] &[mas yr$^-1$] &[mas yr$^{-1}$] & [mas yr$^{-1}$] \\
\hline
FSR0009   & 18:28:30.6 & -31:54:24 & 3.6 & $ -1.39 \pm1.10 $  & $ -5.22 \pm 0.99$ & 2.0\\
FSR1775   & 17:56:05.3 & -36:33:57 &3.0 & $-3.00 \pm 0.80$ & $-5.53 \pm 0.73$ & 1.0\\
FSR1767   & 17:35:43.0 & -36:21:28 & 3.0 & $-3.02 \pm 0.50$ & $-4.85 \pm 0.50$ & 1.0\\
VVVCL110  & 17:22:47.0 & -34:41:17 & 3.0 & $-4.25 \pm 0.48$ & $-6.23 \pm 0.48$ & 1.0 \\
VVVCL119  & 17:30:46.0 & -32:39:05 & 2.4 &$ -3.93 \pm 0.92$ & $-6.07 \pm 0.93$ & 2.0\\
VVVCL128  & 17:39:59.0 & -32:26:27 & 3.0 &$ -4.22 \pm 0.50$ & $-6.22\pm 0.50$ & 1.0\\
VVVCL131  & 17:41:17.0 & -34:34:02 & 3.0 &$-3.24 \pm 0.81$ &$-5.65\pm 0.07$ &1.0\\
VVVCL143  & 17:44:36.0 & -33:44:18 & 3.0 &$-3.18 \pm 0.91$ & $-6.17 \pm 0.85$ &1.0\\
VVVCL150  & 17:50:41.0 & -25:13:06 & 3.0 &$-3.33 \pm 0.48$ & $-5.  87\pm 0.50$ &1.0\\
VVVCL153  & 17:53:32.0 & -25:22:56 &1.8 & $-3.96 \pm 0.67$ &$-6.44 \pm 0.69$ & 1.5\\
VVVCL154  & 17:55:08.0 & -28:06:01 &3.0& $-1.04\pm 0.49$ & $-3.84 \pm 0.48$ &1.0\\
VVVCL160 & 18:06:57.0  &-20:00:40 & 3.0 & $-2.90 \pm 1.28$ & $-16.47\pm 1.31$ & 1.0 \\
ESO393-12 & 17:38:37.6 & -35:39:02 &3.0 & $-2.86 \pm 0.47$&  $-5.39 \pm 0.44$ &1.0\\
ESO456-09 & 17:53:54.3 & -32:27:58 &3.0&$ -3.41 \pm 0.71$ & $ -4.36 \pm  0.75$&1.5\\
DB044     & 17:46:35.0 & -24:53:28 & 3.0 &$-3.68 \pm 1.61$ & $-6.43 \pm 1.14$&2.0\\
Kronberger49    & 18:10:23.9 & -23:20:25 &3.0& $-2.84\pm 0.69$ & $-5.52 \pm 0.71$&1.0\\
\hline\hline
\end{tabular}
\end{adjustbox}
\label{DP}
\end{table}

\begin{table}[th]
\onecolumn
\centering
\caption{PMs comparison between field and cluster stars. }
\begin{tabular}{lcccc}
\hline\hline 
Cluster+Field & $\mu_{\alpha}^{\ast}$ & $\sigma_{\mu_{\alpha}^{\ast}}$ & $\mu_{\delta}$ & $\sigma_{\mu_{\delta}}$\\
 & [mas yr$^{-1}$] &[mas yr$^{-1}$] &[mas yr$^{-1}$] &[mas yr$^{-1}$] \\
\hline
FSR 0009 &  -1.39 &0.93 & -5.1 &1.09 \\
Field 	    & -1.25 & 5.08 &-3.90 & 4.95\\
\hline
FSR 1775 & -3.00  &  0.80 &  -5.53 &0.73\\
Field  	  & -1.61 & 4.86 &   -3.99 & 5.08\\
\hline
FSR 1767 & -3.01  &  0.49 &  -4.85&0.49\\
Field  	  & -2.24& 4.05 &   -3.88 & 4.14\\
\hline
VVV-CL110 & -4.25  &  0.48 &  -6.23 &0.48\\
Field  	  & -3.38& 3.64 &   -4.08 & 3.66\\
\hline
VVV-CL119 & -3.93  &  0.91 &  -6.07 &0.93\\
Field  	  & -3.20& 3.48 &   -3.96 & 3.58\\
\hline
VVV-CL128 & -4.22  &  0.50 &  6.22 &0.50\\
Field  	  & -3.14& 4.43 &   -4.69 & 4.68\\
\hline
VVV-CL131 & -3.55  &  0.78 &  -5.84 &0.82\\
Field  	  & -2.74& 3.31 &   -4.41 & 3.81\\
\hline
VVV-CL143 & -3.18  &  0.91 &  -6.18 &0.85\\
Field  	  & -2.54& 3.24 &   -4.64 & 4.1\\
\hline
VVV-CL150 & -3.33  &  0.5 &  -5.87 &0.5\\
Field  	  & -2.33& 4.40 &   -3.61 & 4.50\\
\hline
VVV-CL153 & -3.96  &  0.67 &  -6.44 &0.68\\
Field  	  & -2.35& 4.49 &  -4.13 & 4.60\\
\hline
VVV-CL154 & -1.04  &  0.49 &  -3.85 &0.48\\
Field  	  & -2.67& 6.53 &  -4.09 & 6.38\\
\hline
VVV-CL160 & -2.90  &  1.21 &  -16.48 &1.26\\
Field  	  &-3.22 & 5.1 & -4.14 & 5.02\\
\hline
ESO 393-12& -2.83  &  0.49 &  -5.37 &0.46\\
Field  	  &-2.06 &2.62 & -4.57& 2.90\\
\hline
ESO 456-09& -3.41  &  0.71 &  -4.36 &0.75\\
Field  	  &-2.43 &3.32 & -5.09& 3.42\\
\hline
DB 044& -3.37  &  1.3 &  -6.1&1.3\\
Field  	  &-2.61 &3.17 & -4.92& 3.33\\
\hline
Kronberger 49 & -2.84  &  0.69  &  -5.51&0.71\\
Field  	  &-1.93 &2.78 & -4.35& 3.02\\
\hline
\hline\hline
\end{tabular}
\label{comparison}
\end{table}

\begin{table}[th]
\onecolumn
\centering 
\caption{Main physical parameters for each star cluster candidate, using the VVV photometry.  The absolute magnitude in $V-$band is calculated from the comparison with the well-known GCs (see Table \ref{tablelum}).  We mark with a question mark those clusters for which we need further data to unveil their nature.}
\begin{adjustbox}{max width=\textwidth}
\begin{tabular}{lccccccccccccc}
\hline\hline
Cluster ID	 &   $K_s$ (RC)   &     $E(J-K_s)$  &   $ A_{Ks} $&   $(m-M)_0$ &  $D_{VVV}$  & $R_G$ & $M_{Ks}$ & $M_V$  & [Fe/H] &[Fe/H]$_{s}$\tablefootmark{a}   & Age & Type  \\
   &[mag] &[mag]&[mag]&[mag] &[kpc] &[kpc] &[mag]&[mag]& dex & dex & Gyr \\
\hline
FSR0009 & -- & $0.25 \pm0.04$ & $0.11\pm 0.03$& $14.20 \pm 0.04 $ & $6.9 \pm 0.2$ & 1.81 & $-5.8 \pm 0.7$ & $-3.4$  & $-1.2\pm 0.3$  & $-1.10$ & $11\pm 2$ & GC \\
FSR1775& -- & $0.37\pm 0.03$ & $0.16 \pm 0.01$ & $14.75 \pm 0.02$ & $8.9 \pm  0.2$ & 1.37&$-8.0\pm1.7$ & $-5.6$ & $-1.1\pm0.2$ & $-1.10$ & $10 \pm 2$ & GC \\
FSR1767& $13.8\pm 0.03$ & $0.66 \pm 0.04$& $0.28\pm 0.03$ &$15.12\pm 0.04$ & $10.6\pm 0.2$ & 2.70& $-8.4\pm 1.5$& $-6.3$ &  $-0.7\pm0.2$   & $-0.73$& $11\pm 2$ & GC \\
VVVCL110& $14.5\pm 0.03$  &$2.0\pm0.04$ & $0.86\pm 0.02 $ & $15.25\pm 0.04$ &$11.2\pm 0.5$ &3.25& $-8.6 \pm 1.6$ & $-6.8$&  $-0.1 \pm 0.2$ & $-0.48$ & $11 \pm 2$ & GC? \\
VVVCL119& $14.5\pm 0.05$ & $1.98\pm0.07$ & $0.85 \pm 0.04$ & $15.26\pm 0.06$ & $11.3\pm 0.6$ &2.24& $-9.3\pm 1.8$ & $-7.5$ &  $-0.1 \pm 0.2$ & $-0.48$& $10\pm 3$ &GC? \\
VVVCL128&$14.2\pm 0.05$ & $1.20\pm 0.05 $ &$0.51 \pm 0.04 $ & $15.29\pm 0.06$ & $11.4 \pm 0.5$ & 3.26&$-7.5\pm1.6$ &$-5.7$& $\ \ \ 0.0\pm 0.3$ & $-0.11$ & $10\pm 3$ & GC? \\
VVVCL131& $13.4\pm 0.03 $ & $0.54\pm 0.04 $ &$0.23\pm 0.03$ &$14.77 \pm 0.04$ & $9.0\pm 0.5 $ &1.17& $-8.2\pm 1.5$ &$-5.9$ & $-0.6\pm 0.2$ & $-0.73$  &  $10\pm3$ &GC \\
VVVCL143& $13.4\pm 0.03 $  & $0.50\pm 0.05$  & $0.21\pm 0.04$ & $14.74\pm 0.05 $& $8.9\pm 0.5$ &1.0& $-8.2 \pm 1.3$& $-5.9$ &$-0.6\pm 0.2$ &$-0.61$  &$10\pm3$ & GC \\
VVVCL150 & -- & $1.32 \pm 0.04$ & $0.56\pm 0.03 $ &$14.54\pm 0.04$ & $8.1\pm 0.5$ &0.56&$-8.7 \pm 1.7$& $-6.5$ & $-1.3\pm 0.1$ & $-1.35$  & $12\pm 2$ & GC? \\
VVVCL153& $14.1\pm 0.02 $ & $1.67\pm 0.04$ &$0.71 \pm 0.03$ & $14.99\pm 0.04$ & $10 \pm 0.4$ & 1.90& $-8.6\pm 1.9$ &$-6.8$ & $-0.1\pm 0.1$ & $-0.23$  & $10\pm 3$ & GC?\\
VVVCL154& $13.5\pm 0.03$  &  $0.85\pm 0.05$ & $0.36 \pm 0.04$ & $14.47\pm 0.05$ & $8.9 \pm 0.5$& 0.78 &$-9.6\pm 1.8$& $-7.2$ &$-0.6\pm 0.2$ & $-0.67$  & $11\pm 2$& GC? \\
VVVCL160 & -- & $1.71\pm 0.04$ & $0.73\pm 0.03$ &$14.17\pm 0.04$ & $ 6.8\pm 0.5 $ & 1.92  &$-7.9\pm1.5$ & $-5.5$ & $-1.4\pm 0.3$ & $-1.47$  & $13\pm 2$ & GC\\
ESO393-12& $13.2\pm 0.02$  &$0.54 \pm 0.03$ & $0.23\pm 0.02$ & $14.57 \pm 0.03$ & $8.2\pm 0.4$  & 0.98& $-7.7\pm1.5$& $-5.3$  & $-0.6\pm 0.2$ & $-0.61$ & $10\pm 2$ &  GC \\
ESO456-09& $13.0\pm 0.02$  &  $0.44\pm 0.04 $& $0.18\pm 0.03$ & $14.42 \pm 0.04$ &$7.6\pm 0.4$  & 0.81& $-8.3\pm 1.5$ & $-6.0$ &  $-0.6\pm 0.2$  &$-0.61$ & $10\pm 2$ & GC \\
DB044    & $13.2\pm0.02$  & $0.69\pm 0.03$ & $0.30\pm 0.02$& $14.51\pm 0.1$ &$8.0 \pm 0.5$ &0.61&$-8.0\pm 1.6 $ & $-6.0$ & $-0.1\pm0.2$ &$-0.11$  & $11\pm2$ & GC? \\
Kronberger49& $13.3\pm 0.03$  &$0.69\pm 0.04$ & $0.30\pm 0.03$ &$14.61\pm 0.04$ &$8.3\pm 0.5$ & 1.14&$-8.5\pm 1.5$ &$-6.7$&$-0.2\pm 0.2$ & $-0.36$   &$11\pm 2$ & GC\\
\hline\hline
\end{tabular}
\end{adjustbox}
\label{parameter}
\tablefoot{
   \tablefoottext{a}{The [Fe/H] estimates are derived from the RGB-slope using the relation by \cite{Cohen_2015}}.}    
\end{table}

\begin{table}[th]
\onecolumn
\centering 
\caption{Optical parameters derived for those GCs on which it was possible to obtain adequate Gaia EDR3 data, used to confirm the VVV results. }
\begin{adjustbox}{max width=\textwidth}
\begin{tabular}{lcccc}
\hline\hline
Cluster ID	& $G$(RC)& $A_G$ &   $(m-M)_0$ &  $D_{GEDR3}$  \\
   &[mag]&[mag]&[mag]&[kpc] \\
\hline
FSR0009 & $15.7\pm 0.003 $ & $1.08\pm 0.3$ & $14.12\pm 0.3$ & $6.7\pm 0.8$ \\
FSR1775&$16.9\pm 0.004$&$1.70\pm 0.6$&$14.70\pm 0.6$&$8.7\pm 1.6$\\
VVVCL131 &$18.1\pm 0.003$&$2.80\pm 0.1$&$14.80\pm 0.1$&$9.1\pm 0.5$\\
VVVCL143 &$18.0\pm 0.003$&$2.80\pm 0.1$&$14.70\pm 0.1$&$8.7\pm 0.5$\\
ESO393-12& $18.0\pm 0.005$ &$3.20\pm 0.7$ & $14.30\pm 0.7 $ & $7.3\pm 1.9 $ \\
ESO456-09& $17.2\pm 0.003 $ &$2.40\pm 0.7$ & $14.31\pm 0.7$ & $7.3\pm 1.9$  \\
DB044 & $18.8\pm 0.02$ & $4.0\pm 0.5$&$14.30\pm 0.5$&$7.3\pm 1.2$\\
Kronberger49& $18.8\pm 0.01$ &$3.80\pm 0.2$ &$14.50\pm 0.2$ &$8.0\pm 0.6$ \\
\hline\hline
\end{tabular}
\end{adjustbox}
\label{optparam}
\end{table}

\begin{table}
\onecolumn
\centering 
\caption{Known and well-characterised GCs used to derive the total luminosity for each GC. }
\begin{tabular}{lccccl}
\hline\hline
Cluster ID &    $[Fe/H]$\tablefootmark{a}  &    $E(B-V)$ & $(m-M)_0$   & $M_{V}$\tablefootmark{b}&  References \\
                  &  [dex] & [mag] & [mag] &[mag]& \\
\hline
NGC 6528 & $-0.11$ & 0.54 & 16.17 & -5.50 &\cite{Harris1996} (H96)\\
NGC 6553 & $-0.18$ & 0.63 & 15.83 & -7.01 &H96\\
Terzan 5    &$-0.23$ & 2.28 &13.85 & -7.08 & H96\\
Liller 1 & $-0.33$ & 3.09 & 14.48 & -9.68&\cite{Valenti2010} (V10)  \\ 
NGC 6440 & $-0.36$ &1.15 & 14.58&-7.70&\cite{Valenti2004} (V04)  \\
NGC 6441 & $-0.46$ & 0.52 & 15.65&-8.45 &V04 \\
NGC 6624 & $-0.44$ &0.31  &17.33 & -8.86 &\cite{Siegel2011} (S11)\\
Terzan 6 & $-0.56$ & 2.35 & 14.13&-5.98&\cite{Valenti2007} (V07)  \\
Terzan 12 & $-0.50$ &2.06 & 12.65 &-4.64& \cite{Ortolani1998}\\
NGC 6637 & $-0.64$ & 0.22  & 17.35& -8.90& S11\\
Terzan 2 & $-0.69$ &1.40 & 14.30&-5.72&\cite{Christian1992}  \\
BH 261 & $-0.76$ &0.36 &13.90 &-3.16& \cite{Ortolani2006}\\
NGC 6569 & $-0.76$ &0.49 &15.40 & -7.05&V07 \\
UKS 1 & $-0.98$ &2.2 & 16.01& -9.20&\cite{Minniti2011}\\
NGC 6638 & $-0.95$ & 0.43&15.07 &-6.10&V07\\
Terzan 9 & $-1.05$ & 1.79 &13.73 & -5.60&V10\\
NGC 6642 & $-1.26$&0.42 &14.30 & -4.70 &\cite{Barbuy2006} \\
NGC 6626 & $-1.32$ & 0.42 &13.70 & -6.04&\cite{Kerber2018}\\
NGC 6540 & $-1.35$ &0.66 &13.57 & -4.37&V10 \\
NGC 6558 & $-1.32$ &0.50  & 14.59& -5.12&\cite{Barbuy2018} \\
NGC 6453 & $-1.50$ & 0.69 & 15.15 & -5.83&V10\\
NGC 6715 & $-1.49$ &0.14 &17.27 & -7.62&S11\\
\hline\hline
\end{tabular}
\label{tablelum}
\tablefoot{
   \tablefoottext{a}{The [Fe/H] values are taken from the 2010 version of the \cite{Harris1996} catalogue.} 
 \tablefoottext{b}{The $M_{V}$ values are calculated using the VVV/VVVX datasets, following the same procedure as done for the clusters analysed in the present work (see Sec. 4). }    
   }
\end{table}

\begin{appendix}
\section{VPMs and CMDs for each cluster}
\label{appendix}

\begin{figure}[!htb]
\centering
\includegraphics[width=7cm, height=7cm]{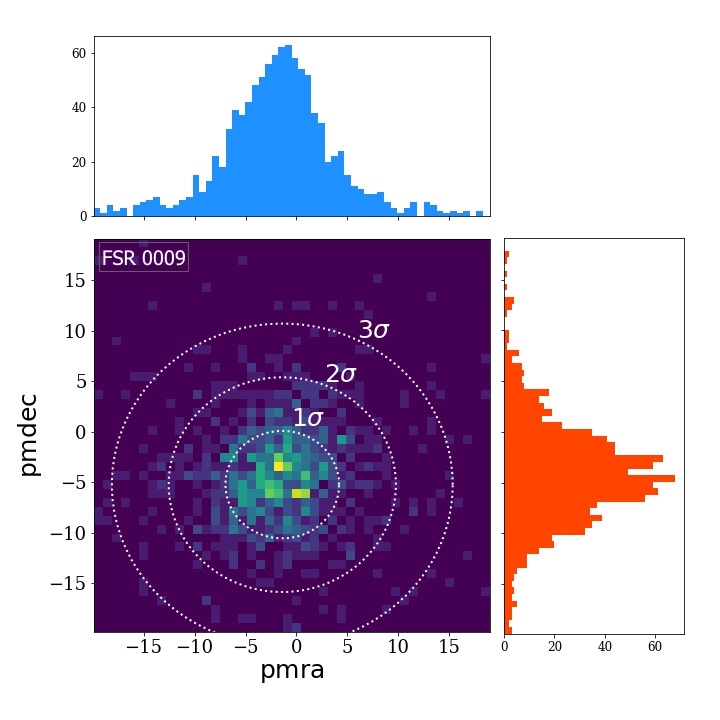} 
\includegraphics[width=7cm, height=7cm]{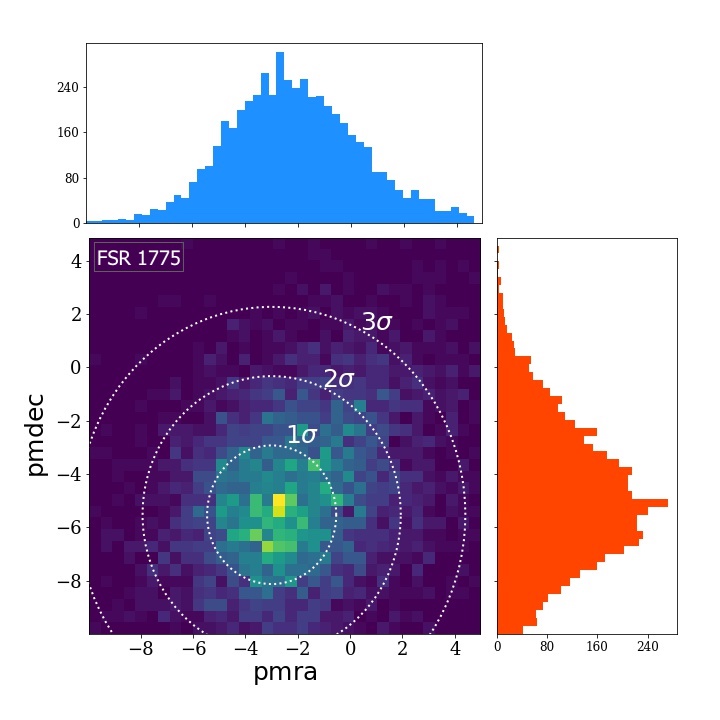} 
\includegraphics[width=7cm, height=7cm]{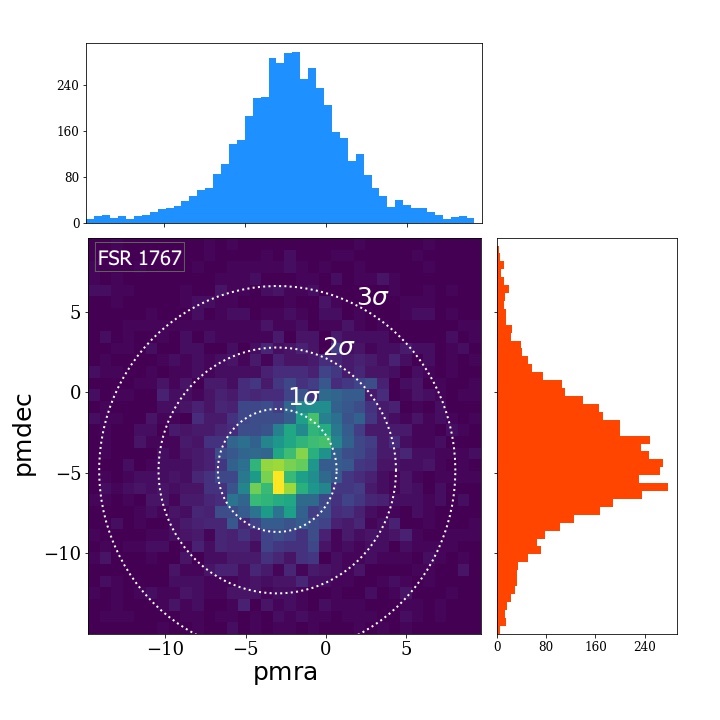} 
\includegraphics[width=7cm, height=7cm]{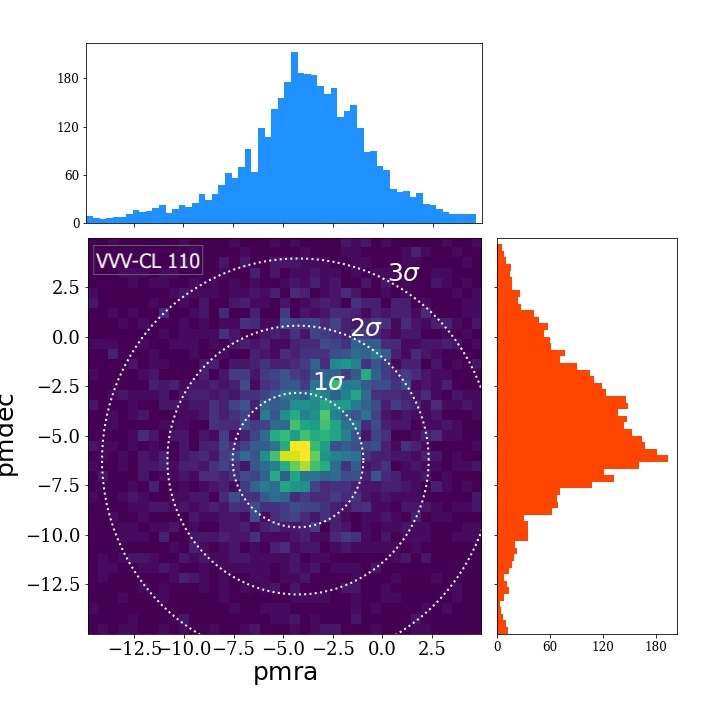} 
\includegraphics[width=7cm, height=7cm]{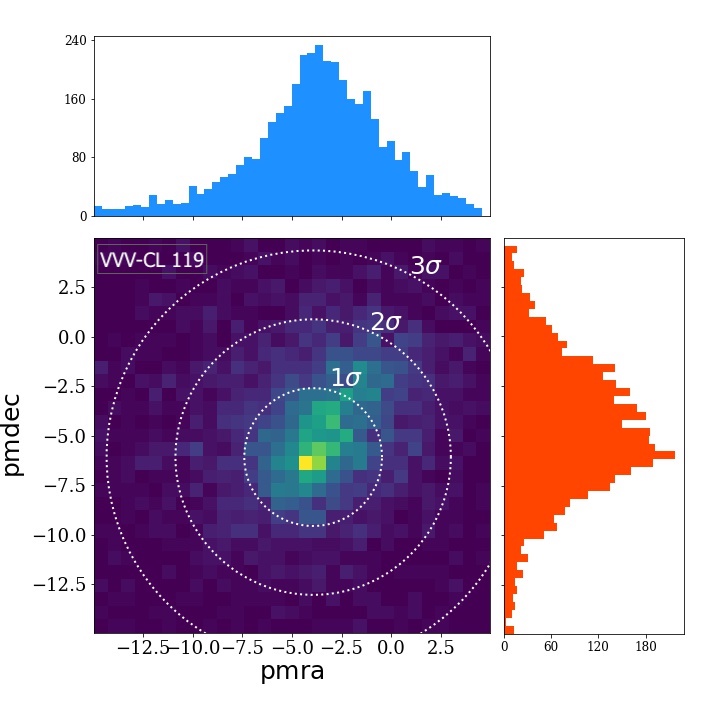} 
\includegraphics[width=7cm, height=7cm]{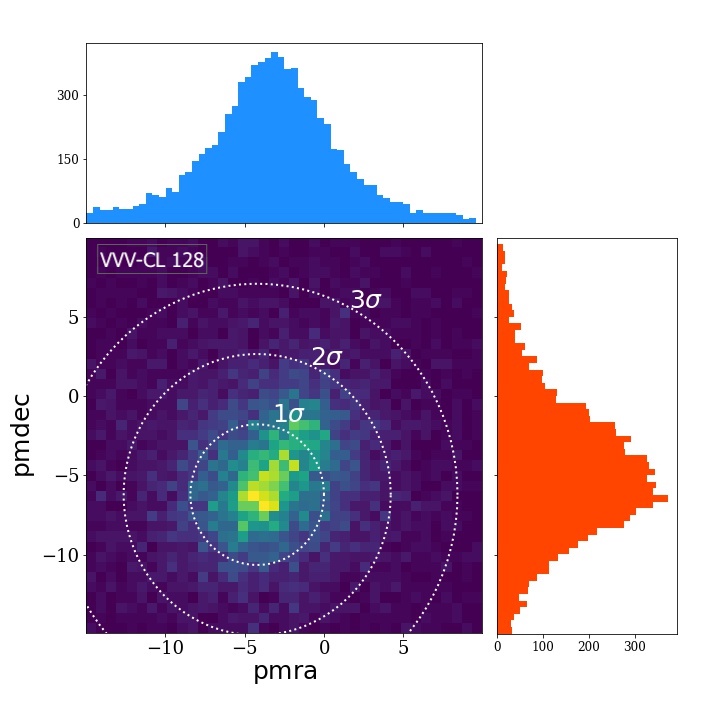} 
\caption{VPM,  as 2D histogram diagram, of the stars in the cluster regions within the given radius $r$ from its centre.  Yellow areas are representative of over-densities,  which become green and blue when the density decreases. White dot circles depict the 1$\sigma$, 2$\sigma$, 3$\sigma$ from the mean cluster PM value.  On the top and on the right panels, we show the corresponding PM histograms, pmra (blue) and pmdec (red), respectively.  We use different bin size, depending on the crowding of the areas.  \textit{Note:} We add an insert for the VVV-CL160 cluster in order to make much more visible the PM over-density.}
\label{vpm}
\end{figure}

\begin{figure}[!htb]
\ContinuedFloat
\captionsetup{list=off, format=continued}
\centering
\onecolumn
\includegraphics[width=7cm, height=7cm]{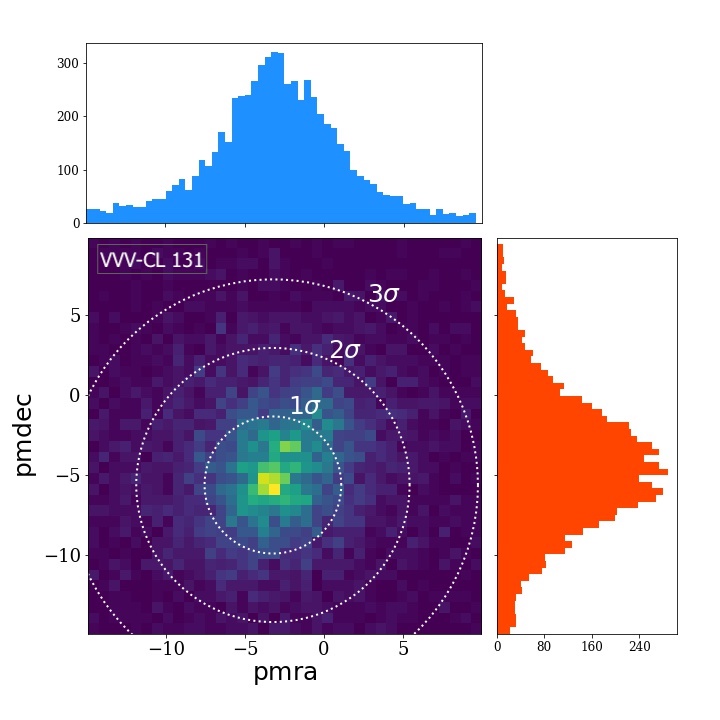} 
\includegraphics[width=7cm, height=7cm]{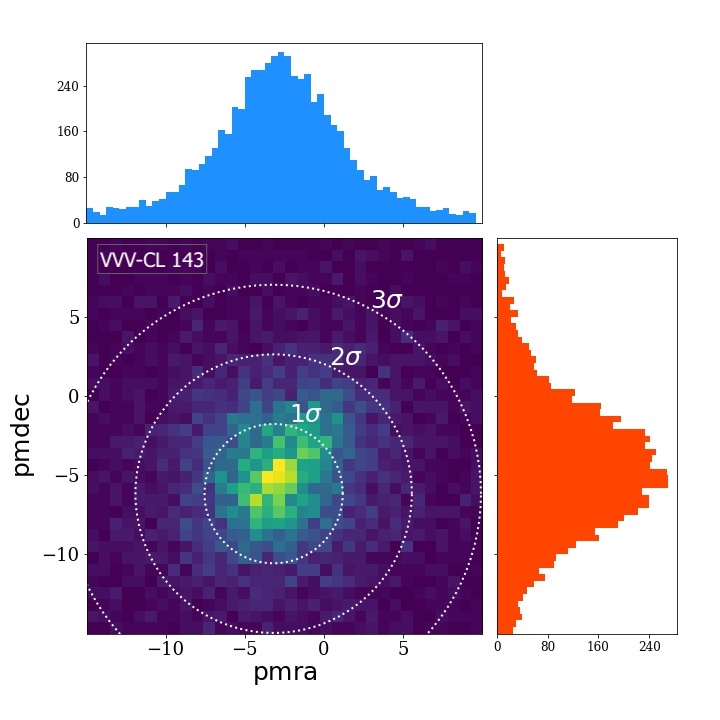} 
\includegraphics[width=7cm, height=7cm]{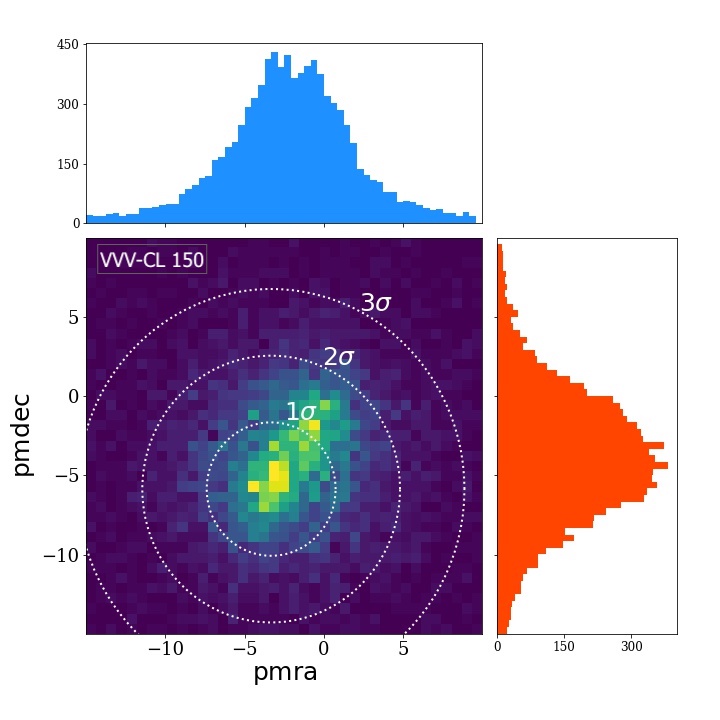} 
\includegraphics[width=7cm, height=7cm]{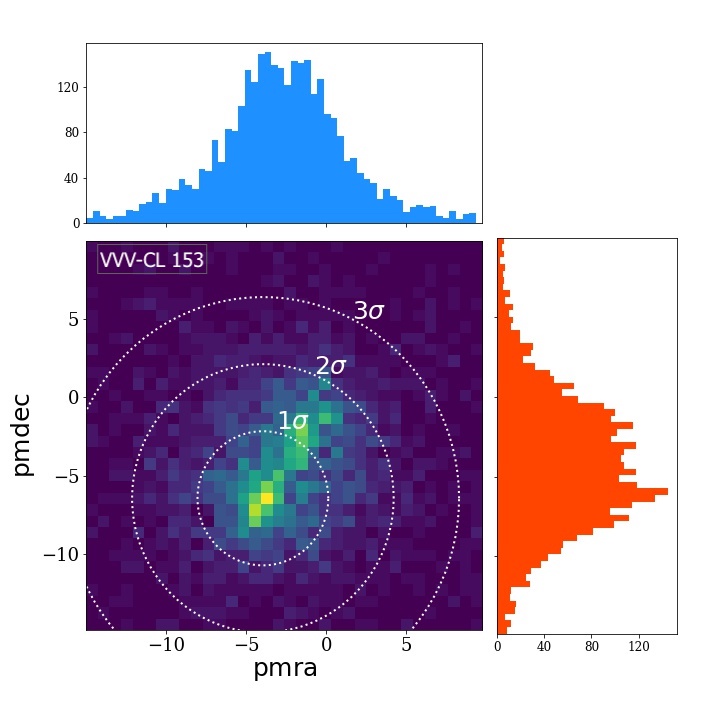} 
\includegraphics[width=7cm, height=7cm]{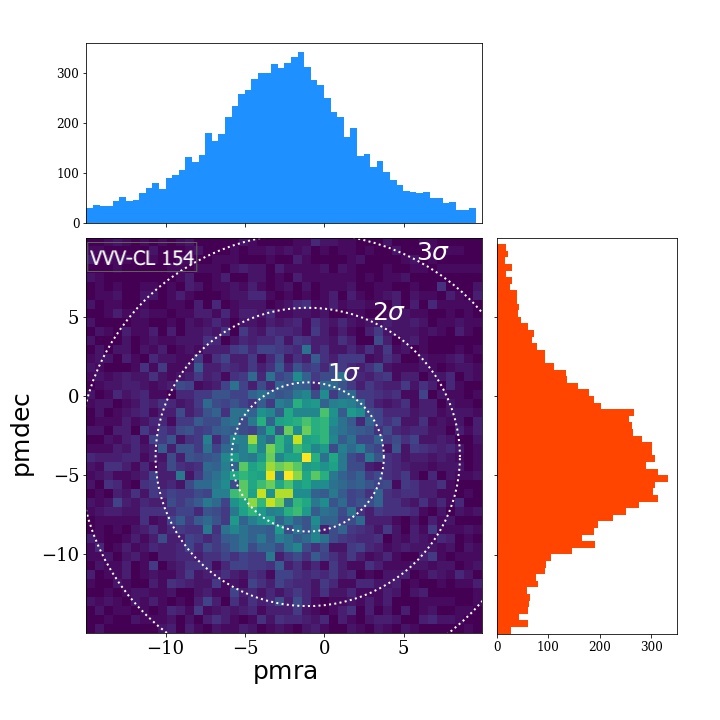} 
\includegraphics[width=7cm, height=7cm]{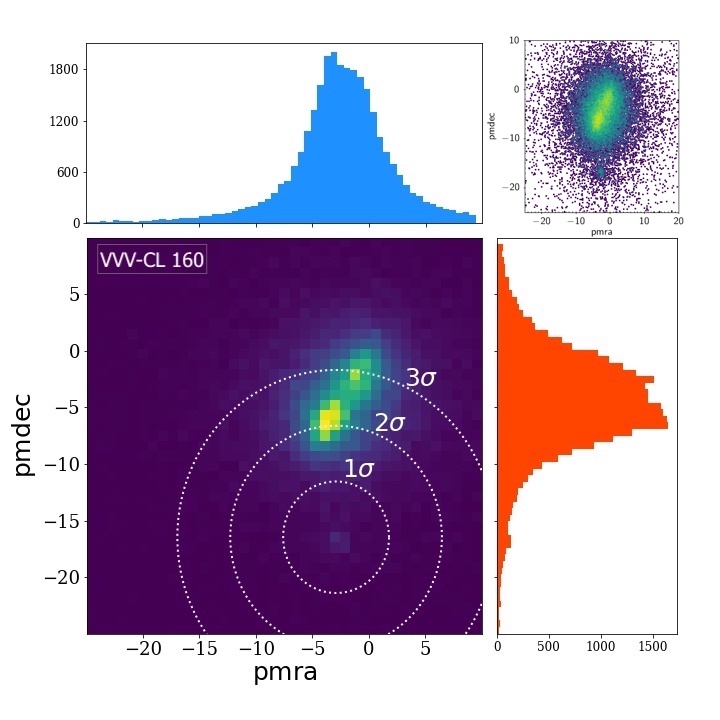} 
\caption{}
\label{vpm}
\end{figure}

\begin{figure}[!htb]
\ContinuedFloat
\captionsetup{list=off, format=continued}
\centering
\onecolumn
\includegraphics[width=8cm, height=8cm]{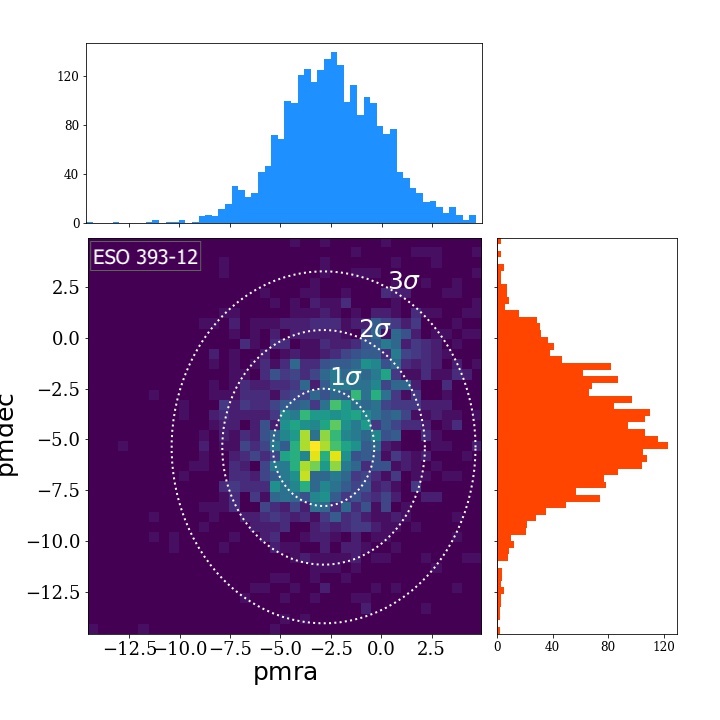} 
\includegraphics[width=8cm, height=8cm]{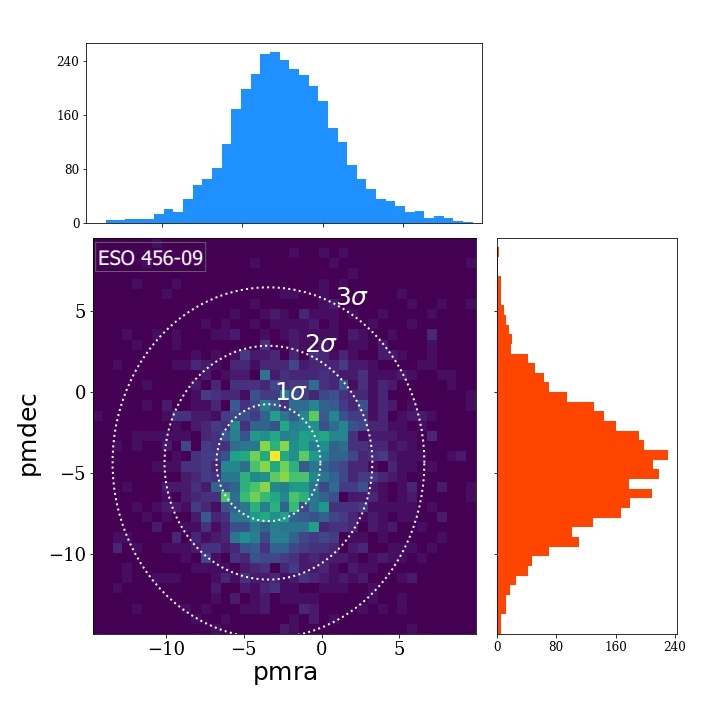} 
\includegraphics[width=8cm, height=8cm]{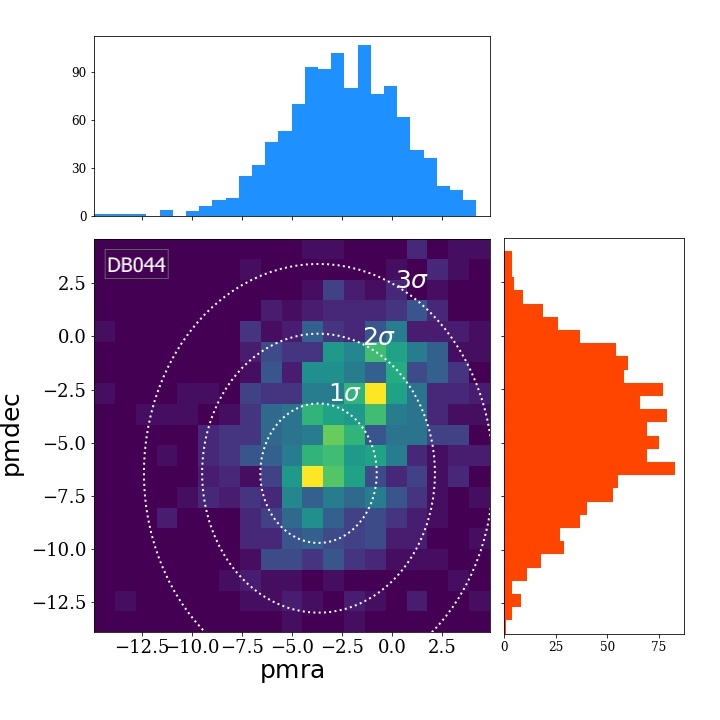} 
\includegraphics[width=8cm, height=8cm]{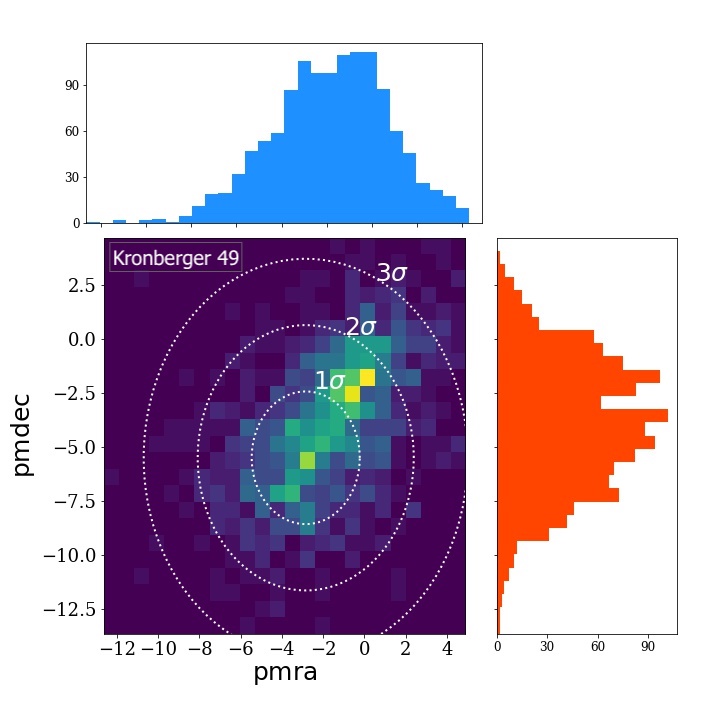} 
\caption{}
\label{vpm}
\end{figure}

\begin{figure}[!htb]
\centering
\includegraphics[width=5cm, height=5cm]{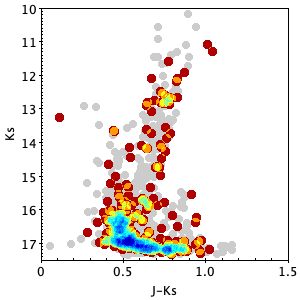} 
\includegraphics[width=5cm, height=5cm]{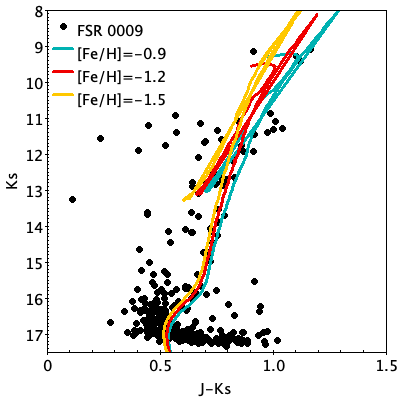} 
\includegraphics[width=5cm, height=5cm]{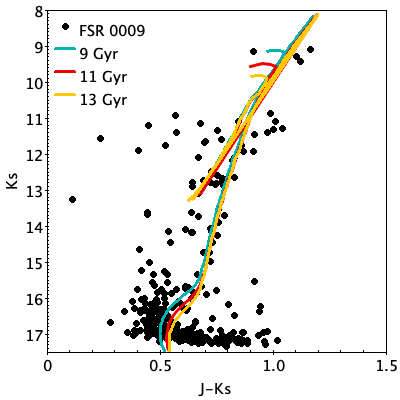} 
\includegraphics[width=5cm, height=5cm]{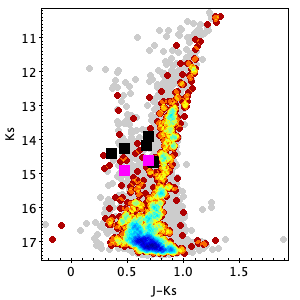} 
\includegraphics[width=5cm, height=5cm]{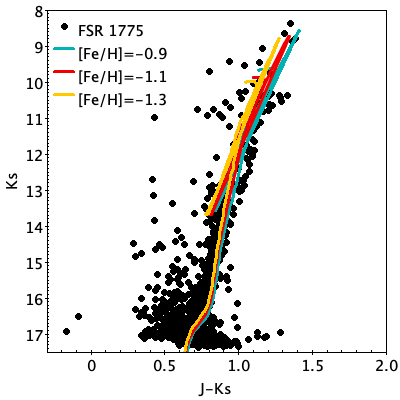} 
\includegraphics[width=5cm, height=5cm]{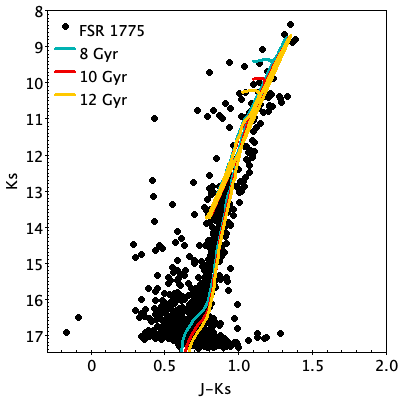} 
\includegraphics[width=5cm, height=5cm]{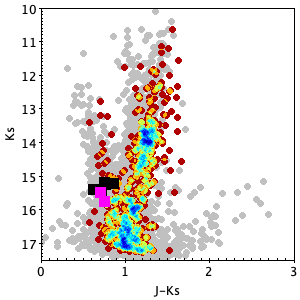} 
\includegraphics[width=5cm, height=5cm]{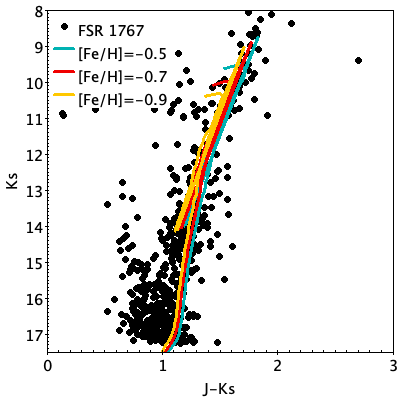} 
\includegraphics[width=5cm, height=5cm]{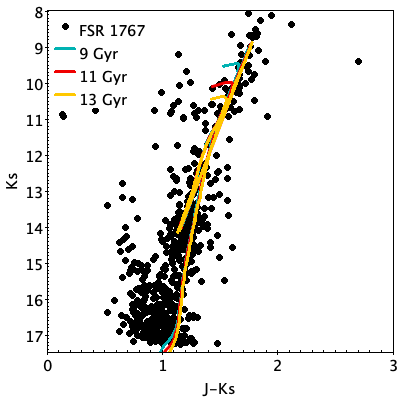} 
\caption{\textit{Left panel:} VVV CMD for contaminated (grey points) and PM-decontaminated samples (Hess diagram).  We show the position in the CMD of the RR Lyrae stars located at $12'$ from the cluster centres (black squared) and those considered cluster members (magenta squared). \textit{Middle and Right panesl:} 
2MASS+VVV CMDs for PM-selected members.  We fit a family of isochrones (cyan, red,  and yellow lines), changing metallicities and ages, respectively. The red line is the PARSEC isochrones best-fit for each cluster (see Table \ref{parameter}). }
\label{plots}
\end{figure}

\begin{figure}[!htb]
\ContinuedFloat
\captionsetup{list=off, format=continued}
\centering
\onecolumn
\includegraphics[width=5cm, height=5cm]{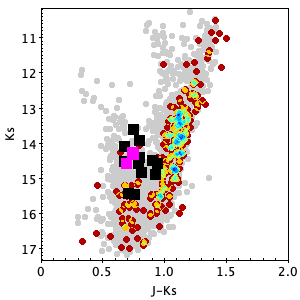} 
\includegraphics[width=5cm, height=5cm]{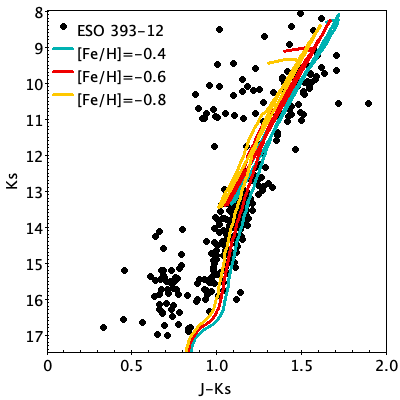} 
\includegraphics[width=5cm, height=5cm]{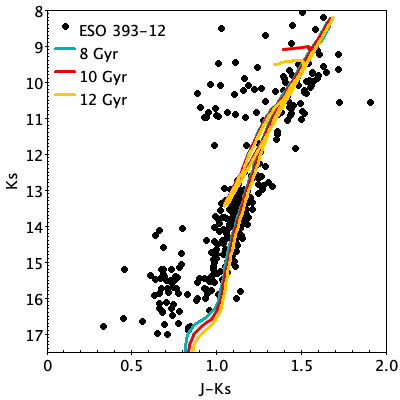} 
\includegraphics[width=5cm, height=5cm]{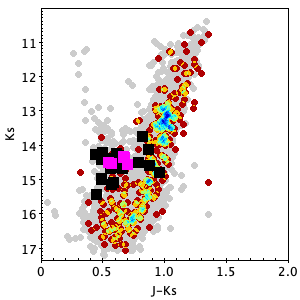} 
\includegraphics[width=5cm, height=5cm]{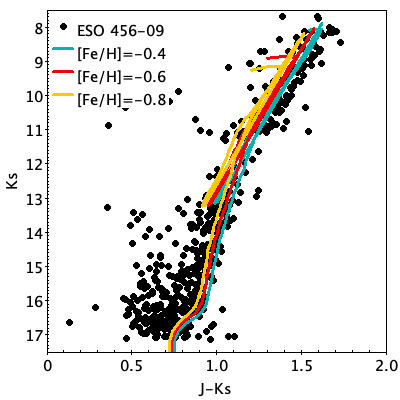} 
\includegraphics[width=5cm, height=5cm]{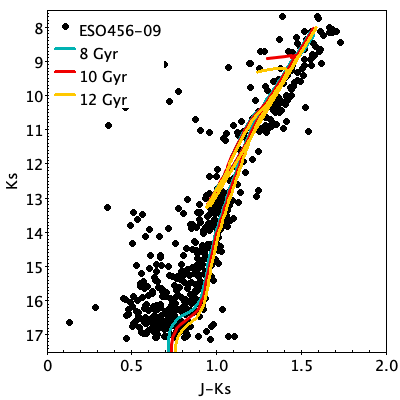} 
\includegraphics[width=5cm, height=5cm]{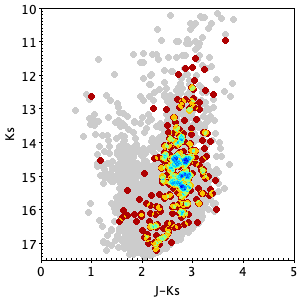} 
\includegraphics[width=5cm, height=5cm]{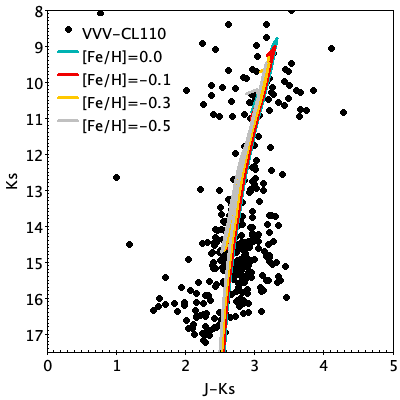} 
\includegraphics[width=5cm, height=5cm]{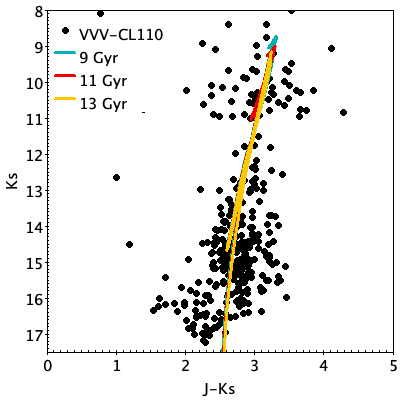} 
\includegraphics[width=5cm, height=5cm]{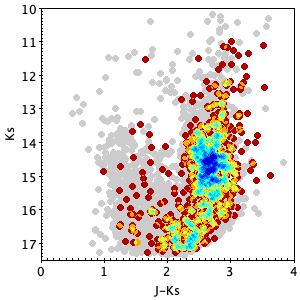} 
\includegraphics[width=5cm, height=5cm]{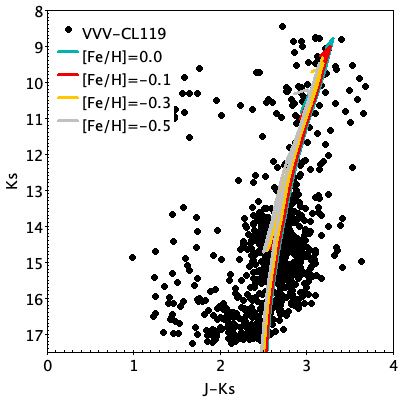} 
\includegraphics[width=5cm, height=5cm]{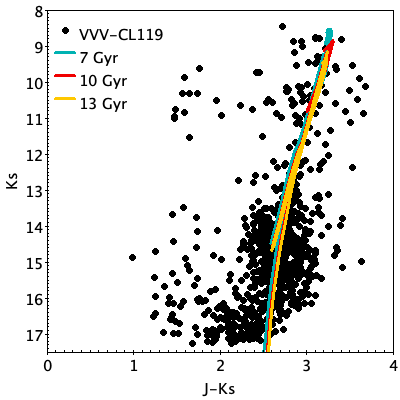} 
\caption{}
\label{lf2}
\end{figure}

\begin{figure}[!htb]
\ContinuedFloat
\captionsetup{list=off, format=continued}
\centering
\onecolumn
\includegraphics[width=5cm, height=5cm]{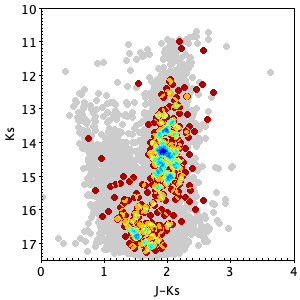} 
\includegraphics[width=5cm, height=5cm]{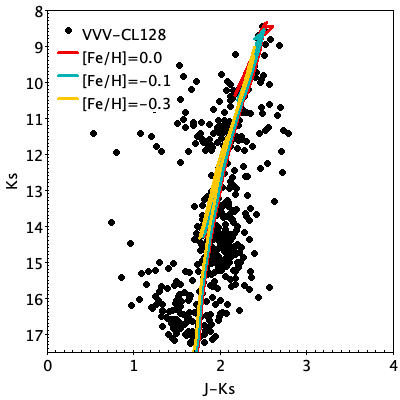} 
\includegraphics[width=5cm, height=5cm]{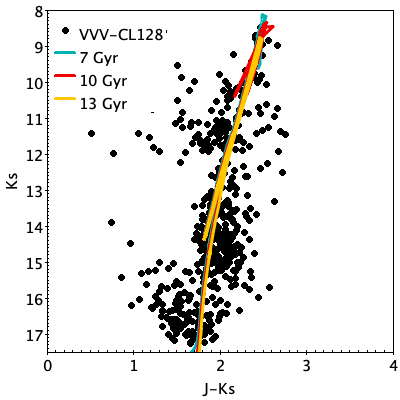} 
\includegraphics[width=5cm, height=5cm]{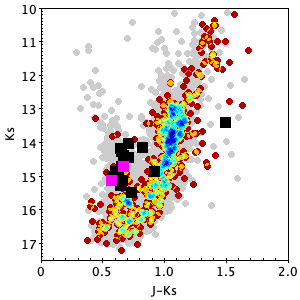} 
\includegraphics[width=5cm, height=5cm]{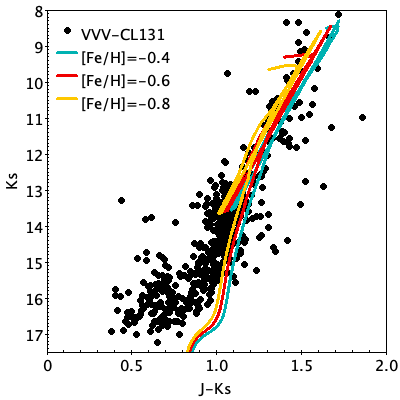} 
\includegraphics[width=5cm, height=5cm]{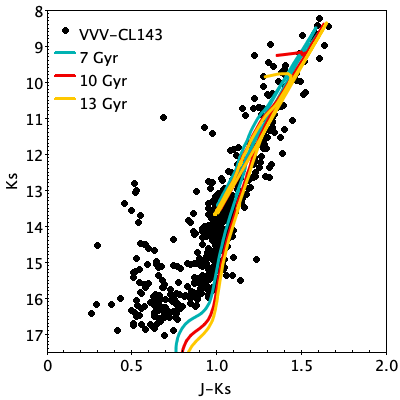} 
\includegraphics[width=5cm, height=5cm]{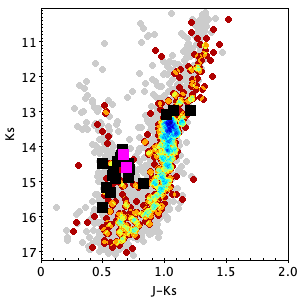} 
\includegraphics[width=5cm, height=5cm]{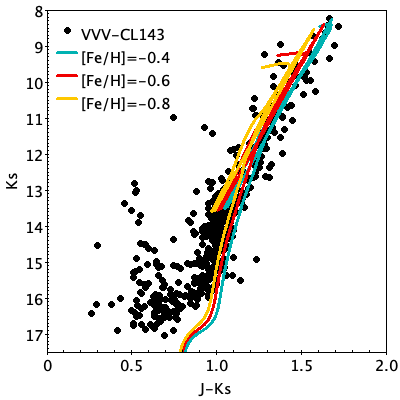} 
\includegraphics[width=5cm, height=5cm]{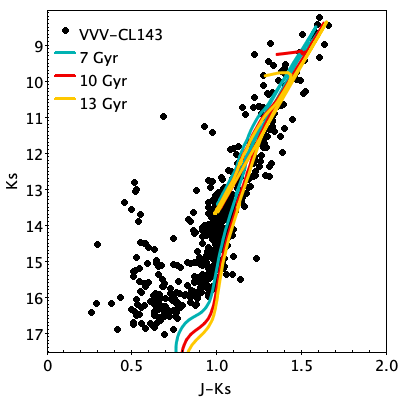} 
\includegraphics[width=5cm, height=5cm]{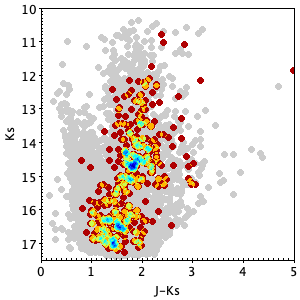} 
\includegraphics[width=5cm, height=5cm]{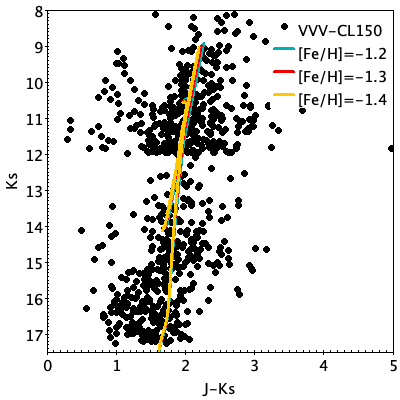} 
\includegraphics[width=5cm, height=5cm]{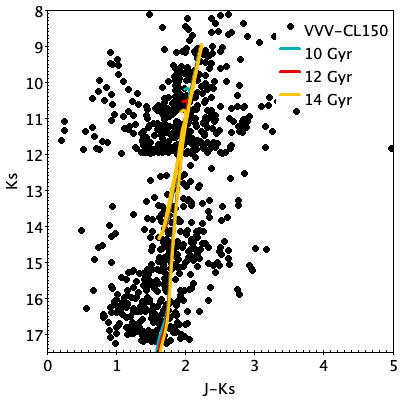} 
\caption{}
\label{lf2}
\end{figure}

\begin{figure}[!htb]
\ContinuedFloat
\captionsetup{list=off, format=continued}
\centering
\onecolumn
\includegraphics[width=5cm, height=5cm]{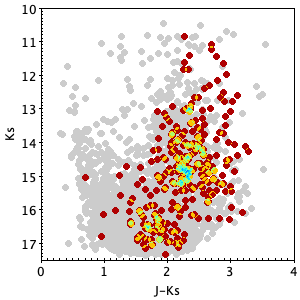} 
\includegraphics[width=5cm, height=5cm]{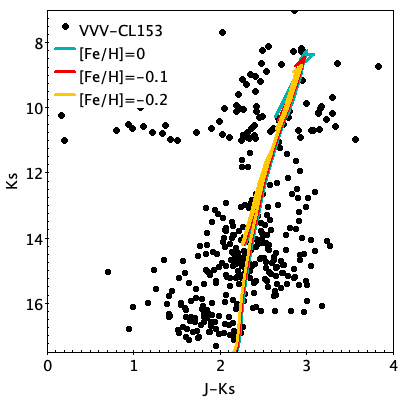} 
\includegraphics[width=5cm, height=5cm]{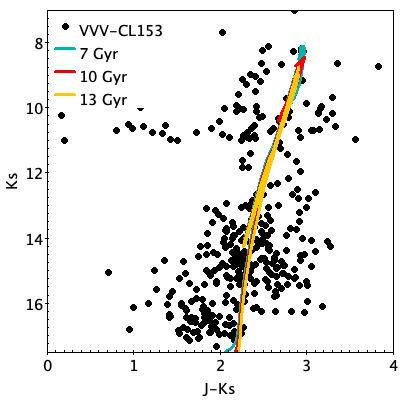} 
\includegraphics[width=5cm, height=5cm]{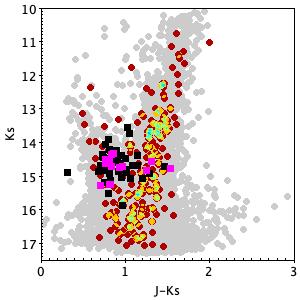} 
\includegraphics[width=5cm, height=5cm]{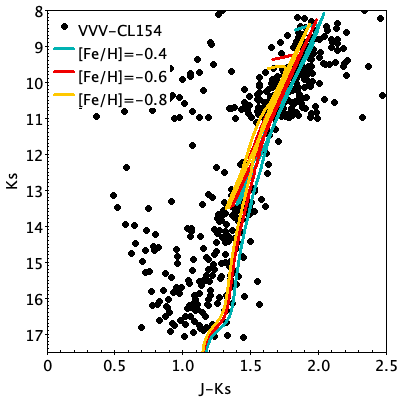} 
\includegraphics[width=5cm, height=5cm]{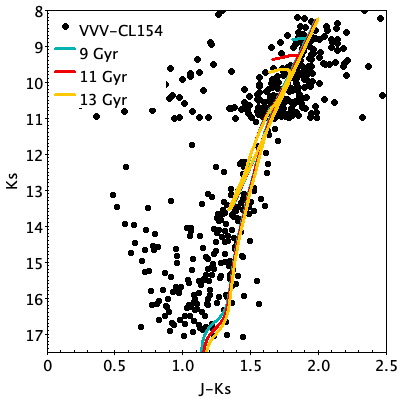} 
\includegraphics[width=5cm, height=5cm]{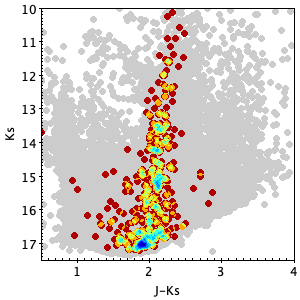} 
\includegraphics[width=5cm, height=5cm]{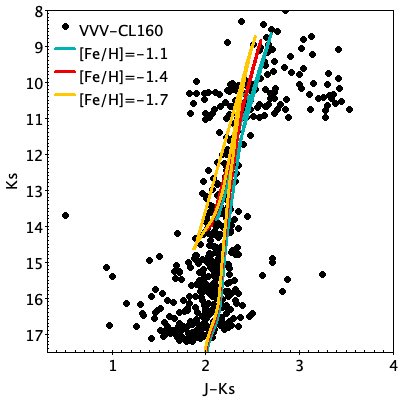} 
\includegraphics[width=5cm, height=5cm]{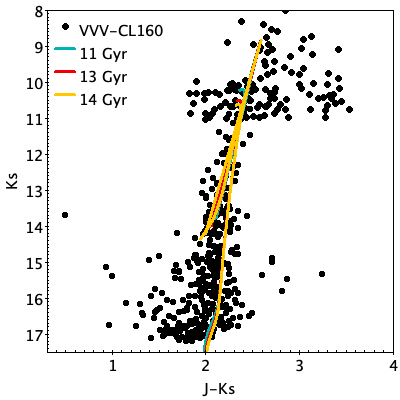} 
\includegraphics[width=5cm, height=5cm]{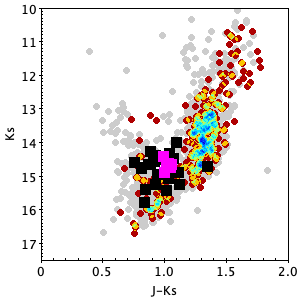} 
\includegraphics[width=5cm, height=5cm]{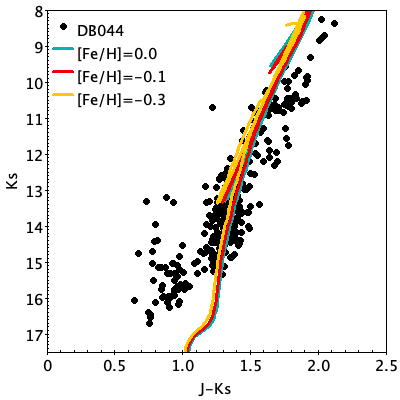} 
\includegraphics[width=5cm, height=5cm]{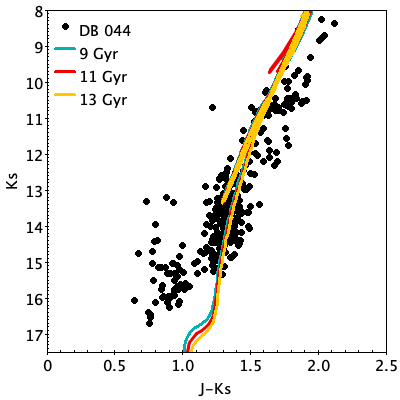} 
\includegraphics[width=5cm, height=5cm]{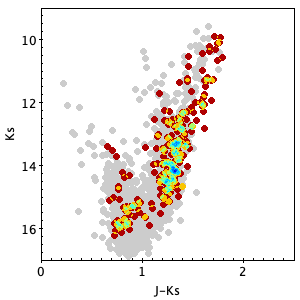} 
\includegraphics[width=5cm, height=5cm]{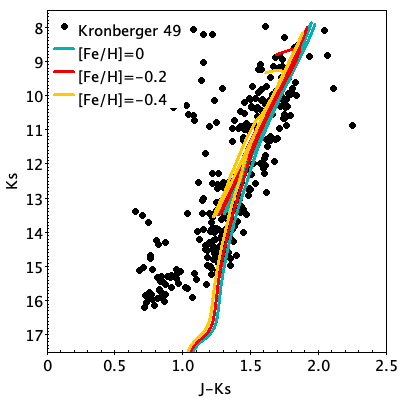} 
\includegraphics[width=5cm, height=5cm]{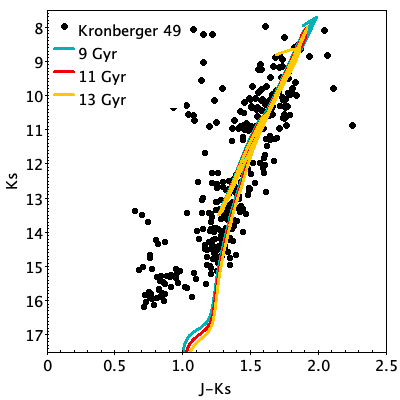} 
\caption{}
\label{lf2}
\end{figure}

\clearpage
\centering
\small
\begin{landscape}
\section{Additional Table}
\begin{longtable}{p{1.5cm}p{4.0cm}p{1.3cm}p{1.3cm}p{1.3cm}p{0.8cm}p{0.8cm}p{0.8cm}p{0.8cm}p{0.8cm}p{1.8cm}p{1.8cm}}
\caption{Properties of RR Lyrae samples in each GC field within $12'$. We highlight with a star symbol all the variables that we consider members of the respective clusters.}
\label{RRL_table}\\
	\hline \hline
	ID &  ID & RA & Dec & P & V & $K_s$& J & $D_{RRLS}$& $D_c$ & $\mu_{\alpha_{\ast}}$ & $\mu_{\delta}$ \\
    Cluster & RR Lyrae & [deg] & [deg] & [day] & [mag] & [mag] & [mag] & [kpc] & [arcmin] &[mas yr$^-1$] &[mas yr$^-1$]  \\
	\hline
	\endfirsthead
	
	\bfseries \tablename \thetable{.continued}\\
	\hline 
	ID &  ID & RA & Dec & P & V & $K_s$& J & $D_{RRLS}$& $D_c$ & $\mu_{\alpha_{\ast}}$ & $\mu_{\delta}$ \\
    Cluster & RR Lyrae & [deg] & [deg] & [day] & [mag] & [mag] & [mag] & [kpc] & [arcmin] &[mas yr$^-1$] &[mas yr$^-1$]  \\
    \hline
    \endhead
    
    \hline \multicolumn{12}{r}{Continue on next page} \\
    \endfoot
    
    \hline\hline
    \endlastfoot
VVVCL154 & OGLE-BLG-RRLYR-31849 &   $268.6284$ &  $-28.0437$ & $0.2957846$ & $ --   $ & $15.08$ &  $16.14$ &  $8.7 $& $ 9.9  $ & $ -6.81 $  &$   -5.119 $ \\
VVVCL154 & OGLE-BLG-RRLYR-31855 &   $268.6331$ &  $-28.0206$ & $0.526845 $ & $ 22.17$ & $14.68$ &  $15.76$ &  $8.4 $& $ 10.2 $ & $ --    $  &$   --     $ \\
VVVCL154 & OGLE-BLG-RRLYR-31867 &   $268.6435$ &  $-28.2037$ & $0.7274215$ & $ 19.49$ & $14.05$ &  $14.82$ &  $7.3 $& $ 10.4 $ & $ -5.545$  &$   -7.643 $ \\
VVVCL154 & OGLE-BLG-RRLYR-31877 &   $268.6523$ &  $-28.2061$ & $0.5223266$ & $ 19.35$ & $14.37$ &  $15.1 $ &  $7.2 $& $ 10.1 $ & $ -1.585$  &$   -5.245 $ \\
VVVCL154 & OGLE-BLG-RRLYR-31898 &   $268.6618$ &  $-28.0111$ & $0.5278617$ & $ --   $ & $14.69$ &  $15.84$ &  $8.4 $& $ 9.0  $ & $ --    $  &$   --     $ \\
VVVCL154 & OGLE-BLG-RRLYR-31902 &   $268.6652$ &  $-28.2087$ & $0.5584574$ & $ 21.98$ & $15.05$ &  $16.01$ &  $10.2 $& $ 9.6  $ & $ -2.194$  &$   -1.874 $ \\
VVVCL154 & OGLE-BLG-RRLYR-31908$\star$ &   $268.6681$ &  $-28.0259$ & $0.6702737$ & $ --   $ & $14.58$ &  $15.89$ &  $9.0 $& $ 8.2  $ & $ 1.636 $  &$   -4.757 $ \\
VVVCL154 & OGLE-BLG-RRLYR-31938 &   $268.6857$ &  $-28.0951$ & $0.6323773$ & $ --   $ & $14.71$ &  $16.17$ &  $9.3 $& $ 5.9  $ & $ --    $  &$   --     $ \\
VVVCL154 & OGLE-BLG-RRLYR-31941 &   $268.6899$ &  $-28.1139$ & $0.2558873$ & $ --   $ & $15.06$ &  $16.26$ &  $8.0 $& $ 5.7  $ & $ 3.04  $  &$   -1.083 $ \\
VVVCL154 & OGLE-BLG-RRLYR-31942$\star$ &   $268.6909$ &  $-27.9855$ & $0.5565757$ & $ 21.9 $ & $14.73$ &  $15.57$ &  $8.8 $& $ 8.8  $ & $ -5.832$  &$   -5.821 $ \\
VVVCL154 & OGLE-BLG-RRLYR-31960 &   $268.7024$ &  $-27.9769$ & $0.2949171$ & $ 20.85$ & $14.89$ &  $15.21$ &  $7.9 $& $ 8.9  $ & $ -2.572$  &$   -3.292 $ \\
VVVCL154 & OGLE-BLG-RRLYR-31962$\star$ &   $268.703 $ &  $-28.0839$ & $0.5279925$ & $ --   $ & $14.84$ &  $16.09$ &  $9.0 $& $ 4.9  $ & $ -3.127$  &$   -6.495 $ \\
VVVCL154 & OGLE-BLG-RRLYR-31966 &   $268.7039$ &  $-28.0552$ & $0.5208234$ & $ --   $ & $14.92$ &  $16.21$ &  $9.3 $& $ 5.5  $ & $ --    $  &$   --     $ \\
VVVCL154 & OGLE-BLG-RRLYR-31967 &   $268.7049$ &  $-28.1295$ & $0.2869678$ & $ 20.4 $ & $14.74$ &  $15.48$ &  $7.3 $& $ 5.0  $ & $ -2.49 $  &$   1.132  $ \\
VVVCL154 & OGLE-BLG-RRLYR-31969 &   $268.7058$ &  $-28.1439$ & $0.484147 $ & $ --   $ & $14.62$ &  $15.55$ &  $7.8$& $ 5.3  $ & $ -5.771$  &$   -4.005 $ \\
VVVCL154 & OGLE-BLG-RRLYR-31971 &   $268.7097$ &  $-28.2149$ & $0.7063366$ & $ 21.37$ & $14.38$ &  $15.34$ &  $8.4 $& $ 8.2  $ & $ 2.047 $  &$   -1.077 $ \\
VVVCL154 & OGLE-BLG-RRLYR-31990$\star$ &   $268.7178$ &  $-28.031 $ & $0.5542389$ & $ --   $ & $14.73$ &  $15.7 $ &  $8.8 $& $ 5.7  $ & $ -5.639$  &$   -8.718 $ \\
VVVCL154 & OGLE-BLG-RRLYR-32004 &   $268.7267$ &  $-28.0066$ & $0.4577286$ & $ 20.48$ & $13.57$ &  $14.59$ &  $4.7 $& $ 6.6  $ & $ -9.049$  &$   -5.778 $ \\
VVVCL154 & OGLE-BLG-RRLYR-32007$\star$ &   $268.7289$ &  $-27.9811$ & $0.2600223$ & $ 21.58$ & $15.26$ &  $16.08$ &  $8.9 $& $ 7.9  $ & $ -4.066$  &$   -9.283 $ \\
VVVCL154 & OGLE-BLG-RRLYR-32024$\star$ &   $268.743 $ &  $-28.1039$ & $0.5293469$ & $ --   $ & $14.77$ &  $16.3 $ &  $8.7$& $ 2.4  $ & $ -2.788$  &$   -4.074 $ \\
VVVCL154 & OGLE-BLG-RRLYR-32029 &   $268.745 $ &  $-28.2024$ & $0.5918788$ & $ --   $ & $15.87$ &  $16.84$ &  $15.3$& $ 6.5  $ & $ --    $  &$   --     $ \\
VVVCL154 & OGLE-BLG-RRLYR-32038 &   $268.7515$ &  $-28.1637$ & $0.5024584$ & $ 19.88$ & $13.9 $ &  $14.7 $ &  $5.7 $& $ 4.3  $ & $ -2.971$  &$   -5.835 $ \\
VVVCL154 & OGLE-BLG-RRLYR-32042$\star$ &   $268.7538$ &  $-28.2099$ & $0.6431962$ & $ 20.42$ & $14.37$ &  $15.21$ &  $8.0 $& $ 6.8  $ & $ -0.482$  &$   -7.594 $ \\
VVVCL154 & OGLE-BLG-RRLYR-32046 &   $268.7578$ &  $-28.2583$ & $0.3334006$ & $ 20.65$ & $14.6 $ &  $15.5 $ &  $7.4 $& $ 9.6  $ & $ 1.244 $  &$   -0.215 $ \\
VVVCL154 & OGLE-BLG-RRLYR-32051 &   $268.7624$ &  $-27.9862$ & $0.6701652$ & $ 22.26$ & $14.39$ &  $15.54$ &  $7.2 $& $ 7.0  $ & $ -3.752$  &$   1.226  $ \\
VVVCL154 & OGLE-BLG-RRLYR-32052$\star$ &   $268.7626$ &  $-28.2543$ & $0.5227533$ & $ 21.4 $ & $14.73$ &  $15.57$ &  $8.3 $& $ 9.3  $ & $ -2.126$  &$   -6.599 $ \\
VVVCL154 & OGLE-BLG-RRLYR-32056 &   $268.7638$ &  $-28.0611$ & $0.2592107$ & $ --   $ & $15.18$ &  $15.93$ &  $8.5 $& $ 2.6  $ & $ --    $  &$   --     $ \\
VVVCL154 & OGLE-BLG-RRLYR-32057 &   $268.7676$ &  $-28.2012$ & $0.6348947$ & $ --   $ & $14.35$ &  $15.25$ &  $8.5$& $ 6.1  $ & $ 1.103 $  &$   -4.045 $\\ 
VVVCL154 & OGLE-BLG-RRLYR-32067$\star$ &   $268.7788$ &  $-27.9733$ & $0.2673053$ & $ 22.28$ & $15.24$ &  $16.05$ &  $8.9 $& $ 7.6  $ & $ -2.387$  &$   -6.92  $\\ 
VVVCL154 & OGLE-BLG-RRLYR-32072$\star$ &   $268.7824$ &  $-27.9565$ & $0.6135911$ & $ 20.81$ & $14.32$ &  $15.18$ &  $7.6 $& $ 8.6  $ & $ -1.541$  &$   -3.611 $ \\
VVVCL154 & OGLE-BLG-RRLYR-32083 &   $268.7906$ &  $-28.2239$ & $0.2812811$ & $ 20.15$ & $14.68$ &  $15.38$ &  $7.1 $& $ 7.4  $ & $ 3.001 $  &$   -3.569 $ \\
VVVCL154 & OGLE-BLG-RRLYR-32085 &   $268.7924$ &  $-28.1722$ & $0.4918871$ & $ 20.28$ & $14.48$ &  $15.36$ &  $7.3 $& $ 4.3  $ & $ -2.983$  &$   -6.053 $ \\
VVVCL154 & OGLE-BLG-RRLYR-32101$\star$ &   $268.8049$ &  $-28.2565$ & $0.5903478$ & $ 19.92$ & $14.52$ &  $15.29$ &  $8.2 $& $ 9.5  $ & $ 0.664 $  &$   -1.73  $ \\
VVVCL154 & OGLE-BLG-RRLYR-32102 &   $268.8051$ &  $-28.1453$ & $0.4806654$ & $ 2.0  $ & $14.48$ &  $15.25$ &  $7.3 $& $ 3.0  $ & $ -7.417$  &$   -9.068 $ \\
VVVCL154 & OGLE-BLG-RRLYR-32105 &   $268.8104$ &  $-28.1677$ & $0.3156331$ & $ 21.68$ & $15.51$ &  $16.31$ &  $10.7 $& $ 4.4  $ & $ -2.697$  &$   -2.515 $ \\
VVVCL154 & OGLE-BLG-RRLYR-32109 &   $268.8143$ &  $-27.9695$ & $0.5341929$ & $ --   $ & $14.59$ &  $15.37$ &  $8.1 $& $ 8.1  $ & $ --    $  &$   --     $ \\
VVVCL154 & OGLE-BLG-RRLYR-32125 &   $268.8239$ &  $-27.943 $ & $0.536193 $ & $ 19.92$ & $14.22$ &  $15.03$ &  $6.8 $& $ 9.7  $ & $ -4.121$  &$   -4.032 $ \\
VVVCL154 & OGLE-BLG-RRLYR-32126 &   $268.8254$ &  $-28.0719$ & $0.2878272$ & $ 21.53$ & $14.86$ &  $15.78$ &  $7.8 $& $ 3.0  $ & $ 4.409 $  &$   -5.655 $ \\
VVVCL154 & OGLE-BLG-RRLYR-32128$\star$ &   $268.8265$ &  $-28.2278$ & $0.5197572$ & $ 20.42$ & $14.64$ &  $15.41$ &  $8.1 $& $ 8.1  $ & $ 2.003 $  &$   -3.581 $ \\
VVVCL154 & OGLE-BLG-RRLYR-32139 &   $268.8335$ &  $-28.2183$ & $0.3100202$ & $ 19.63$ & $14.41$ &  $15.14$ &  $6.5 $& $ 7.7  $ & $ -10.32$  &$   -1.854 $ \\
VVVCL154 & OGLE-BLG-RRLYR-32146 &   $268.8375$ &  $-28.1352$ & $0.5954208$ & $ 19.97$ & $14.23$ &  $15.06$ &  $7.2$& $ 3.9  $ & $ -5.873$  &$   -10.197$ \\
VVVCL154 & OGLE-BLG-RRLYR-32147$\star$ &   $268.8379$ &  $-27.9986$ & $0.7253929$ & $ 21.84$ & $14.74$ &  $15.67$ &  $10.1 $& $ 6.9  $ & $ -2.38 $  &$   -4.776 $ \\
VVVCL154 & OGLE-BLG-RRLYR-32152 &   $268.8392$ &  $-27.9709$ & $0.3874457$ & $ 20.26$ & $14.44$ &  $15.18$ &  $7.4 $& $ 8.5  $ & $ 3.413 $  &$   -6.722 $ \\
VVVCL154 & OGLE-BLG-RRLYR-32161$\star$ &   $268.8439$ &  $-28.1966$ & $0.2575597$ & $ 20.99$ & $15.29$ &  $15.99$ &  $8.9 $& $ 6.8  $ & $ -3.118$  &$   -5.865 $ \\
VVVCL154 & OGLE-BLG-RRLYR-32167 &   $268.8453$ &  $-27.9902$ & $0.3117361$ & $ 20.79$ & $14.93$ &  $15.74$ &  $8.3 $& $ 7.6  $ & $ -0.512$  &$   -9.141 $ \\
VVVCL154 & OGLE-BLG-RRLYR-32171 &   $268.8463$ &  $-28.0338$ & $0.5983338$ & $ 21.97$ & $14.82$ &  $15.82$ &  $9.5 $& $ 5.5  $ & $ -6.168$  &$   -10.743$ \\
VVVCL154 & OGLE-BLG-RRLYR-32185 &   $268.8569$ &  $-27.965 $ & $0.2508808$ & $ 21.09$ & $15.12$ &  $15.94$ &  $8.1 $& $ 9.2  $ & $ -4.886$  &$   -7.319 $ \\
VVVCL154 & OGLE-BLG-RRLYR-32197 &   $268.8652$ &  $-27.9533$ & $0.6042157$ & $ 20.28$ & $14.25$ &  $15.14$ &  $7.4 $& $ 10.1 $ & $ -8.543$  &$   -8.788 $ \\
VVVCL154 & OGLE-BLG-RRLYR-32198 &   $268.8667$ &  $-28.2269$ & $0.2756981$ & $ 20.22$ & $14.88$ &  $15.56$ &  $7.7 $& $ 9.1  $ & $ -2.061$  &$   -8.591 $ \\
VVVCL154 & OGLE-BLG-RRLYR-32212 &   $268.8749$ &  $-27.9691$ & $0.3468166$ & $ 20.43$ & $14.85$ &  $15.6 $ &  $8.5 $& $ 9.6  $ & $ -5.738$  &$   -1.765 $ \\
VVVCL154 & OGLE-BLG-RRLYR-32230 &   $268.8871$ &  $-28.0332$ & $0.3593307$ & $ 20.55$ & $14.67$ &  $15.43$ &  $7.9 $& $ 7.4  $ & $ -3.445$  &$   -11.858$ \\
VVVCL154 & OGLE-BLG-RRLYR-32234 &   $268.8901$ &  $-28.2309$ & $0.5318649$ & $ 20.36$ & $14.56$ &  $15.45$ &  $7.9 $& $ 10.1 $ & $ -5.794$  &$   -5.428 $ \\
VVVCL154 & OGLE-BLG-RRLYR-32238 &   $268.8942$ &  $-28.1385$ & $0.6298719$ & $ 21.01$ & $14.54$ &  $15.45$ &  $8.6 $& $ 7.0  $ & $ -10.81$  &$   -7.612 $ \\
VVVCL154 & OGLE-BLG-RRLYR-32261$\star$ &   $268.9102$ &  $-28.1785$ & $0.6813527$ & $ 21.18$ & $14.57$ &  $15.38$ &  $9.1$& $ 8.9  $ & $ -0.447$  &$   -4.361 $ \\
VVVCL154 & OGLE-BLG-RRLYR-32268 &   $268.9143$ &  $-28.156 $ & $0.3038136$ & $ 20.5 $ & $14.8 $ &  $15.52$ &  $7.8 $& $ 8.5  $ & $ 3.136 $  &$   -2.847 $ \\
VVVCL154 & OGLE-BLG-RRLYR-32283 &   $268.9246$ &  $-28.0078$ & $0.4927651$ & $ 20.4 $ & $13.73$ &  $14.78$ &  $5.2 $& $ 10.1 $ & $ 3.597 $  &$   -9.091 $ \\
VVVCL154 & OGLE-BLG-RRLYR-32297 &   $268.9327$ &  $-28.1388$ & $0.5333822$ & $ 21.27$ & $14.9 $ &  $15.91$ &  $10.0 $& $ 9.3  $ & $ -6.901$  &$   -6.53  $ \\
VVVCL154 & OGLE-BLG-RRLYR-32322 &   $268.9521$ &  $-28.0384$ & $0.4673814$ & $ 20.69$ & $14.47$ &  $15.47$ &  $7.6 $& $ 10.8 $ & $ --    $  &$   --     $ \\
DB044 &  OGLE-BLG-RRLYR-28715 &$266.4744 $&$-24.8868 $&$0.5610397 $&$-- $&$15.07 $&$16.1  $&$9.7  $&$10.3 $&$    -6.753 $&$-1.822$\\
DB044 &  OGLE-BLG-RRLYR-28782 &$266.517  $&$-24.998  $&$0.5207337 $&$21.32  $&$15.43 $&$16.45 $&$11.0 $&$10.0 $&$    -3.266 $&$-5.884$\\
DB044 &  OGLE-BLG-RRLYR-28794 &$266.5275 $&$-24.9285 $&$0.6581612 $&$20.28  $&$14.28 $&$15.17 $&$7.2  $&$7.4  $&$    -4.348 $&$-2.717$\\
DB044 &  OGLE-BLG-RRLYR-28810 &$266.5392 $&$-24.8102 $&$0.810599  $&$21.37  $&$14.01 $&$15.11 $&$7.0  $&$8.0  $&$    -1.376 $&$-7.399$\\
DB044 &  OGLE-BLG-RRLYR-28813 &$266.54   $&$-24.99   $&$0.796195  $&$20.44  $&$14.51 $&$15.45 $&$8.8  $&$8.7  $&$    -7.253 $&$-4.381$\\
DB044 &  OGLE-BLG-RRLYR-28835 &$266.5514 $&$-24.7805 $&$0.469473  $&$21.08  $&$14.41 $&$15.38 $&$6.5  $&$8.7  $&$    -6.122 $&$-5.806$\\
DB044 &  OGLE-BLG-RRLYR-28845${\star}$  &$266.5591 $&$-25.0286 $&$0.6161224 $&$21.04  $&$14.42 $&$15.4  $&$7.5  $&$9.8  $&$    -6.754 $&$-7.669$\\
DB044 &  OGLE-BLG-RRLYR-28857 &$266.5669 $&$-24.9776 $&$0.4712295 $&$20.06  $&$14.39 $&$15.32 $&$6.5  $&$7.0  $&$    -4.568 $&$-1.2  $\\
DB044 &  OGLE-BLG-RRLYR-28870${\star}$ &$266.5759 $&$-24.9575 $&$0.5675825 $&$20.91  $&$14.64 $&$15.67 $&$8.0  $&$5.8  $&$    -1.594 $&$-7.447$\\
DB044 &  OGLE-BLG-RRLYR-28875${\star}$ &$266.5784 $&$-24.9967 $&$0.4820862 $&$20.9   $&$14.91 $&$15.91 $&$8.4  $&$7.5  $&$    -4.073 $&$-7.89 $\\
DB044 &  OGLE-BLG-RRLYR-28876${\star}$ &$266.5787 $&$-24.9653 $&$0.7569609 $&$20.73  $&$14.43 $&$15.42 $&$8.3  $&$6.0  $&$    -2.098 $&$-7.042$\\
DB044 &  OGLE-BLG-RRLYR-28878 &$266.5803 $&$-24.9563 $&$0.5827392 $&$20.11  $&$14.33 $&$15.37 $&$7.0  $&$5.5  $&$    -7.019 $&$-5.054$\\
DB044 &  OGLE-BLG-RRLYR-28889 &$266.5901 $&$-24.9258 $&$0.5984483 $&$20.49  $&$14.65 $&$15.52 $&$8.2  $&$3.9  $&$    -6.95  $&$-5.86 $\\
DB044 &  OGLE-BLG-RRLYR-28893 &$266.5928 $&$-25.0445 $&$0.4890572 $&$20.89  $&$14.57 $&$15.49 $&$7.2  $&$9.7  $&$    -6.943 $&$-3.903$\\
DB044 &  OGLE-BLG-RRLYR-28895 &$266.5934 $&$-25.0144 $&$0.3096242 $&$-99.99 $&$14.74 $&$15.55 $&$6.6  $&$8.0  $&$    -1.793 $&$-9.514$\\
DB044 &  OGLE-BLG-RRLYR-28897 &$266.5938 $&$-24.8594 $&$0.4852231 $&$20.6   $&$14.46 $&$15.37 $&$6.8  $&$3.7  $&$    -4.84  $&$-3.493$\\
DB044 &  OGLE-BLG-RRLYR-28907 &$266.6002 $&$-24.8185 $&$0.2612227 $&$-99.99 $&$15.24 $&$16.17 $&$7.7  $&$5.1  $&$    -6.806 $&$-5.405$\\
DB044 &  OGLE-BLG-RRLYR-28916 &$266.6054 $&$-24.8065 $&$0.4890075 $&$-99.99 $&$14.73 $&$16.08 $&$7.8  $&$5.6  $&$    -4.317 $&$-7.519$\\
DB044 &  OGLE-BLG-RRLYR-28923 &$266.6088 $&$-24.7283 $&$0.2768075 $&$21.53  $&$15.4  $&$16.25 $&$8.6  $&$10.0 $&$    -9.282 $&$-5.984$\\
DB044 &  OGLE-BLG-RRLYR-28937 &$266.6158 $&$-24.8575 $&$0.576656  $&$21.3   $&$14.79 $&$15.6  $&$8.6  $&$2.7  $&$    -1.577 $&$-2.033$\\
DB044 &  OGLE-BLG-RRLYR-28962 &$266.6258 $&$-25.0254 $&$0.2810044 $&$20.51  $&$14.45 $&$15.43 $&$5.6  $&$8.1  $&$    -- $&$--$\\
DB044 &  OGLE-BLG-RRLYR-28970 &$266.6276 $&$-24.7371 $&$0.4247647 $&$-99.99 $&$15.79 $&$16.63 $&$11.8 $&$9.3  $&$    -6.182 $&$-6.7  $\\
DB044 &  OGLE-BLG-RRLYR-28972${\star}$ &$266.6281 $&$-24.8412 $&$0.5042931 $&$20.96  $&$14.76 $&$15.78 $&$8.0 $&$3.2  $&$    -1.122 $&$-4.27 $\\
DB044 &  OGLE-BLG-RRLYR-29003 &$266.6479 $&$-24.9667 $&$0.4983005 $&$20.49  $&$14.61 $&$15.37 $&$7.4  $&$4.5  $&$    -2.112 $&$0.292 $\\
DB044 &  OGLE-BLG-RRLYR-29044 &$266.6738 $&$-24.8928 $&$0.5575422 $&$-99.99 $&$15.24 $&$16.36 $&$10.4 $&$1.7  $&$    -3.544 $&$-10.1 $\\
DB044 &  OGLE-BLG-RRLYR-29085 &$266.7048 $&$-24.7479 $&$0.5266495 $&$21.14  $&$14.65 $&$15.71 $&$7.7  $&$9.3  $&$    -2.186 $&$-4.64 $\\
DB044 &  OGLE-BLG-RRLYR-29101 &$266.7153 $&$-24.7527 $&$0.5680777 $&$21.53  $&$14.89 $&$16.0  $&$8.9  $&$9.3  $&$    -4.58  $&$0.783 $\\
DB044 &  OGLE-BLG-RRLYR-29183 &$266.7604 $&$-25.0117 $&$0.5564618 $&$22.03  $&$14.65 $&$15.69 $&$7.9  $&$10.0 $&$    -4.625 $&$-9.005$\\
DB044 &  OGLE-BLG-RRLYR-29208 &$266.7717 $&$-24.9875 $&$0.2785731 $&$22.06  $&$14.95 $&$15.89 $&$7.0  $&$9.5  $&$    -5.679 $&$-6.981$\\
DB044 &  OGLE-BLG-RRLYR-29221${\star}$ &$266.7792 $&$-24.9384 $&$0.5040744 $&$-99.99 $&$14.76 $&$15.81 $&$8.0  $&$8.5  $&$    -4.669 $&$-7.874$\\
DB044 &  OGLE-BLG-RRLYR-29232 &$266.7872 $&$-24.8763 $&$0.3237754 $&$21.03  $&$14.67 $&$15.56 $&$6.6  $&$8.5  $&$    -3.856 $&$-4.661$\\
DB044 &  OGLE-BLG-RRLYR-29261 &$266.8099 $&$-24.96   $&$0.5964202 $&$21.84  $&$14.48 $&$15.55 $&$7.6  $&$10.7 $&$    -3.333 $&$-4.05 $\\
VVVCL131& OGLE-BLG-RRLYR-25540 &$265.1568$&$ -34.4823 $&$0.4729397$&$ 20.17 $&$15.48 $&$16.21$ &$12.1$&$ 11.1 $&$  -3.886$  &$    -8.546$\\
VVVCL131& OGLE-BLG-RRLYR-25628 &$265.1924$&$ -34.5224 $&$0.4767443$&$ 19.76 $&$15.28 $&$15.92$ &$11.1$&$ 8.2  $&$ -5.057 $  &$   -1.124 $\\
VVVCL131& OGLE-BLG-RRLYR-25655 &$265.2024$&$ -34.6035 $&$0.5628037$&$ 20.67 $&$14.86 $&$15.78$ &$9.9 $&$ 7.4  $&$ -6.032 $  &$   -6.195 $\\
VVVCL131& OGLE-BLG-RRLYR-25672 &$265.2114$&$ -34.6037 $&$0.4413626$&$ 21.34 $&$13.41 $&$14.90$ &$4.5 $&$ 6.9  $&$ -3.129 $  &$   -4.654 $\\
VVVCL131& OGLE-BLG-RRLYR-25782&$265.2531$&$ -34.6631 $&$0.7436588$&$ 18.87 $&$14.14 $&$14.96$ &$8.2 $&$ 7.0  $&$-0.3417 $  &$   -3.671 $\\
VVVCL131& OGLE-BLG-RRLYR-25821${\star}$ &$265.2686$&$ -34.5599 $&$0.2920761$&$ 19.09 $&$15.14 $&$15.71$ &$9.4 $&$ 3.2  $&$ -1.422 $  &$    -4.31 $\\
VVVCL131& OGLE-BLG-RRLYR-25829 &$265.2734$&$ -34.4837 $&$0.4050738$&$ 19.54 $&$15.12 $&$15.78$ &$9.5 $&$ 5.8  $&$ -6.684 $  &$   -3.091 $\\
VVVCL131& OGLE-BLG-RRLYR-25844&$265.2776$&$ -34.5019 $&$0.3265574$&$ 18.95 $&$14.99 $&$15.62$ &$9.3 $&$ 4.7  $&$ -6.927 $  &$   -3.983 $\\
VVVCL131& OGLE-BLG-RRLYR-25850&$265.2807$&$ -34.4797 $&$0.2144367$&$ 19.16 $&$15.02 $&$15.60$ &$7.6 $&$ 5.8  $&$ -3.007 $  &$   -2.984 $\\
VVVCL131& OGLE-BLG-RRLYR-25872 &$265.2895$&$ -34.6289 $&$0.4121736$&$ 19.04 $&$14.73 $&$15.38$ &$8.0 $&$ 4.2  $&$ -5.189 $  &$   -9.796 $\\
VVVCL131& OGLE-BLG-RRLYR-25879 &$265.2913$&$ -34.5464 $&$0.5483412$&$ 18.65 $&$14.44 $&$15.15$ &$8.1 $&$ 2.2  $&$ -5.606 $  &$   -1.293 $\\
VVVCL131& OGLE-BLG-RRLYR-25979 &$265.3297$&$ -34.6453 $&$0.6133653$&$ 18.36 $&$14.27 $&$14.92$ &$7.9 $&$ 4.7  $&$ -7.074 $  &$   -7.687 $\\
VVVCL131& OGLE-BLG-RRLYR-25989 &$265.3362$&$ -34.5353 $&$0.5697337$&$ 18.20 $&$14.38 $&$15.05$ &$8.0 $&$ 2.1  $&$   -7.4 $  &$   -8.295 $\\
VVVCL131& OGLE-BLG-RRLYR-26062 &$265.3587$&$ -34.5604 $&$0.2868744$&$ 18.26 $&$14.88 $&$15.48$ &$8.3 $&$ 2.3  $&$ 0.4802 $  &$   -2.474 $\\
VVVCL131& OGLE-BLG-RRLYR-26148 &$265.3902$&$ -34.5821 $&$0.9635919$&$ 18.21 $&$14.02 $&$14.73$ &$8.9 $&$ 4.3  $&$ -6.718 $  &$  -0.7498 $\\
VVVCL131& OGLE-BLG-RRLYR-26154${\star}$ &$265.3918$&$ -34.5918 $&$0.5227194$&$ 19.05 $&$14.73 $&$15.40$ &$9.0 $&$ 4.5  $&$ -1.806 $  &$   -4.457 $\\
VVVCL131& OGLE-BLG-RRLYR-26323 &$265.4554$&$ -34.5359 $&$0.5784567$&$ 18.23 $&$14.19 $&$14.83$ &$6.7 $&$ 7.4  $&$ -4.292 $  &$   -7.745 $\\
  VVVCL143&  OGLE-BLG-RRLYR-1808 &266.0056 &-33.819 &$  0.3161524 $ &$ 19.4 $  &$ 14.65 $ &$ 15.37$ &$  7.9$  &$  9.9  $  &$    -3.677$ &$  -2.274$ \\
  VVVCL143&  OGLE-BLG-RRLYR-1810 &266.0078 &-33.7866&$  0.6267657 $ &$ 19.12$  &$ 12.98 $ &$ 14.19$ &$  4.5$  &$  9.0  $  &$   --$ &$  --$ \\
  VVVCL143&  OGLE-BLG-RRLYR-1816 &266.0225 &-33.8636&$  0.4897335 $ &$ 20.92$  &$ 15.07 $ &$ 15.9 $ &$  10.3$  &$  10.7 $  &$    0.697 $ &$  -7.379$ \\
  VVVCL143&  OGLE-BLG-RRLYR-1817 &266.0233 &-33.8149&$  0.4998257 $ &$ 18.67$  &$ 12.98 $ &$ 14.05$ &$  4.0$  &$  8.9  $  &$    -1.148$ &$  -2.604$ \\
  VVVCL143&  OGLE-BLG-RRLYR-1840 &266.0564 &-33.7357&$  0.3622305 $ &$ 19.46$  &$ 14.59 $ &$ 15.28$ &$  8.2$  &$  5.6  $  &$    -3.585$ &$  -1.964$ \\
  VVVCL143&  OGLE-BLG-RRLYR-1842 &266.0594 &-33.604 &$  0.5501256 $ &$ 18.82$  &$ 14.48 $ &$ 15.19$ &$  8.3$  &$  9.7  $  &$    -4.875$ &$  -2.943$ \\
  VVVCL143&  OGLE-BLG-RRLYR-1843 &266.0605 &-33.6834&$  0.4862021 $ &$ 19.47$  &$ 14.92 $ &$ 15.52$ &$  9.6$  &$  6.3  $  &$    -0.562$ &$  -9.28 $ \\
  VVVCL143&  OGLE-BLG-RRLYR-1856 &266.0751 &-33.6656&$  0.4861809 $ &$ 19.92$  &$ 14.49 $ &$ 15.2 $ &$  7.9$  &$  6.3  $  &$    -6.761$ &$  -2.469$ \\
  VVVCL143&  OGLE-BLG-RRLYR-1871 &266.0927 &-33.819 &$  0.4876707 $ &$ 19.54$  &$ 14.89 $ &$ 15.6 $ &$  9.5$  &$  5.9  $  &$    -10.98$ &$  -5.435$ \\
  VVVCL143&  OGLE-BLG-RRLYR-1902 &266.1483 &-33.6071&$  0.2530813 $ &$ 18.81$  &$ 14.84 $ &$ 15.42$ &$  7.7$  &$  7.9  $  &$    3.832 $ &$  -5.825$ \\
  VVVCL143&  OGLE-BLG-RRLYR-1921 &266.1687 &-33.7774&$  0.5616153 $ &$ 18.47$  &$ 14.44 $ &$ 15.06$ &$  8.3$  &$  2.6  $  &$    0.135 $ &$  -9.054$ \\
  VVVCL143&  OGLE-BLG-RRLYR-1930 &266.1828 &-33.7096&$  0.2128831 $ &$ 18.72$  &$ 15.32 $ &$ 15.88$ &$  8.8$  &$  2.6  $  &$    -2.31 $ &$  -1.229$  \\
  VVVCL143&  OGLE-BLG-RRLYR-1934 &266.1884 &-33.6093&$  0.5451788 $ &$ 18.28$  &$ 13.1  $ &$ 14.12$ &$  4.4$  &$  8.1  $  &$   --$ &$ --$ \\
  VVVCL143&  OGLE-BLG-RRLYR-1951 &266.212  &-33.6885&$  0.4103589 $ &$ 18.01$  &$ 14.48 $ &$ 14.98$ &$  7.2$  &$  4.8  $  &$    -7.108$ &$  -7.203$ \\
  VVVCL143&  OGLE-BLG-RRLYR-1952 &266.2131 &-33.8177&$  0.5854312 $ &$ 18.23$  &$ 14.25 $ &$ 14.9 $ &$  7.7$  &$  6.1  $  &$    --$ &$  --$ \\
  VVVCL143&  OGLE-BLG-RRLYR-1953$\star$ &266.2132 &-33.8861&$  0.5645519 $ &$ 18.33$  &$ 14.6  $ &$ 15.29$ &$  8.9$  &$  9.6  $  &$    -2.369$ &$  -5.567$  \\
  VVVCL143&  OGLE-BLG-RRLYR-1981 &266.2543 &-33.8232&$  0.266701  $ &$ 18.98$  &$ 15.74 $ &$ 16.24$ &$  12.0$&$  8.1  $  &$    -5.475$ &$  -7.227$ \\
  VVVCL143&  OGLE-BLG-RRLYR-2019$\star$ &266.2908 &-33.7593&$  0.5942394 $ &$ 18.32$  &$ 14.22 $ &$ 14.89$ &$  7.7$  &$  8.5  $  &$    -0.392$ &$  -6.068$ \\
  VVVCL143&  OGLE-BLG-RRLYR-2033 &266.3029 &-33.8348&$  0.3189595 $ &$ 18.39$  &$ 14.71 $ &$ 15.3 $ &$  8.2$  &$  10.8 $  &$    0.363 $ &$  -8.91 $ \\
  VVVCL143&  OGLE-BLG-RRLYR-2083 &266.3386 &-33.6841&$  0.6098489 $ &$ 18.09$  &$ 14.32 $ &$ 14.96$ &$  8.2$  &$  11.8 $  &$    -3.406$ &$  -7.908$ \\
  VVVCL143&  OGLE-BLG-RRLYR-28294$\star$&266.2448 &-33.7902&$  0.2950622 $ &$ 18.77$  &$ 15.18 $ &$ 15.71$ &$  9.7$  &$  6.5  $  &$    -0.488$ &$  -3.132$  \\
  VVVCL143&  OGLE-BLG-RRLYR-28382$\star$&266.2947 &-33.7378&$  0.8623781 $ &$ 17.83$  &$ 14.09 $ &$ 14.75$ &$  8.8$  &$  8.7  $  &$    -7.268$ &$  -2.622$ \\
FSR1775 & OGLE-BLG-RRLYR-07563$\star$ &$  268.9155$ &$  -36.5419$ & $ 0.6295925$ & $ 17.12$ & $ 14.62$ &  $15.31 $& $9.9$ & $6.6 $ & $-2.367  $&$ -4.914 $\\
FSR1775 & OGLE-BLG-RRLYR-07686 &$  268.9768$ &$  -36.6496$ & $ 0.3468452$ & $ 16.44$ & $ 14.26$ &  $14.73 $& $7.2$ & $5.7 $ & $-2.295  $&$ -0.841 $\\
FSR1775 & OGLE-BLG-RRLYR-07854 &$  269.0391$ &$  -36.6278$ & $ 0.4896602$ & $ 16.51$ & $ 14.41$ &  $14.77 $& $7.9$ & $3.9 $ & $-0.138  $&$ -3.219 $\\
FSR1775 & OGLE-BLG-RRLYR-07909$\star$ &$  269.0592$ &$  -36.6091$ & $ 0.2832088$ & $ 16.84$ & $ 14.92$ &  $15.39 $& $8.8$ & $3.4 $ & $-2.154  $&$ -5.334 $\\
FSR1775 & OGLE-BLG-RRLYR-07941 &$  269.0744$ &$  -36.4699$ & $ 0.3388444$ & $ 16.17$ & $ 13.90$ &  $14.59 $& $6.0$ & $6.6 $ & $-0.630  $&$ -12.885$ \\
FSR1775 & OGLE-BLG-RRLYR-08075 &$  269.1244$ &$  -36.6262$ & $ 0.5052062$ & $ 16.92$ & $ 14.66$ &  $15.39 $& $9.0$ & $7.1 $ & $-5.536  $&$  2.243 $ \\
FSR1775 & OGLE-BLG-RRLYR-08080 &$  269.1257$ &$  -36.4505$ & $ 0.5689601$ & $ 17.30$ & $ 14.63$ &  $15.35 $& $9.4$ & $ 9.3$ & $ -0.821 $&$  -5.112$ \\
FSR1775 & OGLE-BLG-RRLYR-32595 &$  269.2184$ &$  -36.5455$ & $ 0.5800306$ & $ 16.30$ & $ 14.17$ &  $14.84 $& $7.7$ & $11.8$ & $ -3.126 $&$  -1.385$  \\
FSR1767 &OGLE-BLG-RRLYR-21653&$ 263.7614$ &$ -36.3636$ & $ 0.5382471$ & $ 19.26$ & $  14.35$ &  $ 15.21$ & $6.7 $ & $10.1 $ & $-3.061$ & $ -4.352  $\\
FSR1767 &OGLE-BLG-RRLYR-22035$\star$&$ 263.9167$ &$ -36.3381$ & $ 0.6364407$ & $ 19.64$ & $  14.80$ &  $ 15.50$ & $9.0 $ & $ 1.4 $ & $-2.334$ & $ -5.879  $\\
FSR1767 &OGLE-BLG-RRLYR-22054&$ 263.9221$ &$ -36.2895$ & $ 0.5399269$ & $ 20.15$ & $  14.45$ &  $ 15.20$ & $7.0 $ & $4.1  $ & $ 4.402$ & $ -10.080 $\\
FSR1767 &OGLE-BLG-RRLYR-22325&$ 264.0311$ &$ -36.4791$ & $ 0.2938189$ & $ 19.34$ & $  14.48$ &  $ 15.25$ & $6.1 $ & $9.5  $ & $-4.658$ & $ -2.448  $\\
FSR1767 &OGLE-BLG-RRLYR-22342$\star$&$ 264.0370$ &$ -36.3327$ & $ 0.6084582$ & $ 20.43$ & $  15.02$ &  $ 15.77$ & $9.7$ & $6.6  $ & $-2.221$ & $ -7.208  $ \\
FSR1767 &OGLE-BLG-RRLYR-22385&$ 264.0524$ &$ -36.4410$ & $ 0.6041570$ & $ 19.45$ & $  14.71$ &  $ 15.43$ & $8.4 $ & $8.9  $ & $-3.999$ & $ -7.417  $ \\
FSR1767 &OGLE-BLG-RRLYR-22500&$ 264.0981$ &$ -36.3759$ & $ 0.3061927$ & $ 19.33$ & $  14.79$ &  $ 15.41$ & $7.2 $ & $10.2 $ & $-0.119$ & $ -6.111  $ \\
ESO393-12 & OGLE-BLG-RRLYR-24232& $ 264.7043$&$ -35.6632$&$ 0.2833770$&$ 16.46$&$ 14.52$&$ 15.20$&$ 7.0 $&$ 3.0 $&$-8.679 $&$-3.248$\\
ESO393-12 & OGLE-BLG-RRLYR-24016& $ 264.6437$&$ -35.6989$&$ 0.5980570$&$ 17.68$&$ 15.46$&$ 16.22$&$ 13.6$&$ 3.0 $&$-3.970 $&$-5.946$\\
ESO393-12 & OGLE-BLG-RRLYR-23892$\star$& $ 264.6068$&$ -35.6957$&$ 0.6708310$&$ 16.63$&$ 14.26$&$ 15.00$&$ 8.3 $&$ 4.0 $&$-3.258 $&$-6.625$\\
ESO393-12 & OGLE-BLG-RRLYR-24326& $ 264.7361$&$ -35.6510$&$ 0.8650970$&$ 15.74$&$ 13.60$&$ 14.35$&$ 6.9 $&$ 4.8 $&$-2.407 $&$-9.103$\\
ESO393-12 & OGLE-BLG-RRLYR-23793& $ 264.5673$&$ -35.6081$&$ 0.2892410$&$ 17.89$&$ 15.43$&$ 16.14$&$ 10.7 $&$ 5.9 $&$-6.695 $&$-6.716$ \\
ESO393-12 & OGLE-BLG-RRLYR-24380& $ 264.7588$&$ -35.6008$&$ 0.4282180$&$ 15.90$&$ 14.10$&$ 14.78$&$ 6.1 $&$ 6.8 $&$-3.239 $&$ 2.426$ \\
ESO393-12 & OGLE-BLG-RRLYR-24366& $ 264.7513$&$ -35.7116$&$ 0.6268510$&$ 17.11$&$ 14.57$&$ 15.51$&$ 9.2 $&$ 6.8 $&$--     $&$   -- $ \\
ESO393-12 & OGLE-BLG-RRLYR-24146& $ 264.6791$&$ -35.7649$&$ 0.7082520$&$ 17.13$&$ 14.83$&$ 15.64$&$ 11.0$&$ 7.0 $&$0.839  $&$ -6.124$  \\
ESO393-12 & OGLE-BLG-RRLYR-23808$\star$& $ 264.5715$&$ -35.5553$&$ 0.5479670$&$ 16.43$&$ 14.33$&$ 15.08$&$ 7.7 $&$ 7.7 $&$-3.690 $&$ -6.456$\\
ESO393-12 & OGLE-BLG-RRLYR-24391& $ 264.7647$&$ -35.7579$&$ 0.7508770$&$ 16.93$&$ 14.40$&$ 15.20$&$ 9.3 $&$ 9.1 $&$-1.188 $&$ -4.613$\\
ESO393-12 & OGLE-BLG-RRLYR-23846$\star$& $ 264.5873$&$ -35.5193$&$ 0.3340570$&$ 16.82$&$ 14.58$&$ 15.27$&$ 7.8 $&$ 8.9 $&$-0.903 $&$ -5.977$\\
ESO393-12 & OGLE-BLG-RRLYR-23596& $ 264.4851$&$ -35.6136$&$ 0.5379749$&$ 16.86$&$ 14.50$&$ 15.40$&$ 8.2$&$ 10.5$&$ -1.604$&$ 0.090 $\\
ESO393-12 & OGLE-BLG-RRLYR-23789& $ 264.5656$&$ -35.7756$&$ 0.6234820$&$ 16.89$&$ 14.57$&$ 15.36$&$ 9.2 $&$ 9.3 $&$-9.734 $&$ -5.777$ \\
ESO393-12 & OGLE-BLG-RRLYR-23717& $ 264.5355$&$ -35.5366$&$ 0.6613420$&$ 16.13$&$ 13.93$&$ 14.73$&$ 7.0 $&$ 10.0$&$ -10.553$&$ -2.703$ \\
ESO393-12 & OGLE-BLG-RRLYR-23580& $ 264.4784$&$ -35.5976$&$ 0.5232100$&$ 17.72$&$ 14.88$&$ 15.81$&$ 9.7 $&$ 11.2$&$ -8.638$&$  -8.047$ \\
ESO456-09 &  OGLE-BLG-RRLYR-6584  &$268.506 $&$ -32.469  $&$0.300182$&$ 17.149 $&$14.28 $&$14.72   $ &$5.8  $&$1.79  $&$   -3.947 $&$ -18.124  $\\
ESO456-09 &  OGLE-BLG156.7-119095 &$268.4357$&$ -32.4728 $&$0.315771$&$ 17.861 $&$14.57 $&$15.15   $ &$6.8  $&$2.47  $&$   -5.932 $&$ -1.305   $\\
ESO456-09 &  OGLE-BLG156.7-63639  &$268.4594$&$ -32.5136 $&$0.311188$&$ 17.748 $&$14.67 $&$15.23   $ &$7.1  $&$3.02  $&$   -2.437 $&$ -9.79    $\\
ESO456-09 &  OGLE-BLG-RRLYR-6416  &$268.4456$&$ -32.5115 $&$0.278815$&$ 18.016 $&$14.98 $&$15.47   $ &$7.8  $&$3.29  $&$   -6.086 $&$ -10.437  $\\
ESO456-09 &  OGLE-BLG-RRLYR-6615$\star$  &$268.5186$&$ -32.5158 $&$0.610167$&$ 17.915 $&$14.32 $&$14.99   $ &$7.8  $&$3.92  $&$   -0.695 $&$ -5.337   $\\
ESO456-09 &  OGLE-BLG-RRLYR-6298$\star$  &$268.4025$&$ -32.5029 $&$0.316274$&$ 17.996 $&$14.55 $&$15.25   $ &$6.8  $&$4.95  $&$   -5.133 $&$ -6.141   $\\
ESO456-09 &  OGLE-BLG-RRLYR-6695$\star$  &$268.5483$&$ -32.5099 $&$0.351545$&$ 17.584 $&$14.5  $&$15.05   $ &$7.0  $&$5.06  $&$   -4.435 $&$ -2.27    $\\
ESO456-09 &  OGLE-BLG-RRLYR-6751  &$268.57  $&$ -32.4531 $&$0.622606$&$ 17.963 $&$14.48 $&$15.12   $ &$8.5  $&$5.68  $&$   0.919  $&$ -8.021   $\\
ESO456-09 &  OGLE-BLG-RRLYR-6490$\star$  &$268.4704$&$ -32.5564 $&$0.571457$&$ 17.714 $&$14.35 $&$15.03   $ &$7.7  $&$5.43  $&$   -3.512 $&$ -4.738   $\\
ESO456-09 &  OGLE-BLG155.2-118495$\star$ &$268.3918$&$ -32.5255 $&$0.540373$&$ 18.036 $&$14.54 $&$15.11   $ &$8.2  $&$6.2   $&$   -4.244 $&$ -4.546   $\\
ESO456-09 &  OGLE-BLG-RRLYR-6728  &$268.5599$&$ -32.5266 $&$0.506702$&$ 17.735 $&$14.41 $&$14.9    $ &$7.23   $&$7.5  $&$   -0.407 $&$ 2.563    $\\
ESO456-09 &  OGLE-BLG-RRLYR-6411$\star$  &$268.4411$&$ -32.5591 $&$0.564064$&$ 17.837 $&$14.37 $&$15.04   $ &$7.7  $&$5.96  $&$   -6.211 $&$ -3.367   $\\
ESO456-09 &  OGLE-BLG-RRLYR-6252  &$268.3881$&$ -32.5391 $&$0.304918$&$ 18.609 $&$15.1  $&$15.69   $ &$8.6  $&$6.87  $&$   -4.113 $&$ -1.387   $\\
ESO456-09 &  OGLE-BLG-RRLYR-6837  &$268.6056$&$ -32.4733 $&$0.567208$&$ 17.551 $&$14.24 $&$14.85   $ &$7.3  $&$7.77  $&$   -6.73  $&$ -3.89    $\\
ESO456-09 &  OGLE-BLG-RRLYR-6150  &$268.3549$&$ -32.5128 $&$0.38559 $&$ 17.908 $&$14.51 $&$15.16   $ &$7.3  $&$7.81  $&$   -6.775 $&$ -5.161   $\\
ESO456-09 &  OGLE-BLG156.6-83017&$268.5593$&$ -32.3778 $&$0.552816$&$ 18.354 $&$14.69 $&$15.35   $ &$8.9 $&$7.27  $&$   -2.005 $&$ -5.778   $\\
ESO456-09 &  OGLE-BLG-RRLYR-6167  &$268.3616$&$ -32.4072 $&$0.359044$&$ 18.238 $&$14.4  $&$14.95   $ &$6.7  $&$7.74  $&$   4.06   $&$ -8.893   $\\
ESO456-09 &  OGLE-BLG156.6-82844  &$268.5573$&$ -32.3724 $&$0.604871$&$ 17.76  $&$14.12 $&$14.99   $ &$7.5 $&$7.43  $&$   -7.102 $&$ -9.897   $\\
ESO456-09 &  OGLE-BLG-RRLYR-6132  &$268.3475$&$ -32.5136 $&$0.425433$&$ 17.97  $&$14.54 $&$--$ &$7.1 $&$8.23  $&$   -4.151 $&$ -2.051   $\\
ESO456-09 &  OGLE-BLG-RRLYR-6858  &$268.615 $&$ -32.4994 $&$0.362771$&$ 18.539 $&$15.44 $&$15.89   $ &$10.9 $&$8.56  $&$   -1.862 $&$ -7.522   $\\
ESO456-09 &  OGLE-BLG-RRLYR-6129  &$268.3473$&$ -32.5331 $&$0.453186$&$ 17.684 $&$14.66 $&$15.27   $ &$7.9  $&$8.72  $&$   -3.085 $&$ -10.76   $\\
ESO456-09 &  OGLE-BLG155.3-115521 &$268.3891$&$ -32.3584 $&$0.586622$&$ 17.97  $&$13.76 $&$14.58   $ &$5.9  $&$8.32  $&$   -1.075 $&$ -6.056   $\\
ESO456-09 &  OGLE-BLG-RRLYR-6909  &$268.6336$&$ -32.4844 $&$0.373116$&$ 18.085 $&$15.0  $&$15.49   $ &$9.0 $&$9.5   $&$   -0.379 $&$ -9.96    $\\
ESO456-09 &  OGLE-BLG-RRLYR-6876  &$268.6214$&$ -32.5274 $&$0.450459$&$ 17.195 $&$14.2  $&$14.7    $ &$6.4  $&$9.45  $&$   -2.977 $&$ -7.047   $\\
ESO456-09 &  OGLE-BLG-RRLYR-6946  &$268.6491$&$ -32.4545 $&$0.333187$&$ 18.153 $&$14.5  $&$15.12   $ &$6.8  $&$10.39 $&$   -3.125 $&$ -6.396   $\\
ESO456-09 &  OGLE-BLG-RRLYR-6037  &$268.3141$&$ -32.5283 $&$0.552234$&$ 19.01  $&$15.15 $&$15.72   $ &$11.0 $&$10.42 $&$   -3.69  $&$ -7.714   $\\
ESO456-09 &  OGLE-BLG-RRLYR-6770  &$268.5753$&$ -32.3411 $&$0.328324$&$ 18.026 $&$14.6  $&$15.24   $ &$7.1$&$9.57  $&$   -4.925 $&$ -6.035   $\\
ESO456-09 &  OGLE-BLG-RRLYR-6191  &$268.3691$&$ -32.3428 $&$0.612561$&$ 19.268 $&$14.59 $&$15.47   $ &$8.9  $&$9.81  $&$   -2.602 $&$ -10.055  $\\
ESO456-09 &  OGLE-BLG-RRLYR-6665  &$268.5381$&$ -32.3206 $&$0.44545 $&$ 18.322 $&$14.51 $&$15.3    $ &$7.4$&$9.49  $&$   3.671  $&$ -10.405  $\\
ESO456-09 &  OGLE-BLG156.7-220529 &$268.6605$&$ -32.4684 $&$0.567581$&$ 17.512 $&$14.3  $&$14.93   $ &$7.5  $&$11.05 $&$   -2.41  $&$ -4.895   $\\
ESO456-09 &  OGLE-BLG-RRLYR-6176  &$268.3644$&$ -32.3409 $&$0.523935$&$ 19.07  $&$14.81 $&$15.77   $ &$9.1  $&$10.08 $&$   -6.319 $&$ -5.821   $\\
Kronberger49 &  OGLE-BLG-RRLYR-34383 & $272.6635$ &$-23.2415$ &$0.5003454$ &$21.59 $& $15.76$&$ 16.9  $&$13.6$&$ 7.1$   &$   -2.327 $&$  -5.113   $\\
 Kronberger49 &  OGLE-BLG-RRLYR-34323 & $272.5573$ &$-23.2705$ &$0.7306612$ &$20.87 $& $14.33$&$ 15.39 $&$8.5 $&$ 4.9$   &$   -8.198 $&$  -10.083  $\\
 Kronberger49 &  OGLE-BLG-RRLYR-34315 & $272.5391$ &$-23.2814$ &$0.3609994$ &$20.0  $& $13.85$&$ 14.8  $&$5.5 $&$ 5.1$   &$   -7.674 $&$  -4.365   $\\
 Kronberger49 &  OGLE-BLG-RRLYR-34337$\star$ & $272.5835$ &$-23.3502$ &$0.5724149$ &$20.51 $& $14.51$&$ 15.39 $&$8.2 $&$ 1.1$   &$   -7.552 $&$  -7.456   $\\
 Kronberger49 &  OGLE-BLG-RRLYR-34377 & $272.6478$ &$-23.4044$ &$0.5973248$ &$20.3  $& $14.23$&$ 15.19 $&$7.3 $&$ 4.8$   &$   -5.479 $&$  -18.806  $\\
 Kronberger49 &  OGLE-BLG-RRLYR-34382 & $272.6582$ &$-23.3706$ &$0.6176361$ &$20.48 $& $14.7 $&$ 15.66 $&$9.3$&$ 4.0$   &$   -6.111 $&$  -9.491   $\\
 Kronberger49 &  OGLE-BLG-RRLYR-34334$\star$ & $272.5812$ &$-23.4668$ &$0.6591129$ &$21.4  $& $14.33$&$ 15.34 $&$8.1$&$ 7.7$   &$   -1.381 $&$  -2.298   $\\
 Kronberger49 &  OGLE-BLG-RRLYR-34340 & $272.5896$ &$-23.5045$ &$0.50217  $ &$--$    & $14.67$&$ 15.89 $&$8.2 $&$ 9.9$   &$   --     $&$   --      $\\
 Kronberger49 &  OGLE-BLG-RRLYR-34416$\star$ & $272.7239$ &$-23.429 $ &$0.2826901$ &$20.92 $& $15.04$&$ 15.96 $&$8.5 $&$ 9.2$   &$   -5.338 $&$  -7.442   $\\                                                                                                  
\end{longtable}
\end{landscape}

\end{appendix}

\end{document}